\documentstyle[10pt,epsf,epsfig,dp_delphititle,oldlfont,lineno]{dp_delphi}
\makeindex
\def\DpPaperGroup{EP}
\def\DpPaperRef{2001-087}
\def\DpDate{16-November-2001}
\def\DpAuthors{DELPHI Collaboration}
\def\DpSubmit{(Accepted by Eur.Phys.J.C)}
\def\DpTitle{{\bf Searches for neutral Higgs bosons in \\
       \boldmath ${\mathrm e}^+{\mathrm e}^-$ collisions from \\
        $\sqrt{s}$~=~191.6 to 201.7~GeV}}
\def\DpComment{ }
\def\DpEMail{ }
\newcommand{\Zz} {\mbox{Z}}

\newcommand{\W} {\mbox{$ {\mathrm W}^{\pm} \,$}}
\newcommand{\Hz} {\mbox{H}}

\newcommand{\tol}{\mbox{$\tau$ }}

\newcommand{\MA} {\mbox{$ m_{\mathrm A} $}}

\newcommand{\MH} {\mbox{$ m_{\mathrm H} \, $}}

\newcommand{\mh} {\mbox{$ m_{\mathrm h} $}}

\newcommand{\mtop} {\mbox{$ m_{\mathrm top} $}}
\newcommand{\ee}{\mbox{${\mathrm e}^+{\mathrm e}^-$}}

\newcommand{\qqbar}{\mbox{${\mathrm q}\bar{\mathrm q}$} }
\newcommand{\bbbar}{\mbox{${\mathrm b}\bar{\mathrm b}$} }
\newcommand{\ccbar}{\mbox{${\mathrm c}\bar{\mathrm c}$} }

\newcommand{\toto}{\mbox{$\tau^+ \tau^-$}}

\newcommand{\xb}{\mbox{$x_{\mathrm b}$}}
\newcommand{\xbi}{\mbox{$x_{\mathrm b}^{i}$}}
\newcommand{\hmm}{\mbox{\Hz$ \mu^+ \mu^-$}}
\newcommand{\hee}{\mbox{\Hz${\mathrm {e^+ e^-}}$}}
\newcommand{\hnn}{\mbox{\Hz$ \nu \bar{\nu}$}}
\newcommand{\hqq}{$\Hz{\mathrm {q \bar{q}}}$ }

\newcommand{\ttqq}{$\tau^+\tau^- {\mathrm q \bar{q}}$ }

\newcommand{\Abb}{A$\rightarrow $\bbbar}
\newcommand{\Acc}{A$\rightarrow $\ccbar}
\newcommand{\hAA} {${\mathrm{h}}\rightarrow {\mathrm{AA}}$}
\newcommand{\Zqq} {${\mathrm{Z}}\rightarrow$ \qqbar}

\newcommand{\qqg} {\mbox{$ {\mathrm q}\bar{\mathrm q}(\gamma) $}}

\newcommand{\gaga}{\mbox{$\gamma \gamma$ }}

\newcommand{\WW} {\mbox{${\mathrm W}^+{\mathrm W}^-$}}

\newcommand{\ZZ} {${\mathrm {ZZ}}$}
\newcommand{\ZH} {\mbox{${\mathrm {HZ}}$}}
\newcommand{\hA} {\mbox{$ {\mathrm h} {\mathrm A}$}}
\newcommand{\hZ} {\mbox{$ {\mathrm h} {\mathrm Z}$}}

\newcommand{\eeqq} {\mbox{\ee \qqbar }}
\newcommand{\mmqq} {\mbox{$\mu^+ \mu^- $\qqbar }}
\newcommand{\nnqq} {\mbox{\qqbar$\nu \bar{\nu}$}}
\newcommand{\tautauqq}{\mbox{$\tau^+ \tau^- $\qqbar }}

\newcommand{\llqq} {\mbox{${\ell^+\ell^- }$\qqbar }}

\newcommand{\tauvqq} {\mbox{$\tau\nu$${\mathrm q'}\bar{\mathrm q}$ }}

\newcommand{\tbeta} {\mbox{$\tan \beta$}}

\newcommand{\MeVc} {\mbox{${\mathrm{MeV}}/c $}}

\newcommand{\GeV} {\mbox{${\mathrm{GeV}} $}}
\newcommand{\GeVc} {\mbox{${\mathrm{GeV}}/c $}}
\newcommand{\GeVcc} {\mbox{${\mathrm{GeV}}/c^2 $}}

\newcommand{\dgree} {\mbox{$^{\circ}$}}
\newcommand{\mydeg} {$^{\circ}$}

\newcommand{\pbinv} {\mbox{pb$^{-1}$}}

\newcommand{\sqrts }{\mbox{$ \sqrt{s} \,$}}
\newcommand{\like}{\mbox{$\cal L$}}
\newcommand{\likear}{\mbox{$\cal Q$}}

\newcommand{\zee} {\mbox{$ {{\mathrm Ze}}^+{{\mathrm e}}^- $}}

\newcommand{\rs}{\mbox{$\sqrt{s}$}}

\begin{document}
%%%%%%%%%%%%%%%%%%%%%%%%%% They are a problem with Coll.Sty ?
\makeatletter
%\input{dp_system:coll.sty}
% Collapse citation numbers to ranges.  Non-numeric and undefined labels
% are handled.  No sorting is done.  E.g., 1,3,2,3,4,5,foo,1,2,3,?,4,5
% gives 1,3,2-5,foo,1-3,?,4,5
\newcount\@tempcntc
\def\@citex[#1]#2{\if@filesw\immediate\write\@auxout{\string\citation{#2}}\fi
  \@tempcnta\z@\@tempcntb\m@ne\def\@citea{}\@cite{\@for\@citeb:=#2\do
    {\@ifundefined
       {b@\@citeb}{\@citeo\@tempcntb\m@ne\@citea\def\@citea{,}{\bf ?}\@warning
       {Citation `\@citeb' on page \thepage \space undefined}}%
    {\setbox\z@\hbox{\global\@tempcntc0\csname b@\@citeb\endcsname\relax}%
     \ifnum\@tempcntc=\z@ \@citeo\@tempcntb\m@ne
       \@citea\def\@citea{,}\hbox{\csname b@\@citeb\endcsname}%
     \else
      \advance\@tempcntb\@ne
      \ifnum\@tempcntb=\@tempcntc
      \else\advance\@tempcntb\m@ne\@citeo
      \@tempcnta\@tempcntc\@tempcntb\@tempcntc\fi\fi}}\@citeo}{#1}}
\def\@citeo{\ifnum\@tempcnta>\@tempcntb\else\@citea\def\@citea{,}%
  \ifnum\@tempcnta=\@tempcntb\the\@tempcnta\else
   {\advance\@tempcnta\@ne\ifnum\@tempcnta=\@tempcntb \else \def\@citea{--}\fi
    \advance\@tempcnta\m@ne\the\@tempcnta\@citea\the\@tempcntb}\fi\fi}
 
\makeatother
%%%%%%%%%%%%%%%%%%%%%%%%%% ??????????????????????????????????
% Generate the title page
\begin{titlepage}
\pagenumbering{roman}
\CERNpreprint{\DpPaperGroup}{\DpPaperRef} % Reference of the paper
\date{{\small\DpDate}} % Date of the paper
\title{\DpTitle} % Title of the paper
\address{\DpAuthors} % General name of the author(s)
\begin{shortabs} % Start the abstract
\noindent
%===================> Abstract     =====> To be filled <=====%
Neutral Higgs bosons of the Standard Model ({\sc SM}) and the 
Minimal Supersymmetric Standard Model ({\sc MSSM}) 
were searched for in the data collected in 1999 by the 
DELPHI experiment at centre-of-mass energies between 
191.6 and 201.7~GeV with a total integrated luminosity of 
228 pb$^{-1}$. 
These analyses, in combination with our results at lower 
energies, set 95\% confidence level lower mass bounds on the 
Standard Model Higgs boson (107.3~GeV/$c^2$) and on the lightest neutral 
scalar (85.9~GeV/$c^2$) and neutral pseudoscalar (86.5~GeV/$c^2$) Higgs
bosons in representative scans of the {\sc MSSM} parameter space.
% excluding  $\tan\beta$ in the range [0.9-1.6]
An extended scan of the {\sc MSSM} parameter space was also performed to test
the robustness of these limits.

\end{shortabs}
\vfill
\begin{center}
\DpSubmit \ \\ % Horrible hack to allow to have DpSubmit empty
\DpComment \ \\
\DpEMail \ \\
\end{center}
\vfill
\clearpage
\headsep 10.0pt
\addtolength{\textheight}{10mm}
\addtolength{\footskip}{-5mm}
\begingroup
% Commands to process the author names
%
\newcommand{\DpName}[2]{\hbox{#1$^{\ref{#2}}$},\hfill}
\newcommand{\DpNameTwo}[3]{\hbox{#1$^{\ref{#2},\ref{#3}}$},\hfill}
\newcommand{\DpNameThree}[4]{\hbox{#1$^{\ref{#2},\ref{#3},\ref{#4}}$},\hfill}
\newskip\Bigfill \Bigfill = 0pt plus 1000fill
\newcommand{\DpNameLast}[2]{\hbox{#1$^{\ref{#2}}$}\hspace{\Bigfill}}
%
%\small
\footnotesize
\noindent
\DpName{J.Abdallah}{LPNHE}
\DpName{P.Abreu}{LIP}
\DpName{W.Adam}{VIENNA}
\DpName{P.Adzic}{DEMOKRITOS}
\DpName{T.Albrecht}{KARLSRUHE}
\DpName{T.Alderweireld}{AIM}
\DpName{R.Alemany-Fernandez}{CERN}
\DpName{T.Allmendinger}{KARLSRUHE}
\DpName{P.P.Allport}{LIVERPOOL}
\DpName{S.Almehed}{LUND}
% ...
\normalsize
\endgroup
\titlefoot{Department of Physics and Astronomy, Iowa State
     University, Ames IA 50011-3160, USA
    \label{AMES}}
\titlefoot{Physics Department, Universiteit Antwerpen,
     Universiteitsplein 1, B-2610 Antwerpen, Belgium \\
     \indent~~and IIHE, ULB-VUB,
     Pleinlaan 2, B-1050 Brussels, Belgium \\
     \indent~~and Facult\'e des Sciences,
     Univ. de l'Etat Mons, Av. Maistriau 19, B-7000 Mons, Belgium
    \label{AIM}}
\titlefoot{Physics Laboratory, University of Athens, Solonos Str.
     104, GR-10680 Athens, Greece
    \label{ATHENS}}
\titlefoot{Department of Physics, University of Bergen,
     All\'egaten 55, NO-5007 Bergen, Norway
    \label{BERGEN}}
\titlefoot{Dipartimento di Fisica, Universit\`a di Bologna and INFN,
     Via Irnerio 46, IT-40126 Bologna, Italy
    \label{BOLOGNA}}
\titlefoot{Centro Brasileiro de Pesquisas F\'{\i}sicas, rua Xavier Sigaud 150,
     BR-22290 Rio de Janeiro, Brazil \\
     \indent~~and Depto. de F\'{\i}sica, Pont. Univ. Cat\'olica,
     C.P. 38071 BR-22453 Rio de Janeiro, Brazil \\
     \indent~~and Inst. de F\'{\i}sica, Univ. Estadual do Rio de Janeiro,
     rua S\~{a}o Francisco Xavier 524, Rio de Janeiro, Brazil
    \label{BRASIL}}
\titlefoot{Coll\`ege de France, Lab. de Physique Corpusculaire, IN2P3-CNRS,
     FR-75231 Paris Cedex 05, France
    \label{CDF}}
\titlefoot{CERN, CH-1211 Geneva 23, Switzerland
    \label{CERN}}
\titlefoot{Institut de Recherches Subatomiques, IN2P3 - CNRS/ULP - BP20,
     FR-67037 Strasbourg Cedex, France
    \label{CRN}}
\titlefoot{Now at DESY-Zeuthen, Platanenallee 6, D-15735 Zeuthen, Germany
    \label{DESY}}
\titlefoot{Institute of Nuclear Physics, N.C.S.R. Demokritos,
     P.O. Box 60228, GR-15310 Athens, Greece
    \label{DEMOKRITOS}}
\titlefoot{Dipartimento di Fisica, Universit\`a di Genova and INFN,
     Via Dodecaneso 33, IT-16146 Genova, Italy
    \label{GENOVA}}
\titlefoot{Institut des Sciences Nucl\'eaires, IN2P3-CNRS, Universit\'e
     de Grenoble 1, FR-38026 Grenoble Cedex, France
    \label{GRENOBLE}}
\titlefoot{Helsinki Institute of Physics, HIP,
     P.O. Box 9, FI-00014 Helsinki, Finland
    \label{HELSINKI}}
\titlefoot{Joint Institute for Nuclear Research, Dubna, Head Post
     Office, P.O. Box 79, RU-101 000 Moscow, Russian Federation
    \label{JINR}}
\titlefoot{Institut f\"ur Experimentelle Kernphysik,
     Universit\"at Karlsruhe, Postfach 6980, DE-76128 Karlsruhe,
     Germany
    \label{KARLSRUHE}}
\titlefoot{Institute of Nuclear Physics,Ul. Kawiory 26a,
     PL-30055 Krakow, Poland
    \label{KRAKOW1}}
\titlefoot{Faculty of Physics and Nuclear Techniques, University of Mining
     and Metallurgy, PL-30055 Krakow, Poland
    \label{KRAKOW2}}
\titlefoot{Universit\'e de Paris-Sud, Lab. de l'Acc\'el\'erateur
     Lin\'eaire, IN2P3-CNRS, B\^{a}t. 200, FR-91405 Orsay Cedex, France
    \label{LAL}}
\titlefoot{School of Physics and Chemistry, University of Landcaster,
     Lancaster LA1 4YB, UK
    \label{LANCASTER}}
\titlefoot{LIP, IST, FCUL - Av. Elias Garcia, 14-$1^{o}$,
     PT-1000 Lisboa Codex, Portugal
    \label{LIP}}
\titlefoot{Department of Physics, University of Liverpool, P.O.
     Box 147, Liverpool L69 3BX, UK
    \label{LIVERPOOL}}
\titlefoot{LPNHE, IN2P3-CNRS, Univ.~Paris VI et VII, Tour 33 (RdC),
     4 place Jussieu, FR-75252 Paris Cedex 05, France
    \label{LPNHE}}
\titlefoot{Department of Physics, University of Lund,
     S\"olvegatan 14, SE-223 63 Lund, Sweden
    \label{LUND}}
\titlefoot{Universit\'e Claude Bernard de Lyon, IPNL, IN2P3-CNRS,
     FR-69622 Villeurbanne Cedex, France
    \label{LYON}}
\titlefoot{Dipartimento di Fisica, Universit\`a di Milano and INFN-MILANO,
     Via Celoria 16, IT-20133 Milan, Italy
    \label{MILANO}}
\titlefoot{Dipartimento di Fisica, Univ. di Milano-Bicocca and
     INFN-MILANO, Piazza della Scienza 2, IT-20126 Milan, Italy
    \label{MILANO2}}
\titlefoot{IPNP of MFF, Charles Univ., Areal MFF,
     V Holesovickach 2, CZ-180 00, Praha 8, Czech Republic
    \label{NC}}
\titlefoot{NIKHEF, Postbus 41882, NL-1009 DB
     Amsterdam, The Netherlands
    \label{NIKHEF}}
\titlefoot{National Technical University, Physics Department,
     Zografou Campus, GR-15773 Athens, Greece
    \label{NTU-ATHENS}}
\titlefoot{Physics Department, University of Oslo, Blindern,
     NO-0316 Oslo, Norway
    \label{OSLO}}
\titlefoot{Dpto. Fisica, Univ. Oviedo, Avda. Calvo Sotelo
     s/n, ES-33007 Oviedo, Spain
    \label{OVIEDO}}
\titlefoot{Department of Physics, University of Oxford,
     Keble Road, Oxford OX1 3RH, UK
    \label{OXFORD}}
\titlefoot{Dipartimento di Fisica, Universit\`a di Padova and
     INFN, Via Marzolo 8, IT-35131 Padua, Italy
    \label{PADOVA}}
\titlefoot{Rutherford Appleton Laboratory, Chilton, Didcot
     OX11 OQX, UK
    \label{RAL}}
\titlefoot{Dipartimento di Fisica, Universit\`a di Roma II and
     INFN, Tor Vergata, IT-00173 Rome, Italy
    \label{ROMA2}}
\titlefoot{Dipartimento di Fisica, Universit\`a di Roma III and
     INFN, Via della Vasca Navale 84, IT-00146 Rome, Italy
    \label{ROMA3}}
\titlefoot{DAPNIA/Service de Physique des Particules,
     CEA-Saclay, FR-91191 Gif-sur-Yvette Cedex, France
    \label{SACLAY}}
\titlefoot{Instituto de Fisica de Cantabria (CSIC-UC), Avda.
     los Castros s/n, ES-39006 Santander, Spain
    \label{SANTANDER}}
\titlefoot{Inst. for High Energy Physics, Serpukov
     P.O. Box 35, Protvino, (Moscow Region), Russian Federation
    \label{SERPUKHOV}}
\titlefoot{J. Stefan Institute, Jamova 39, SI-1000 Ljubljana, Slovenia
     and Laboratory for Astroparticle Physics,\\
     \indent~~Nova Gorica Polytechnic, Kostanjeviska 16a, SI-5000 Nova Gorica, Slovenia, \\
     \indent~~and Department of Physics, University of Ljubljana,
     SI-1000 Ljubljana, Slovenia
    \label{SLOVENIJA}}
\titlefoot{Fysikum, Stockholm University,
     Box 6730, SE-113 85 Stockholm, Sweden
    \label{STOCKHOLM}}
\titlefoot{Dipartimento di Fisica Sperimentale, Universit\`a di
     Torino and INFN, Via P. Giuria 1, IT-10125 Turin, Italy
    \label{TORINO}}
\titlefoot{Dipartimento di Fisica, Universit\`a di Trieste and
     INFN, Via A. Valerio 2, IT-34127 Trieste, Italy \\
     \indent~~and Istituto di Fisica, Universit\`a di Udine,
     IT-33100 Udine, Italy
    \label{TU}}
\titlefoot{Univ. Federal do Rio de Janeiro, C.P. 68528
     Cidade Univ., Ilha do Fund\~ao
     BR-21945-970 Rio de Janeiro, Brazil
    \label{UFRJ}}
\titlefoot{Department of Radiation Sciences, University of
     Uppsala, P.O. Box 535, SE-751 21 Uppsala, Sweden
    \label{UPPSALA}}
\titlefoot{IFIC, Valencia-CSIC, and D.F.A.M.N., U. de Valencia,
     Avda. Dr. Moliner 50, ES-46100 Burjassot (Valencia), Spain
    \label{VALENCIA}}
\titlefoot{Institut f\"ur Hochenergiephysik, \"Osterr. Akad.
     d. Wissensch., Nikolsdorfergasse 18, AT-1050 Vienna, Austria
    \label{VIENNA}}
\titlefoot{Inst. Nuclear Studies and University of Warsaw, Ul.
     Hoza 69, PL-00681 Warsaw, Poland
    \label{WARSZAWA}}
\titlefoot{Fachbereich Physik, University of Wuppertal, Postfach
     100 127, DE-42097 Wuppertal, Germany
    \label{WUPPERTAL}}
\addtolength{\textheight}{-10mm}
\addtolength{\footskip}{5mm}
\clearpage
\headsep 30.0pt
\end{titlepage}
%%%%%%%%%%%%%%%%%%%%%%%%%
%
% Change for the document body
%%\pagestyle{heading} % for page numbering
\pagenumbering{arabic} % page numbering in number
\setcounter{footnote}{0} %
\large
%\linenumbers %%%CD
%   document.tex

\section{Introduction}

In the framework of the Standard Model ({\sc SM}) there is one physical 
Higgs boson, H, which is a neutral CP-even scalar. At LEP2 the main 
production process is through the s-channel,
\ee $\rightarrow \mbox{Z}^* \rightarrow $\ZH, but there are additional
t-channel diagrams in the \hnn\ and \hee\ final states, which proceed
through \WW\ and \ZZ\ fusion, respectively. 
With the data taken previously up to $\sqrt{s}$ = 188.7~\GeV, 
DELPHI excluded a  {\sc SM} Higgs boson with mass less
than 94.6~\GeVcc~\cite{pap98} at the 95\% confidence level (CL). 
The other LEP collaborations reached similar results~\cite{lep}.
The present analysis concentrates on masses between 85 and 115~\GeVcc.
The results obtained in the same mass range with the data taken by 
DELPHI in the last year of LEP operation and analysed with preliminary
calibration constants can be found in~\cite{pap00}. Although the
emphasis is on high masses, the analysis described in this paper is
also applied to lower masses, down to the \bbbar\ threshold, in order
to derive a constraint on the production cross-section of a {\sc SM}-like
Higgs boson as a function of its mass.

%The results of the search for the {\sc SM} Higgs boson are also 
%interpreted in terms of the lightest scalar Higgs boson, h, in 
%the Minimal Supersymmetric Standard Model ({\sc MSSM}). 
In the Minimal Supersymmetric Standard Model ({\sc MSSM}), the
production of the lightest scalar Higgs boson, h, proceeds 
through the same processes as in the {\sc SM}.
%, but 
The results of the search for the {\sc SM} Higgs boson thus also apply
to the h boson. However, in the {\sc MSSM}  
the production cross-section is reduced with respect to the {\sc SM} one 
and even vanishes in part of the {\sc MSSM} parameter space.
This model predicts also a CP-odd
pseudo-scalar, A, that would be produced mostly in the
\ee $\rightarrow \mbox{Z}^* \rightarrow {\mathrm h} {\mathrm A}$
process at LEP2. The {\sc MSSM} parameters are such that when
the single h production is suppressed, the associated \hA\ production
is enhanced. This channel is thus also considered in this paper.
Previous 95\% CL limits from DELPHI on the masses of h and
A of the  {\sc MSSM} were 82.6~\GeVcc\ and 
84.1~\GeVcc\ respectively~\cite{pap98}. 
The results of the other LEP collaborations are described in 
Ref.~\cite{lep}. The present analysis in the \hA\ channel 
concentrates on masses between 80 and 95~\GeVcc.

In the \ZH\ channel, all known decays of the Z boson
have been taken into account (hadrons, charged leptons and neutrinos)
while the analyses have been optimised either for decays of the Higgs 
particle into \mbox{${\mathrm b}\bar{\mathrm b}$},
making use of the expected high branching fraction of this mode,
or for Higgs boson decays into a pair of \mbox{$\tau$}'s, which is
the second main decay channel in the  {\sc SM} and in most of the  
{\sc MSSM} parameter space.
A dedicated search for the Higgs boson invisible decay modes will be 
reported separately. 
The \hA\ production has been searched for in the two main decay channels, 
namely the 4b and \mbox{${\mathrm b}\bar{\mathrm b}\tau^+ \tau^-$} 
final states.

\section{Data samples and detector overview}

In 1999 LEP ran at centre-of-mass energies ranging 
from 191.6~\GeV\ to 201.7~\GeV. DELPHI recorded 
%25.9$\pm$1.6~pb$^{-1}$ at  191.6~\GeV,
%76.9$\pm$4.6~pb$^{-1}$ at  195.6~\GeV,
%84.3$\pm$5.0~pb$^{-1}$ at  199.6~\GeV\
%and 41.1$\pm$2.5~pb$^{-1}$ 
25.9~pb$^{-1}$ at  191.6~\GeV,
76.9~pb$^{-1}$ at  195.6~\GeV,
84.3~pb$^{-1}$ at  199.6~\GeV\
and 41.1~pb$^{-1}$ at  201.7~\GeV.
The requirement of full detector performance reduces the luminosities
in the \hee\ and \hnn\ searches by at most 3\%.
The detector was unchanged from the previous data taking period.
Ref.~\cite{pap97} provides a short
description while more details can be found in Ref.~\cite{delsim,perfo}
for the original setup and in Ref.~\cite{upgrade} for the LEP2 upgrade
of the silicon tracking detector.

Large numbers of background and signal events have been produced by
Monte Carlo simulation and then passed through the DELPHI detector 
simulation program~\cite{delsim}. 
These samples typically correspond to about 100 times the
luminosity of the collected data. Backgrounds were generated with 
{\tt PYTHIA}~\cite{pythia} for hadronic two-fermion final states
(hereafter denoted as \qqg) 
and with {\tt KORALZ}~\cite{koralz}  for leptonic two-fermion 
final states.
%($\mbox{${\mathrm e}^+{\mathrm e}^-$} \rightarrow$ 
%\mbox{${\mathrm f}\bar{\mathrm f}$} $\gamma$%), {\tt PYTHIA} and 
The four-fermion background, which is a coherent sum of many processes
whose main components are referred as $\Zz\gamma^*$, 
\WW\ and \ZZ\ in the following,
was generated with
{\tt EXCALIBUR}~\cite{excalibur} in most of the phase space, but
{\tt GRC4F}~\cite{grc4f} and {\tt KORALW}~\cite{koralw} were used to 
complete the {\tt EXCALIBUR} samples in the case of very forward electrons
or low mass hadronic resonances, respectively.
{\tt TWOGAM}~\cite{twogam} and {\tt BDK}~\cite{bdk} 
were used for two-photon processes (hereafter denoted as $\gamma \gamma$)
and {\tt BHWIDE}~\cite{bhwide} for
Bhabha events in the main acceptance region.

Signal events were produced using the {\tt HZHA}~\cite{hzha} generator. 
As the enhancement in the production cross-section due to \WW\ fusion
and to its interference with the \ZH\ process
is significant in the \hnn\ channel~\cite{fusion}
(for \MH\ around the \ZH\ kinematic threshold),  
signal events in this channel were generated using a 
version of {\tt HZHA}
%~\cite{hzfu} 
modified to include also fusion
and interference between the \ZH\ and \WW\ fusion diagrams.
For the \ZH\ process, the H mass was varied 
from 12 to 115~\GeVcc, while for \hA\
the range for the A mass was 12 to 95~\GeVcc.
A step of 5~\GeVcc\ was used above 80~\GeVcc, since the analyses
were optimized at high mass. Wider steps were used at lower
masses. 
Moreover, the \hA\ signal events were simulated
for three values of \tbeta\ 
(the ratio of the vacuum expectation values of the two Higgs field
doublets of the {\sc MSSM}) 
equal to 2, 20 or 50. This fixes 
the h mass, almost equal to \MA\ for \tbeta~=~20,~50 and 
lower than \MA\ by around 20~\GeVcc\ if \tbeta~=~2. The 
h and A widths are lower than 1~\GeVcc\ for \tbeta\ below 20 and 
increase rapidly to reach several~\GeVcc\ at \tbeta~=~50, 
thus going above the experimental mass resolution which is
typically of 3~\GeVcc\ in the \hA\ channels.

The \ZH\ simulated samples were classified
according to the Higgs and Z boson decay modes. For \hee, \hmm\ and 
\hnn\ the natural  {\sc SM} mix of \Hz\ decay modes into fermions was 
permitted. 
As final states with hadrons and two $\tau$'s benefit from a dedicated
analysis, the $\tau\tau$ decay mode was removed in the 
\hqq\ channel simulations and we generated
separately the two \ZH\ channels involving $\tau$ leptons for which one 
of the
bosons is forced to decay to $\tau$'s and the other hadronically. Finally,
the \hA\ simulations cover final states involving either four b quarks 
or two b
quarks and two $\tau$'s, irrespective of which Higgs boson decays into 
$\tau$'s. Efficiencies are defined relative
to these states. The sizes of these samples vary from 2000 to 3000
events and they were produced at the four centre-of-mass energies. 

  Although the above signal simulations cover most of the 
expected final states in the {\sc SM}  and  {\sc MSSM}, they were 
completed by two 
additional sets at~199.6~GeV. We generated \hA\ samples with large mass 
differences between the h and A bosons, as expected when scanning the  
{\sc MSSM} 
parameter space more widely than in the representative scans, and 
\hZ\ samples with \hAA, as
expected in restricted regions of the  {\sc MSSM} parameter space. 
In these two sets, the A (h) mass was varied from 12~\GeVcc\ (50~\GeVcc)
up to the kinematic limit and only the main decays were simulated;
the \hA\ samples were restricted to four b final states, 
and the (\hAA)Z samples to hadronic decays of the Z boson and
either four b or four c quarks from the A pair.
The results obtained
from these samples were assumed to be valid also at the three other
centre-of-mass energies.

\section{Features common to all analyses}
\label{sec:common}
\subsection{Particle selection}

In all analyses, charged particles are selected if their momentum is greater
than 100~\MeVc\ and if they originate from the interaction region 
(within  4 cm in the transverse plane and within
4 cm / $\sin\theta$ along the beam direction, where $\theta$ is the 
particle polar angle). Neutral
particles are defined either as energy clusters 
in the calorimeters not associated to charged particle tracks,
or as reconstructed vertices of photon conversions, interactions of 
neutral hadrons
or decays of neutral particles in the tracking volume.
All neutral clusters of energy greater than 200 or 300~MeV (depending on
the calorimeter) are used.
The $\pi^{\pm}$ mass is used for all charged particles except identified 
leptons,
while  zero mass is used for electromagnetic clusters and the K$^0$ mass is
assigned to neutral hadronic clusters. 

\subsection{b-quark identification }
\label{sec:btag}

The method of separation of  b quarks from other flavours is described 
in~\cite{btag_combi}, where the various differences between B-hadrons and
other particles are accumulated in a single variable, hereafter denoted 
\xb\ for an event and \xbi\ for jet $i$. A major input to this combined 
variable is the probability $P_{i}^+$ that all tracks with a positive 
lifetime-signed impact parameter\footnote{Throughout the paper,
all impact parameters are defined with respect to the 
reconstructed primary vertex.}
in the jet lead to a product of
track significances as large as that observed, if these tracks do
originate from the interaction point. 
%The probability is also defined at the event level from all 
%These probabilities are called $P_{E}^+$ 
%tracks with a positive lifetime-signed impact parameter in the event.
A low value of this probability is a signature for a B-hadron. 
%for an event and $P_{i}^+$ for a jet.
The likelihood ratio technique is then used to construct
\xbi\ by combining $P_{i}^+$ with information from any secondary vertex
found in the jet
(the mass computed from the particles assigned to the secondary vertex,
the rapidity of those particles, and the fraction of the jet momentum 
carried by them) and with the transverse momentum 
(with respect to the jet axis) of any lepton belonging to the jet.
The event variable, \xb, is a linear combination of the 
jet variables.
Increasing values of \xb\ (or \xbi) correspond to increasingly 
`b-like' events (or jets). 

\begin{figure}[htbp]
\begin{center}
\epsfig{figure=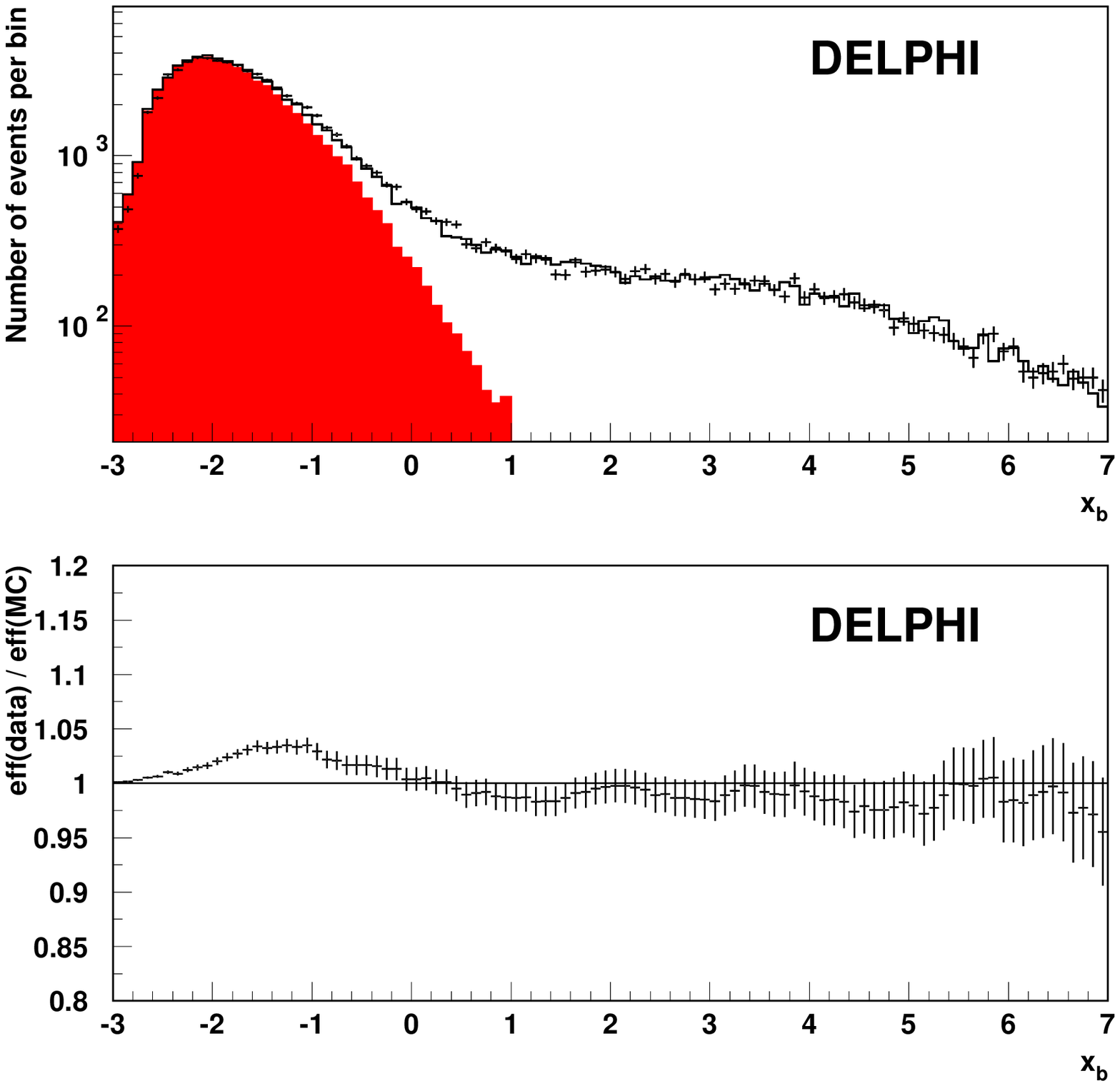,width=13cm} \\
\vspace{-1.3cm}
\epsfig{figure=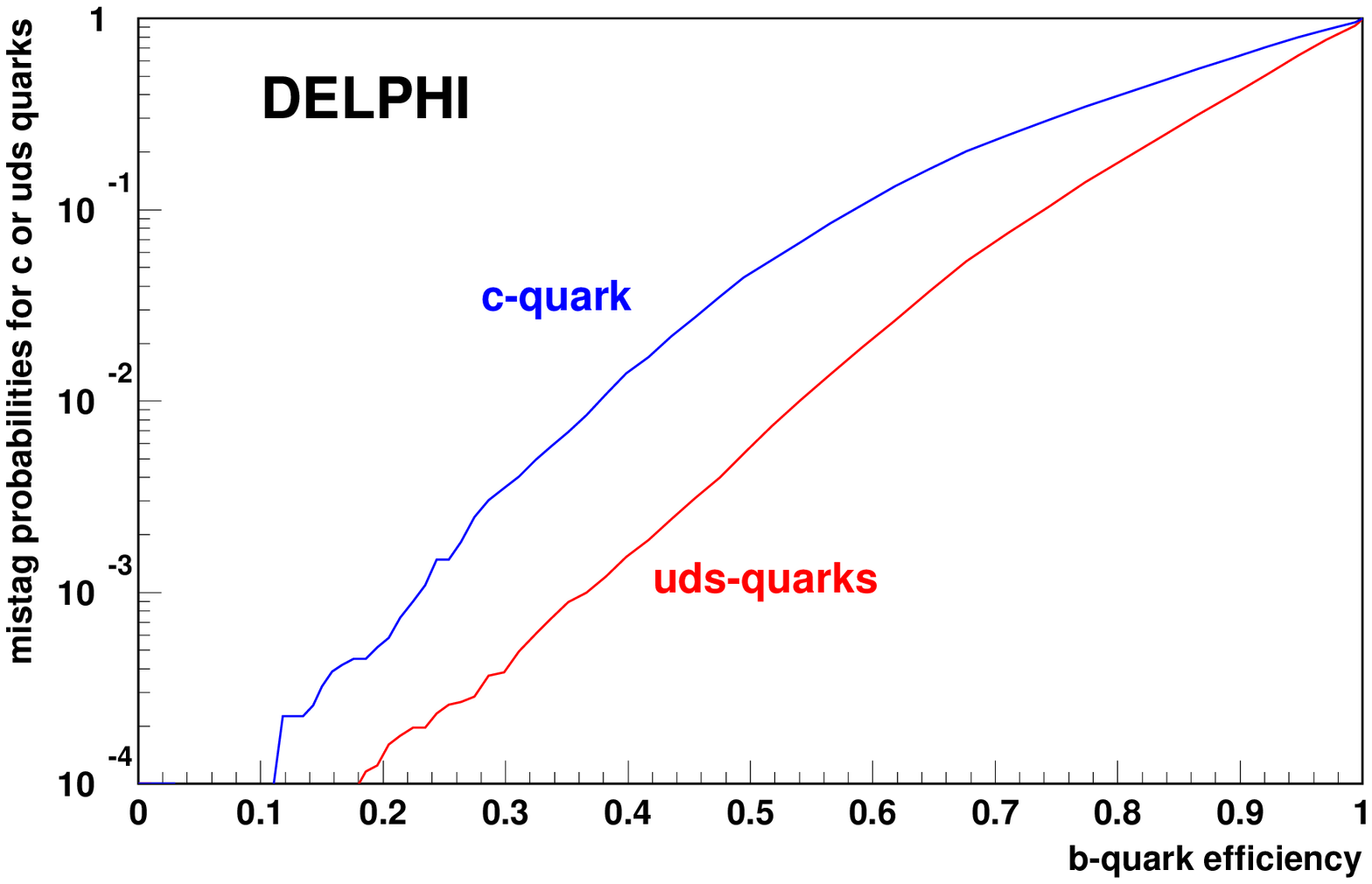,width=13cm}
\caption[]{b-tagging: 
Top: distributions of the combined b-tagging variable \xb, in 1999 
\Zz\ data
(dots) and simulation (histogram). 
The contribution of udsc-quarks is shown
as the dark histogram. Middle: ratio of integrated tagging rates in 
\Zz\ data
and simulation as a function of the cut in \xb.
Bottom: mistag probabilities for c or uds-quarks as a function of the
efficiency for b-quarks, estimated from simulated \Zz\ data.}
\label{fig:btagging}
\end{center}
\end{figure}

The procedure is calibrated on events recorded in the same experimental 
conditions at the Z resonance. The performance of the combined b-tagging 
is 
described in Ref.~\cite{Rb} and that of the impact parameter tagging 
alone 
in Ref.~\cite{pap96}. The overall performance of the combined b-tagging
for 1999 Z data is illustrated in Fig.~\ref{fig:btagging}. 
Data agree with simulation to better than 5\% in the whole range of 
cut values. 
%Thus, an uncertainty of 5\% is considered in all analysese
%as the systematic uncertainty due to b-tagging.

%A careful study of possible systematic effects, including data versus 
%simulation agreement at the Z pole (checked inclusively, per flavour and 
%for multi-jet events), leads to an overall relative b-tagging 
%uncertainty 
%below 5\%, varying slightly with the exact tagging value used. 
%At high energy, an inclusive comparison of
%data with simulation confirms this number. 

%However, in the four-jetS
%topologies at high energy, the b-tagging efficiency is not reproduced by
%simulation in the intermediate region where c- and b-quarks are expected to 
%dominate. This difference has been attributed partly to an imperfect 
%knowledge of the prompt b-quark rate but mostly to incorrect
%gluon splitting rates into \bbbar and \ccbar.
%The gluon splitting rates into \bbbar and \ccbar have been rescaled 
%in the simulation
%according to the LEP Heavy Flavour Working Group~\cite{gsplit}
% prescription. 
%In addition, a 50\% uncertainty on these splitting
%rates is applied to the \qqbar($\gamma$) background estimate.

\subsection{Constrained fits}

In most channels a constrained fit~\cite{pufitc} is performed to 
reconstruct 
the Higgs boson mass, and often to reject background processes as well.
In order to allow the removal of most of the events involving
radiative return to the \Zz,
an algorithm has been developed~\cite{sprime} in order to estimate the
effective energy of the \ee\ collision. This algorithm makes use of 
a three-constraint kinematic fit in order to test the presence of an 
initial state photon 
along one of the beam directions and hence lost in the beam pipe. 
This effective 
centre-of-mass energy is called \mbox{$ \sqrt{s'}$} throughout this paper.

\subsection{Confidence level definitions and calculations}
\label{sec:limits}
The confidence level definitions rely on a test-statistic built with the
likelihood ratio technique~\cite{alex}. Let \likear\ be the ratio of the
likelihood of the observed candidates assuming signal plus background
to that found using the background-only hypothesis. 
\likear\ classifies the result of an observation between the 
background-like
and signal plus background-like situations.
We then define the confidence level for the background hypothesis, 
CL$_{\rm b}$, as the probability, in background-only experiments,
to obtain equal or smaller values of \likear\ (that is more 
background-like results)
than that observed. Similarly,
the confidence level for the signal plus background hypothesis,
CL$_{\rm s+b}$, is the probability, in signal plus background experiments,
to obtain more background-like results than those observed. 
The pseudo-confidence level
for the signal hypothesis, CL$_{\rm s}$, is conservatively defined 
as the ratio of these
two probabilities, CL$_{\rm s+b}$/CL$_{\rm b}$.
CL$_{\rm s}$ measures the confidence with which the
signal hypothesis can be rejected and must fall below 5\% for an
exclusion confidence of 95\%. More technical details about how the
confidence levels are calculated or how uncertainties are taken into 
account
can be found in~\cite{pap97}. 
 
  In the definition of the test-statistic \likear,  
two-dimensional discriminant information is used in all channels, 
as in our previous publication~\cite{pap98}.
The first variable is the reconstructed Higgs boson mass (or the sum of
the reconstructed h and A masses in the \hA\ channels),
the second one is channel-dependent, as specified in the 
following sections.
In order to make full use of the information contained in the second
variable, the final selections are loose: the method used
for deriving the confidence levels ensures that adding regions of
lower signal and higher background can only enhance the performance
relative to a tighter selection, provided the systematic errors are
small.
 
 The distributions are represented as two-dimensional histograms which
are derived from the simulation samples. These distributions are
then smoothed using a two-dimensional kernel, which is essentially 
Gaussian
but with a small component of a longer tail. The width of the kernel
varies from point to
point, such that the statistical error on the estimated background
is never more than 30\%. 
%The same width is applied to background and
%all signal samples to eliminate the possibility of the smearing itself
%increasing the estimated signal to background ratio. 
Finally the distribution is
reweighted so that when projected onto either axis it has the same
distribution 
as would have been observed if the smoothing had been only in one 
dimension.
This makes better use of the simulation statistics if there are features 
which are essentially one dimensional, such as mass peaks.
  A check for residual statistical fluctuations was made by dividing the
simulation into sub-samples, and comparing the expected results; 
no significant effects were observed.

%%%%%%%%%%%%%%%%%%%%%%%%%%%%%%%%%%%%%%%%%%%%%%%%%%%%%%%%%%%%%%%%

\section{Higgs boson searches in events with jets and electrons}

 The analysis is based upon the same electron identification algorithm and 
discriminant variables as in~\cite{pap97,pap98}
and is briefly described in the following.
The preselection requires at least 8 charged particles,
a total energy above 0.12\rs\ and at least a pair of 
loose electron candidates of energies above 10~GeV and impact parameters
below 2~mm (1~cm) in the transverse plane (along the beam direction).
The Bhabha veto and the modified selections allowing for the tau decays of
the Higgs boson are as described in~\cite{pap98}.
To reduce the $\Zz\gamma^*$ and \qqg\ backgrounds,
the sum of the di-electron and hadronic system masses
must be above 50~\GeVcc, while the missing
%event 
momentum is required to be below 50~\GeVc\ if its direction is 
within 10$^\circ$ to the beam axis. The jet reconstruction and selection
proceed as in~\cite{pap97}.

  After this preselection, each pair of electron candidates with opposite
charges is submitted to further cuts. The electron identification is
first tightened, allowing at most one electron candidate in the 
insensitive  regions of the calorimeter.
The two electrons are required to have energies above 20~GeV and 
15~GeV. 
Electron isolation angles with respect to the closest jet are 
required to be more than 20$^\circ$ for the more isolated electron 
and more than 8$^\circ$ for the other one. 
A five-constraint kinematic fit is performed to test the compatibility of
the \ee\ invariant mass with the \Zz\ mass; the fit imposes
energy and momentum conservation and takes into account the Breit-Wigner
shape of the \Zz\ resonance~\cite{pap98}. Events with a fit probability
below 10$^{-8}$ are rejected. 
%A kinematic fit is performed, imposing energy and momentum conservation, 
%and constraining the invariant mass of the \ee\ system to \MZ~\cite{pap98}.
%The unfitted masses are kept if the fit does not converge.
As the search is restricted to high mass Higgs bosons produced in
association with a \Zz\ particle, the sum of the
fitted masses of the electron pair and of the hadronic system 
is required to be above 150~\GeVcc\ and their difference
in the range from -100~\GeVcc\ to 50~\GeVcc.
The fitted hadronic mass and the b-tagging variable
$x_b$ are used in the two-dimensional calculation of the confidence 
levels.
 
\begin{figure}[htbp]
\hspace{-1.cm}
\epsfig{figure=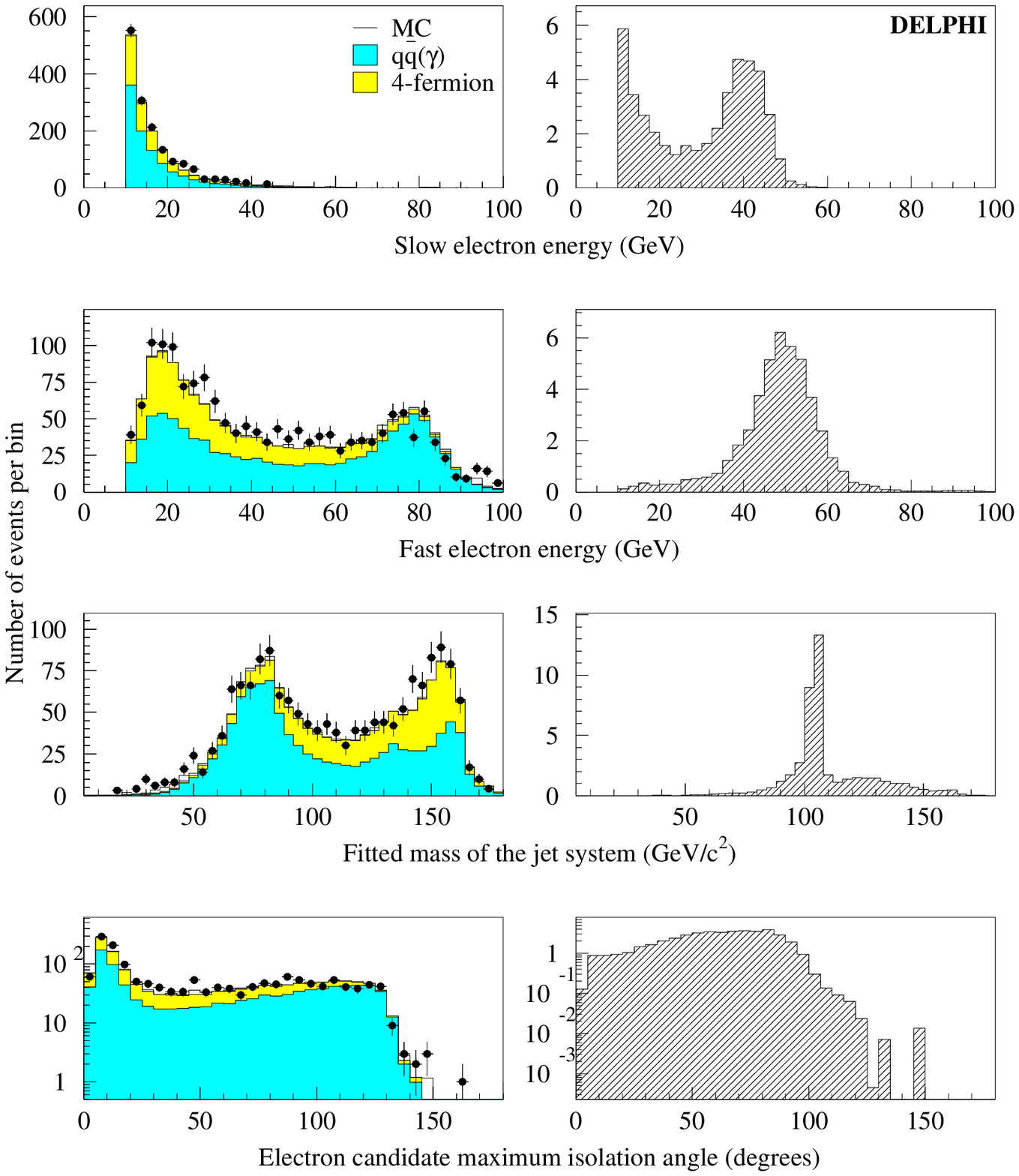,width=17cm}
\caption[]{\hee channel: 
  distributions of four analysis variables,
  as described in the text, at preselection level. Data at 
  \rs~=~191.6-201.7~\GeV\ (dots) are compared with  {\sc SM} background 
  expectations (left-hand side histograms) and with the expected distribution
  for a 105~\GeVcc\ signal (right-hand side histogram, 
  normalised to 50 times the expected rate).}
\label{fig:hee}
\end{figure}

   The effect of the selections on data and simulated samples  
are detailed in Tables~\ref{ta:hzsum192} to~\ref{ta:hzsum202},
while the efficiencies at the end of the analysis are reported in
Tables~\ref{ta:hzeff} and~\ref{ta:hzeff_lowm} as a function of \MH.
The agreement between data and simulation at the preselection level
is illustrated in Fig.~\ref{fig:hee} which shows
the distributions of the electron energies, the fitted mass of the jet 
system and the isolation angle of the more isolated electron candidate.
At the end of the analysis, 11 events are selected in the data for 
a total expected background of 
\mbox{$11.5 \pm 0.2 (stat.)$} events coming mainly from the \eeqq\ 
process.

  The systematic uncertainties on background and efficiency estimates
are mainly due to the imperfect simulation of the detector response
and were estimated as described in~\cite{pap97}. 
The relative error on the efficiencies is $\pm 3\%$
%$^{+1.2\%}_{-2.8\%}$, 
while that on the background estimates at each centre-of-mass energy  
is $\pm 7\%$.

\section{Higgs boson searches in events with jets and muons}

 The analysis follows that published in~\cite{pap97,pap98}, with
slight modifications in the preselection to adapt to somewhat different
beam conditions in 1999. The preselection requires at least 9 
charged particles with two of them in the central part of the detector 
($40~\dgree < \theta <140~\dgree $) and
at least two high quality tracks of particles with a transverse momentum
greater than 5~\GeVc. For high quality tracks, impact
parameters less than 100~$\mu$m in the transverse plane and 
less than 500~$\mu$m along the beam direction are required.
The rest of the preselection is unchanged and requires at least 
two particles of opposite charges 
and momenta greater than 15~\GeVc.

The rest of the analysis is based upon the same muon identification
algorithm and discriminant variables as in~\cite{pap97}, but 
the selection criteria have been re-optimised~\cite{pap97}. 
As a result, the level of muon identification corresponds now to an 
efficiency of 88\% and a misidentification probability of 8.8\% per
pair of muon candidates. At least two muons are required with 
opposite charges, 
an opening angle  larger than 10\dgree, and momenta greater than 
34~\GeVc\ and 21~\GeVc. The jet reconstruction and selection proceed
as in~\cite{pap97}. Finally,
the angle with respect to the closest jet axis
must be greater than 
9\dgree\ for the more isolated muon and greater than 7\dgree\ 
for the other one. 
A five-constraint kinematic fit taking into account 
energy and momentum conservation and the Breit-Wigner shape of
the Z resonance is then performed to test 
the compatibility of the di-muon mass with the Z mass 
in a window of $\pm$ 30~GeV around the \Zz\ pole.
Events are kept only if 
the fit converges in this mass window.
%the fit probability is above 10$^{-30}$.
As in the electron channel, 
the fitted mass of the hadronic system and the b-tagging variable \xb\
are chosen as the discriminant variables 
for the two-dimensional calculation of the confidence levels.

\begin{figure}[htbp]
\begin{center}
\epsfig{figure=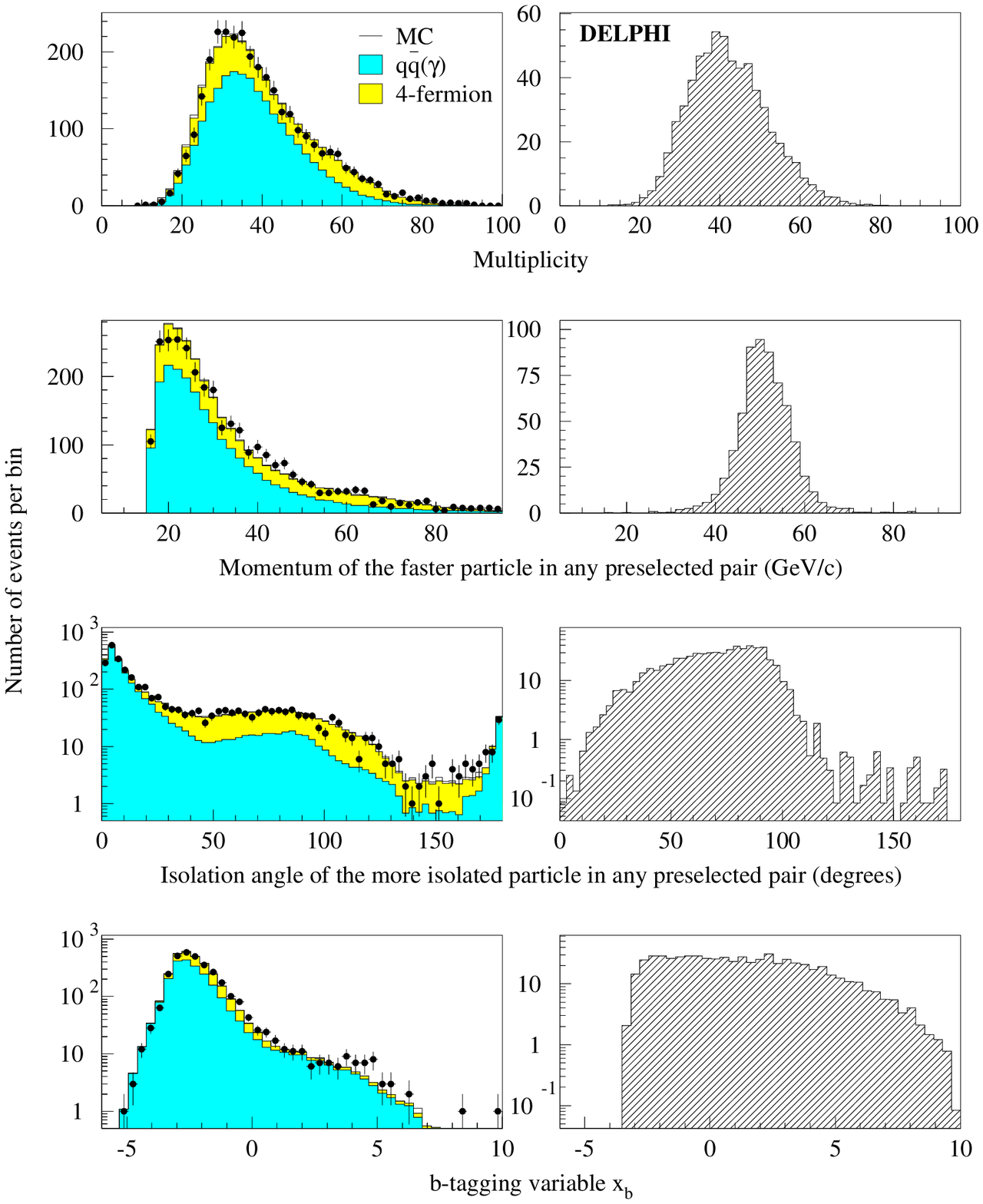,width=17cm} 
\caption[]{\hmm\ channel: 
  distributions of four analysis variables,
  as described in the text, at preselection level. Data at 
  \rs~=~191.6-201.7~\GeV\ (dots) are compared with  {\sc SM} background 
  expectations (left-hand side histograms) and with the expected distribution
  for a 105~\GeVcc\ signal (right-hand side histogram,  
  normalised to one thousand times the expected rate).}
\label{fig:hmumu}
\end{center}
\end{figure}

  The effect of the selections on data and simulated samples 
are detailed in Tables~\ref{ta:hzsum192} to~\ref{ta:hzsum202},
while the efficiencies at the end of the analysis are reported in
Tables~\ref{ta:hzeff} and~\ref{ta:hzeff_lowm} as a function of \MH.
The agreement of simulation with data is quite good, as illustrated 
at preselection level in Fig.~\ref{fig:hmumu},
which shows the multiplicity of the charged particles, 
the momentum of the higher-momentum particle in any preselected pair,
the isolation angle of the more isolated particle in any preselected pair
and the b-tagging variable \xb.
At the end of the analysis, 8 events are selected in data in 
agreement with the total expected background of 
\mbox{$9.4 \pm 0.1 (stat.)$} events coming mainly from the \mmqq\
process.

The imperfect simulation of the detector response leads to
systematic errors in background and efficiency evaluation. As explained
in \cite{pap97}, each of the momentum and angular cuts was varied in a 
range
given by the difference between the mean values of the simulated and
real data distributions of the corresponding variable at preselection 
level. 
The muon pair identification level, which is a discrete variable, 
was modified randomly with a probability of 5\%, corresponding to 
the maximum difference observed in muon identification results 
when comparing data with simulation. This is the main
source of systematic uncertainty in this channel.
As a result, a relative error of $\pm 2\%$
is quoted for the efficiencies, independent of \MH,
while the relative error on the expected backgrounds at  
each centre-of-mass energy is $\pm 3\%$.

%{\it Systematiques detaillees}\\
%192 GeV $BG = 0.94^{+0.01}_{-0.03} $\\
%196 GeV $BG = 3.09^{+0.06}_{-0.08} $\\
%200 GeV $BG = 3.63^{+0.05}_{-0.14} $\\
%202 GeV $BG = 1.90^{+0.01}_{-0.06} $ \\

\section{Higgs boson searches in events with jets and taus} 
\label{sec:htau}

Three channels are covered by these searches, two for the \ZH\ channel,
depending on which boson decays into \toto, and one for the
\hA\ channel.
The analysis, identical to that described in~\cite{pap98}, 
selects hadronic events by requiring at least ten charged particles, 
a total reconstructed energy greater than $0.4\sqrts$, a reconstructed 
charged energy above $0.2\sqrts$ and 
%an effective centre-of-mass energy, 
\mbox{$ \sqrt{s'}$} greater than 120~\GeV.

\begin{figure}[htbp]
\begin{center}
\begin{tabular}{c}
\epsfig{figure=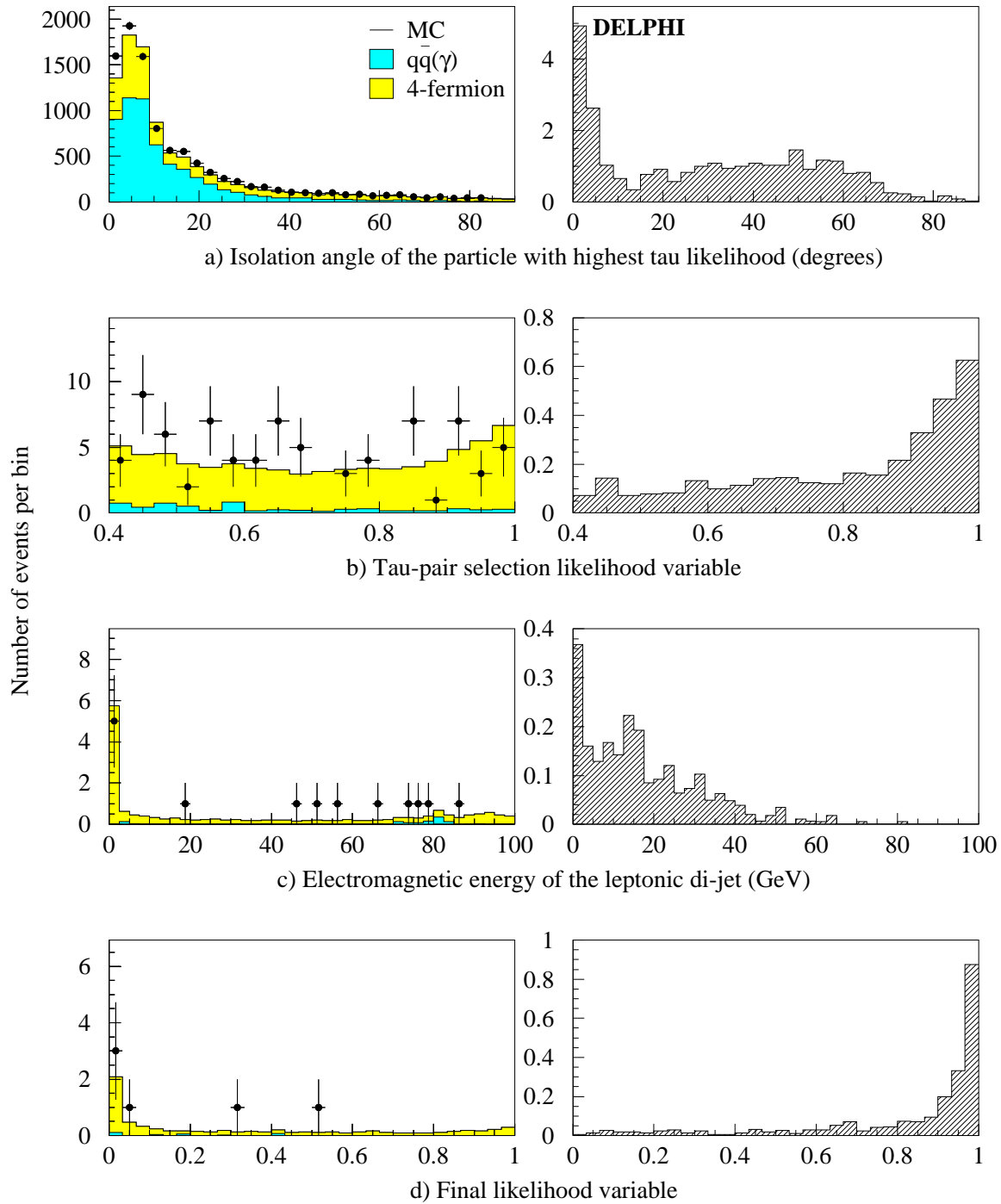,width=17cm}
\end{tabular}
\caption{\tautauqq\ channel: 
  distributions of four analysis variables at different levels of the
  selection, as described in the text. Data at 
  \rs~=~191.6-201.7~\GeV\ (dots) are compared with  {\sc SM} background 
  expectations (left-hand side histograms) and with the expected distribution
  for a 105~\GeVcc\ signal in the 
  (h$\rightarrow$\toto)(Z$\rightarrow$\qqbar) 
  (right-hand side histogram, normalised to one hundred times
   the expected rate).}
\label{fig:ttqq}
\end{center}
\end{figure}

A search for $\tau$ lepton candidates is then performed using a 
likelihood ratio
technique. 
%Clusters of one or three charged particles are preselected  
Single charged particles are preselected 
if they are isolated from all other charged particles by more than 
10\mydeg, 
%if the cluster momentum is above 2~\GeVc\ and if all particles in a 10\mydeg\ 
if their momentum is above 2~\GeVc\ and if all neutral particles in a 
10\mydeg\ 
%cone around the cluster direction make an invariant mass below 2~\GeVcc. 
cone around their direction make an invariant mass below 2~\GeVcc.
%The likelihood variable is calculated for the preselected clusters using
The likelihood variable is calculated for the preselected particles using
%distributions of the cluster momentum, of its isolation angle and of 
distributions of the particle momentum, of its isolation angle and of 
%the probability that the tracks forming the cluster come from the 
the probability that it comes from the primary vertex. 
As an illustration of the agreement between data and simulation at this
level of the analysis, Fig.~\ref{fig:ttqq}a shows the distribution of
the isolation angle of the preselected charged particle with the highest
$\tau$ likelihood variable in the event.
%Pairs of $\tau$ candidates are selected using a cut on the product of the
%$\tau$ likelihoods, and requiring opposite charges and an opening angle
%greater than 90\mydeg.
Pairs of $\tau$ candidates are then selected requiring opposite charges, 
an opening angle greater than 90\mydeg\ and a product of the
$\tau$ likelihood variables above 0.45. If more than one pair is
selected, only the pair with the highest product is kept. 
The distribution of the highest product of two $\tau$ likelihood 
variables 
in the event is given in Fig.~\ref{fig:ttqq}b. The discrimination
between the Higgs signal and the {\sc SM} background is clearly visible.
Moreover, the percentage of $\tau$ pairs correctly identified is over
90\% in simulated Higgs events.
%restricting only to the topology where the two leptons
%decay into one prong. Opposite charges and an opening angle greater than
% 90\mydeg\ are also required. The distribution of the 

Two slim jets are then reconstructed with all neutral particles  
inside a 10\mydeg\ cone around the directions of the $\tau$ candidates.
%cluster directions. 
The rest of the event is forced into two jets using the DURHAM algorithm.
The slim jets are required to be in the
\mbox{20\mydeg$\le~\theta_{\tau}~\le$ 160\mydeg} polar angle region to 
reduce the \zee\ background, while the hadronic di-jet
invariant mass is required to be between 20 and 110~\GeVcc\ in order
to reduce the \qqg\ and \mbox{${\mathrm Z}\gamma^*$} backgrounds.
The jet energies and masses are then rescaled, imposing energy and 
momentum conservation, to give a better estimate of the masses 
of both di-jets (\toto\ and \qqbar), that are required to have 
a rescaled mass above 20~\GeVcc, and below \sqrts\ to discard
unphysical solutions of the rescaling procedure. 
Each hadronic jet must have a rescaling factor in the range 0.4 to 1.5.

The remaining background comes from genuine \llqq\ events. In order 
to reject 
the \eeqq\ and \mmqq\ backgrounds the measured mass of 
the leptonic system is required to be between 10 and 80~\GeVcc\ and its 
electromagnetic energy to be below 60~\GeV\ (see Fig.~\ref{fig:ttqq}c). 
This terminates the selection procedure. 
The effect of the selections on data and simulated samples  
are detailed in Tables~\ref{ta:hzsum192} to~\ref{ta:hzsum202},
while the efficiencies at the end of the analysis in the three
\tautauqq\ channels are reported in 
Tables~\ref{ta:hzeff}, \ref{ta:hzeff_lowm} 
and~\ref{ta:haeff} as a function of the Higgs boson masses.
At the end of the analysis, 6 events are selected in data for a 
total expected background of 
\mbox{$6.9 \pm 0.2 (stat.)$} events coming mainly from the 
\tautauqq\ and \tauvqq\ processes. 

Systematic uncertainties from the imperfect modelling of the detector
response were estimated by moving each selection cut according to
the resolution in the corresponding variable. The main contributions arise
from the cuts on the \toto\ invariant mass and electromagnetic energy.
The total relative systematic uncertainties amount to $\pm$6\% 
on signal efficiencies and $\pm$11\% on the background estimates at
each centre-of-mass energy.

The two-dimensional calculation of the confidence levels uses the
reconstructed mass given by the sum of the \toto\ and 
\qqbar di-jet masses after rescaling and 
a likelihood variable built from the distributions of the 
rescaling factors of the $\tau$ jets, the $\tau$ momenta and the global  
b-tagging variable, \xb. 
The distribution of this likelihood variable at the end of the analysis
is shown in Fig.~\ref{fig:ttqq}d to illustrate the discrimination between the
Higgs signal and the {\sc SM} backgrounds. 
Since the three possible \tautauqq\ signals are covered by the same 
analysis, the three channels cannot be considered as independent in the
confidence level computation. For this computation,
they are combined into one global \tautauqq\ channel:
at each test point, the signal expectations (rate, two-dimensional 
distribution) in this channel are obtained by summing the contributions 
from the three original signals weighted by their expected rates.

\section{Higgs boson searches in events with missing
energy and jets } \label{sec:hnn}

  The signal topology in this channel is characterized by two acollinear jets 
and a large imbalance in the energy collected by the detector compared 
to the collision energy, due to neutrinos coming either from the
decay of a Z boson or from the fusion process. In addition to the
irreducible \nnqq\ four-fermion background, several other backgrounds can
lead to similar topologies, like beam-related backgrounds or the
\qqg\ process with initial state radiation photons emitted along the beam
axis. Thus, a correct description of the initial state radiation in the 
\qqg\ generator particularly matters in this channel. 
As differences of a few \% were observed in the $\sqrt{s'}$ 
distributions obtained with 
the current \qqg\ generator~\cite{pythia} and with an analytical 
calculation~\cite{zfitter}, the simulated \qqg\ events have been 
reweighted\footnote{Events with $\sqrt{s'}$ above (resp. below) 
$0.85\sqrt{s}$ are reweighthed by a factor 0.96 (resp. 1.025).} 
to reproduce the analytical result. This correction has been
applied throughout the analysis which is described below.

  Events due to 
%off-momentum particles
particles of the beam with momenta far from the nominal values 
%background from the beam 
are first excluded by requiring
at least two charged particles with impact parameters 
less than 1~mm  in the transverse plane
and less than 3~mm along the beam direction, and  
with a transverse momentum greater than 2~\GeVc.
A loose hadronic preselection is then applied, requiring
at least nine charged particles, 
a total charged energy greater than $0.16\sqrt{s}$,
a transverse energy greater than  $0.15\sqrt{s}$
and the sum of the magnitudes of all particle momenta resolved along the 
thrust axis to be greater than  $0.25\sqrt{s}$. Finally, events with 
an electromagnetic shower exceeding $0.45\sqrt{s}$ are rejected.
These criteria remove 97\% of the  $\gamma \gamma$ background and veto 
completely the Bhabha background.

In order to reject events coming from a radiative return to the \Zz\
with photons emitted in the beam pipe,
%the effective centre-of-mass energy 
$\sqrt{s'}$ 
is required to be greater than 115~GeV when the polar angle
of the missing
%total 
momentum is within 40\mydeg\ to the beam axis.
To reduce the contamination of radiative return events with photons 
in the detector acceptance,
events are rejected if their total electromagnetic energy 
within 30\mydeg\ to the beam axis is greater than $0.16\sqrt{s}$
or if the total energy in the small angle luminosity monitor is greater 
than $0.08\sqrt{s}$. A veto based on the hermeticity 
counters of DELPHI as described in~\cite{pap96} is also applied  
to reject events with 
%an on-shell Z and 
photons crossing the small insensitive 
regions of the electromagnetic calorimeters. 
To reduce the two-fermion background outside the radiative 
return peak as well as four-fermion backgrounds without missing energy, 
%the effective centre-of-mass energy 
$\sqrt{s'}$ must not exceed  $0.96\sqrt{s}$. 
Two-fermion events with jets pointing to the insensitive regions of the 
electromagnetic calorimeters are also a potential background 
due to mismeasurements of the jet properties. To reject such a background,
events are forced into a two-jet configuration using the DURHAM algorithm
and are rejected if the jet polar angles are within $\pm$5\mydeg\
of 40\mydeg\ for one jet and of 140\mydeg\ for the other jet,
unless the acoplanarity\footnote{The acoplanarity is defined
as the supplement of the angle between the transverse momenta 
(with respect to the beam axis) of the two jets.}
is greater than 10\mydeg.
At this stage, 88\% of the total \qqg\ background is removed. 
In order to reduce most of the contamination from semi-leptonic decays 
of \WW\ pairs, the energy of the most energetic particle of 
the event must not exceed $0.2\sqrt{s}$. To reinforce the rejection
of those decays containing a \tol\ lepton, there must be no
charged particle in the event with a transverse momentum with respect to 
its jet axis greater than 10~\GeVc\ when forcing the event 
into the two-jet configuration.
The final selection of signal-like events requires
the total visible energy to be lower than $0.70\sqrt{s}$. 
All the above criteria define the preselection. 

The final discrimination between signal and background is achieved 
through a multidimensional variable built with the likelihood ratio method.
A short description of the algorithms needed in this step is given below.
As already mentioned, events are forced into two jets with the DURHAM 
algorithm (the so called ``two-jet configuration'') but for each
event jets are also reconstructed with the same algorithm 
using a distance of $y_{cut}=0.005$ (the so called ``free-jet configuration'')
%, but 
%each event is also forced into two  jets using the same
%algorithm (the so called ``two-jet configuration'') 
and general variables of each jet 
(like multiplicities, momenta) are calculated in both configurations.
In order to tag remaining isolated particles from semi-leptonic decays 
of \WW\ pairs, the energies collected between two cones with half 
opening angles of 5\mydeg\ and 25\mydeg\ 
around the most isolated and the most energetic particles are calculated
and normalised to the corresponding particle energies. 
The lesser of these two normalised energies defines the 
anti-\WW\ isolation variable. 
%The largest transverse momentum with respect to its 
%jet axis of any charged particle in the two-jet configuration defines 
%the anti-\WW\ \tol\ decay variable.

The likelihood multidimensional variable combines the following
discriminant variables: 
the angle between the missing momentum and the closest jet in the free-jet 
configuration, 
the polar angle of the more forward jet in the two-jet configuration, 
the polar angle of the 
%total 
missing momentum, 
the acoplanarity in the two-jet configuration, 
the ratio between 
%the effective centre-of-mass energy 
\mbox{$\sqrt{s'}$} and the centre-of-mass energy, 
the missing mass of the event, 
the anti-\WW\ isolation variable, 
the largest transverse momentum with respect to its 
jet axis of any charged particle in the two-jet configuration,
the DURHAM distance for the transition betwen the two-jet and three-jet
configurations,
the minimum jet charged multiplicity in the free-jet configuration,
the event lifetime probability $P_{E}^+$ that all tracks with a positive 
lifetime-signed impact parameter in the event give a product
of track significances as large as that observed if they do come 
from the interaction point,
and the global b-tagging variable \xb.
The first five variables discriminate the signal from the \qqg\
background and the other variables provide a discrimination against
\WW\ pairs.
For each variable, probability density functions (p.d.f.s) at each 
centre-of-mass energy are obtained from simulated events, 
using half of the statistics available 
in all backgrounds and in signals of masses 95, 100 and 105~\GeVcc\ at
\rs\ below 198~\GeV, and 
100, 105, 110 and 115~\GeVcc\ for \rs\ above 198~\GeV. The whole samples
are used to derive the final results, in order to improve limited 
statistics in some bins of the two-dimensional discriminant information
used to derive the confidence levels.

\begin{figure}[htbp]
\begin{center}
\epsfig{figure=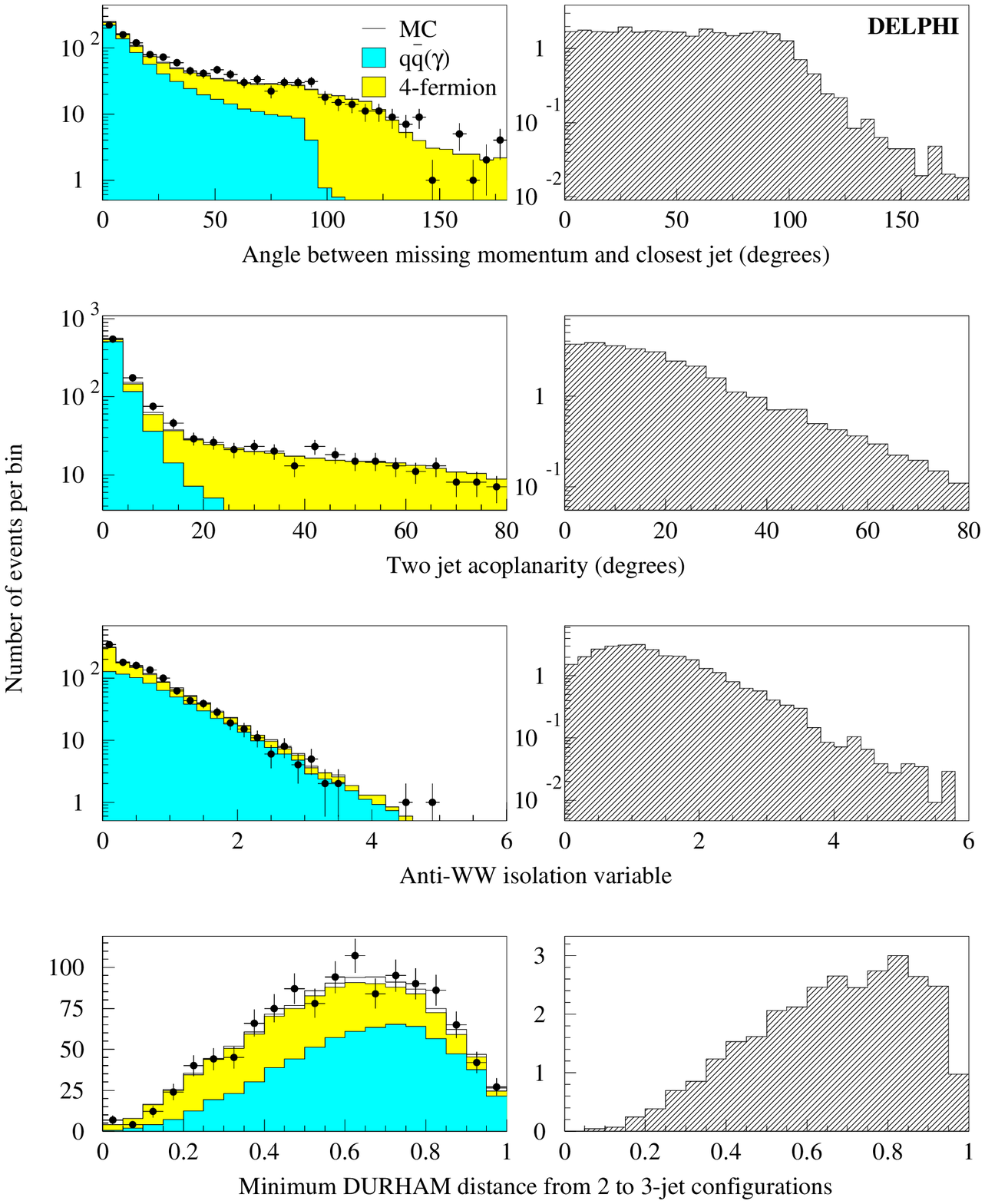,width=17cm}
\caption{ \hnn\ channel: 
  distributions of four analysis variables,
  as described in the text, at preselection level. Data at 
  \rs~=~191.6-201.7~\GeV\ (dots) are compared with  {\sc SM} background 
  expectations (left-hand side histograms) and with the expected distribution
  for a 105~\GeVcc\ signal (right-hand side histogram, 
  normalised to 10 times the expected rate).}
\label{fig:hnunu}
\end{center}
\end{figure}

\begin{figure}[htb]
\begin{center}
\epsfig{figure=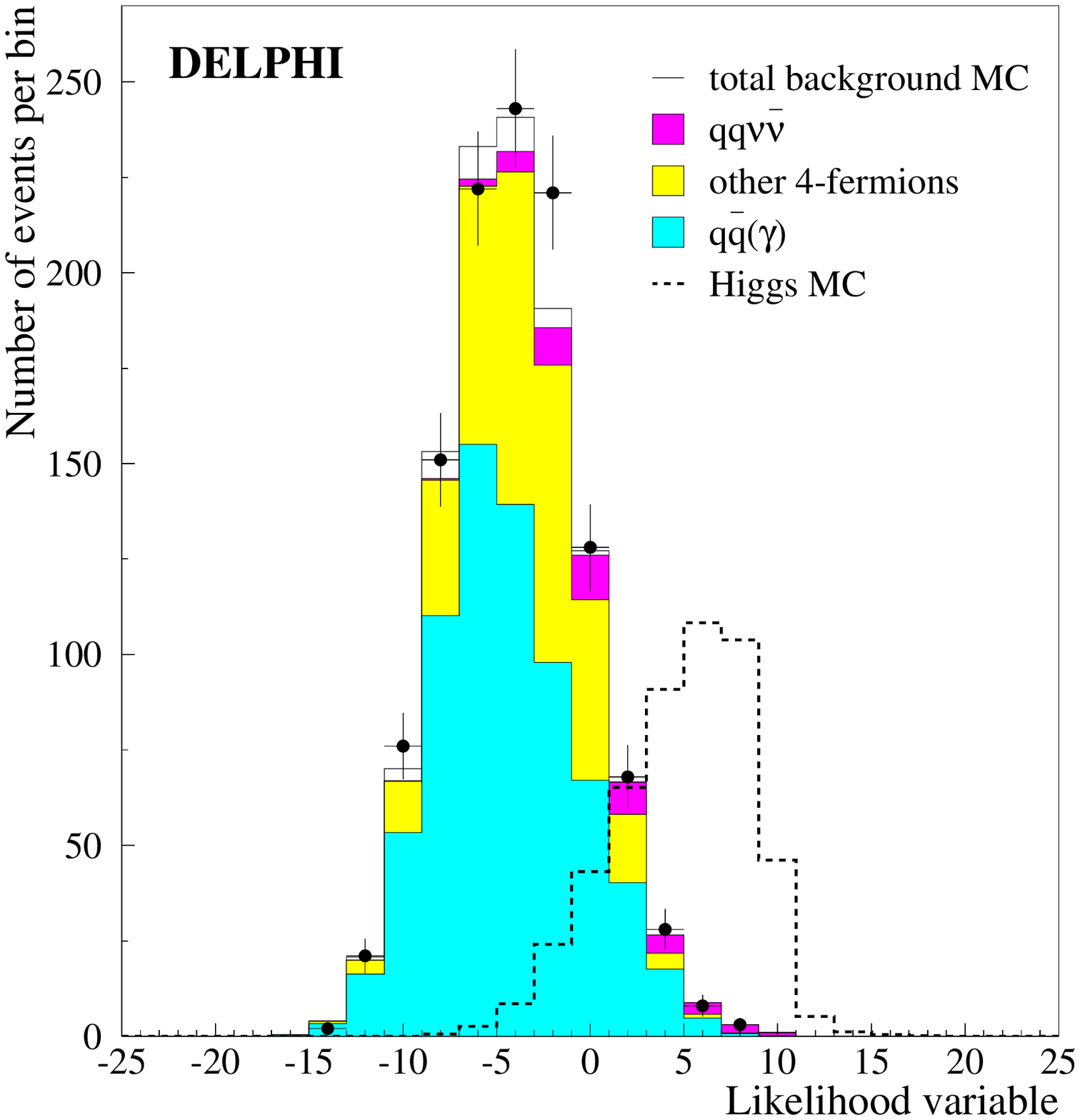,width=11cm} \\
\vspace{-0.5cm}
\epsfig{figure=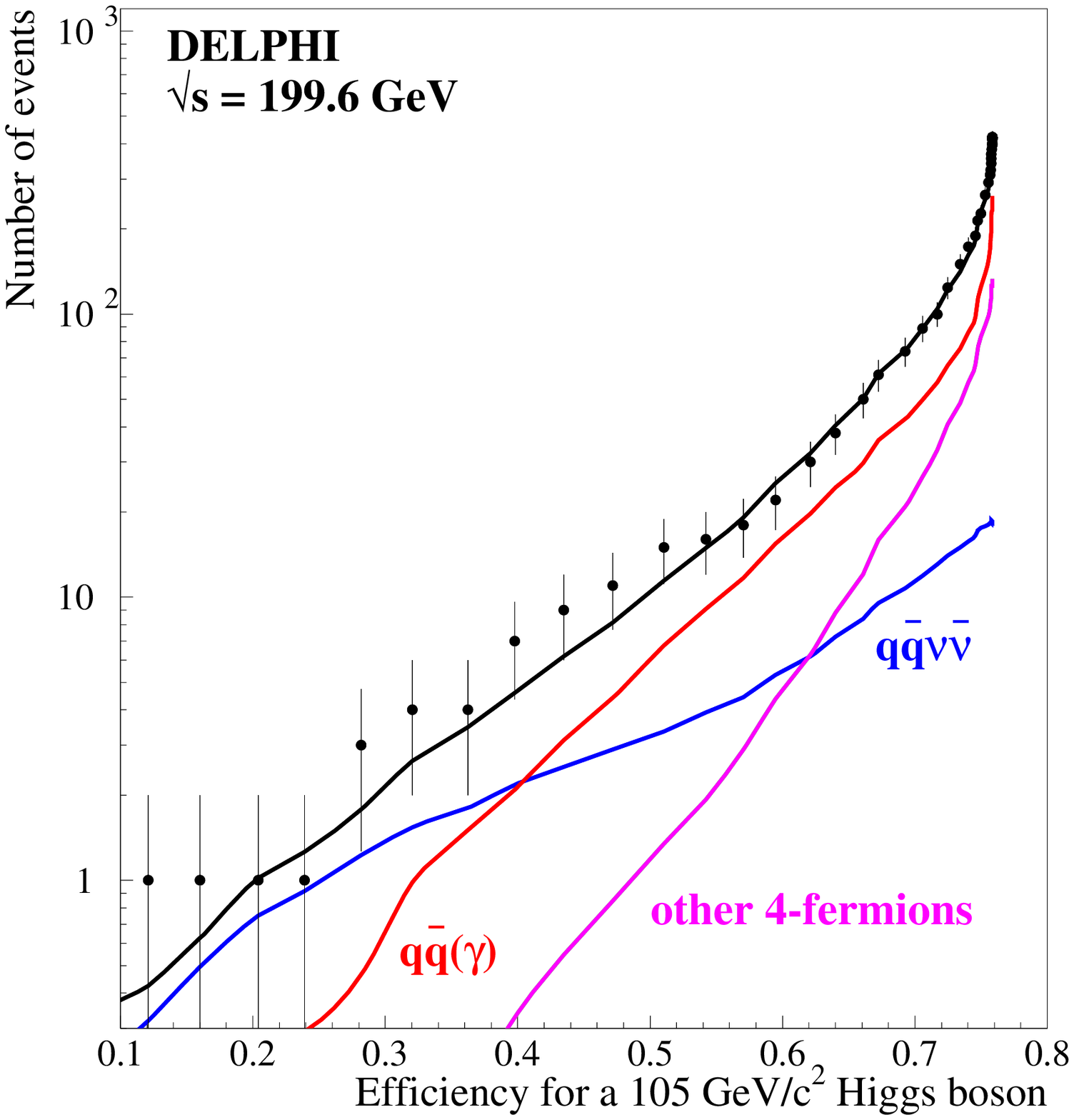,width=11cm}
\caption{ \hnn\ channel: 
  Top: distributions of the likelihood variable 
  for the expected  {\sc SM} backgrounds (full histograms), 
  191.6-201.7~\GeV\ data (dots) and the 
  expected Higgs signal at 105~\GeVcc\ (dashed histogram, 
  normalised to 100 times the expected rate).
  Bottom: curve of the expected  {\sc SM} background rate at \rs~=~199.6~\GeV\ as a 
  function of the efficiency for a 105~\GeVcc\ Higgs signal when
  varying the cut on the likelihood variable. 
  The different background contributions are shown separately. 
  The dots represent the data. }
\label{fig:hnunu_disc}
\end{center}
\end{figure}

The distributions of four of the input variables are shown at preselection
level in Fig.~\ref{fig:hnunu}, while Fig.~\ref{fig:hnunu_disc} shows the 
distribution of the likelihood discriminant variable. The comparison
between the observed and expected rates in the signal-like tail of
this distribution is illustrated further in Fig.~\ref{fig:hnunu_disc}, which 
shows the observed and expected background rates at \rs~=~199.6~\GeVcc\
as a function of the efficiency for a Higgs signal of 105~\GeVcc\ 
when varying the cut on the likelihood variable. 
As a final selection, a minimal value of 1.0 is required, 
leaving 108 events in data for a total expected background of 
\mbox{$105.7 \pm 1.2 (stat.)$}.
The two-dimensional calculation of the confidence levels uses the
likelihood variable and the recontructed Higgs boson mass defined as 
the visible mass given by a one-constraint fit where the recoil system 
is an on-shell Z boson.
  The effect of the selections on data and simulated samples  
are detailed in Tables~\ref{ta:hzsum192} to~\ref{ta:hzsum202},
while the efficiencies at the end of the analysis are reported in
Tables~\ref{ta:hzeff} and~\ref{ta:hzeff_lowm} as a function of \MH.

   Systematic uncertainties due to the use of non-independent samples
in the definition of the p.d.f.s and in the final result derivation were
estimated by comparing the results when running the analysis on the reference
samples and on the complementary samples. The differences between
the two sets of results are then quoted as systematics if they are
higher than the statistical uncertainties. These systematics 
amount to $\pm 2.0\%$ for the efficiencies and 
to $\pm 4.5\%$ for the background
estimates at \rs~=~191.6 and 201.7~\GeVcc\ while no significant
difference is observed for the background estimates at the other two
energies. Systematic uncertainties due to the imperfect modelling of 
the detector response were derived by rescaling the bin contents of each 
p.d.f. from simulation to those in data, restricting to bins where 
the deviation between data and simulation exceeded two standard deviations.
The analysis was then repeated with the rescaled p.d.f. for each
variable in turn and
the largest difference with respect to the initial result taken as systematics.
These amount to $\pm 2.0\%$ on the efficiencies and $\pm 10.0\%$ on 
the background estimates and come from the p.d.f. of the acoplanarity. 
It was checked that these differences remained similar
with tighter selections in the likelihood variable. Thus the overall 
uncertainties are $\pm 3.0\%$ on the efficiencies,  
$\pm 10.0\%$ ($\pm 11.0\%$) on the background estimates
at \rs~=~195.6 and 199.6~\GeVcc\ (191.6 and 201.7~\GeVcc). 

A second analysis using the same preselection criteria followed by
an Iterated nonlinear Discriminant Analysis (IDA) as described 
in~\cite{pap97} 
gave similar results.

\section{Higgs boson searches in pure hadronic events}
\label{sec:4jet}
  Higgs boson searches in pure hadronic final states start with a
common four-jet preselection, which eliminates $\gamma \gamma$ events and 
reduces the \qqg\ and $\Zz\gamma^*$ backgrounds. 
As this step did not change since the
previous analysis, the reader is referred to~\cite{pap97,pap98} 
for the exact description of the cuts and only the
important features are briefly mentioned here.
After a selection of multi-hadron events
excluding those with  
an energetic photon in the calorimeters or lost in the beam pipe, 
topological criteria are applied to select multi-jet events. All 
selected events are then forced into a four-jet topology with the 
DURHAM algorithm and a minimal multiplicity and mass is required 
for each jet. After the preselection, different analysis 
procedures are applied in the \ZH\ and \hA\ channels.

\subsection{The HZ four-jet channel}
\label{sec:1}

  After the common four-jet preselection, events are selected using a 
discriminant variable defined as the output of an artificial neural 
network~\cite{jetnet} which combines four variables. 
Three of them are introduced to reduce the four-fermion contamination. 
The first relies on b-tagging and is the maximum b-tagging 
variable of any di-jet in the event, 
a di-jet b-tagging variable being defined as the sum of the two jet
b-tagging variables, \xbi.
The second  and third variables  rely on mass information and
test the compatibility of the event with the hypotheses of \WW\ and 
\ZZ\ pair-production, respectively. First, constrained fits are used 
to derive the probability density function measuring the compatibility 
of the event kinematics with the production of two objects of any masses. 
This two-dimensional probability, called the ideogram probability, is then
folded with the expected mass distributions for the \WW\ and \ZZ\
processes, respectively. More about the ideogram technique can be 
found in~\cite{ww183}.
Finally, the fourth input variable to the neural network is
intended to reduce the \qqg\ contamination and
is the output of another neural network~\cite{mlpfit} (anti-QCD neural network)
constructed from eight variables. These are mostly shape or jet variables:
the sum of the second and fourth Fox-Wolfram moments,
the product of the minimum jet energy and the minimum opening angle between
any two jets,
the maximum and minimum jet momenta,
the sum of the multiplicities of the two jets with lowest multiplicity and
the sum of the masses of the two jets with lowest masses.
For the last two variables, the six possible pairings of the jets are
considered and the variables are defined as 
the minimum di-jet mass and
the minimum sum of the cosines of the opening angles of the two dijets
in any pairing.
As the discrimination between the \qqg\ background and the \hqq\ signal
provided by these variables depends mainly on the difference \rs$-$\MH,
the anti-QCD neural network was trained with simulations 
at \rs~=~189~GeV, using \qqg\ events and 95~\GeVcc\ \hqq\ events.
In order to minimize the risk of overtraining,
the neural network used for the final discrimination 
was trained with fractions of the available
simulated samples at \rs~=~195.6~\GeV\ in \qqg\ background (10\%), 
four-fermion background (50\%), and 105~\GeVcc\ \hqq\ signal (50\%).
The whole samples were used to derive the final results.

\begin{figure}[htbp]
\begin{center}
\begin{tabular}{cc}
\epsfig{figure=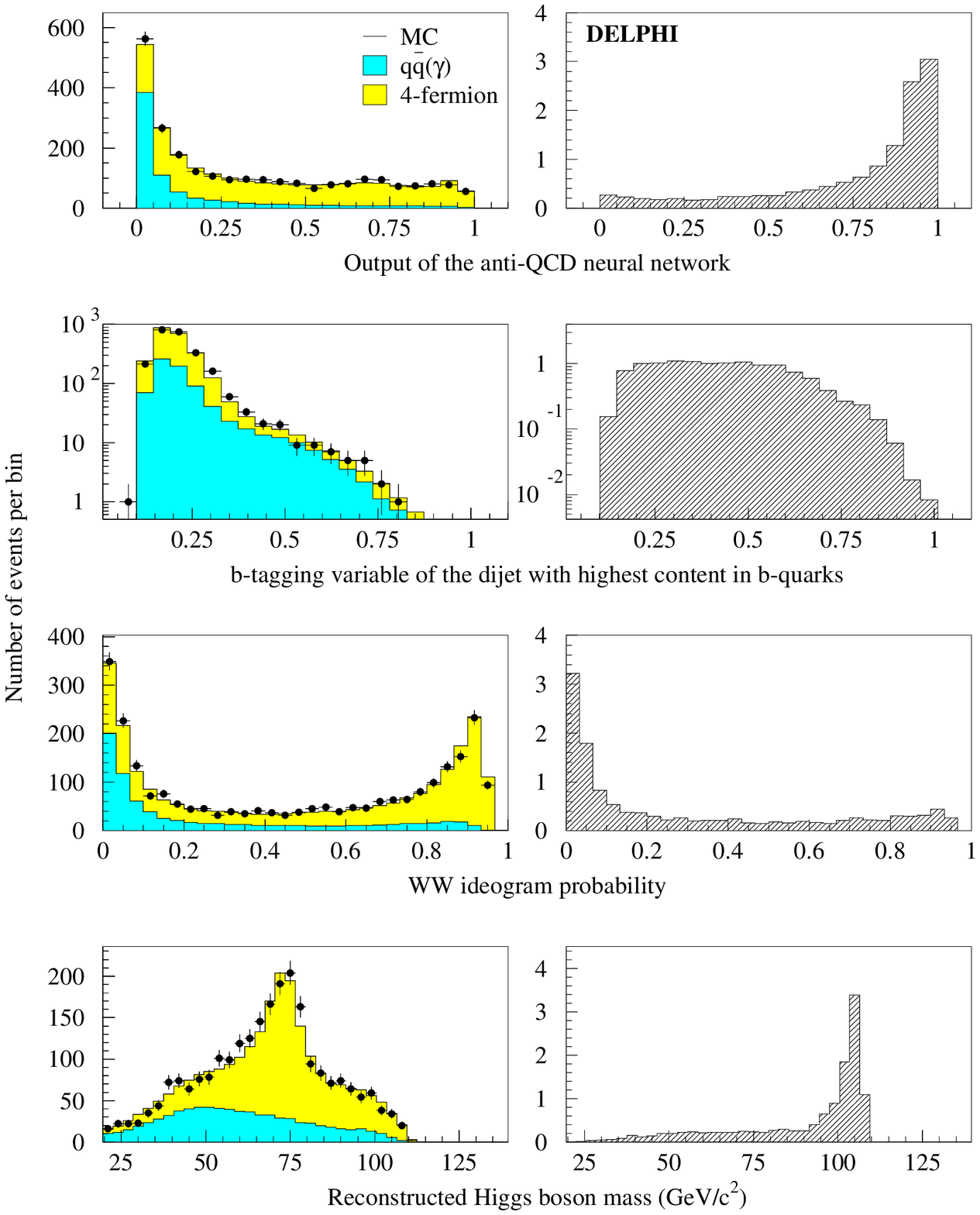,width=17cm} 
\end{tabular}
\caption[]{\hqq\ channel: 
  distributions of four analysis variables,
  as described in the text, at preselection level. Data at 
  \rs~=~191.6-201.7~\GeV\ (dots) are compared with  {\sc SM} background 
  expectations (left-hand side histograms) and with the expected distribution
  for a 105~\GeVcc\ signal (right-hand side histogram, 
  normalised to the expected rate).}
\label{fig:hqq}
\end{center}
\end{figure}

\begin{figure}[htbp]
\begin{center}
\epsfig{figure=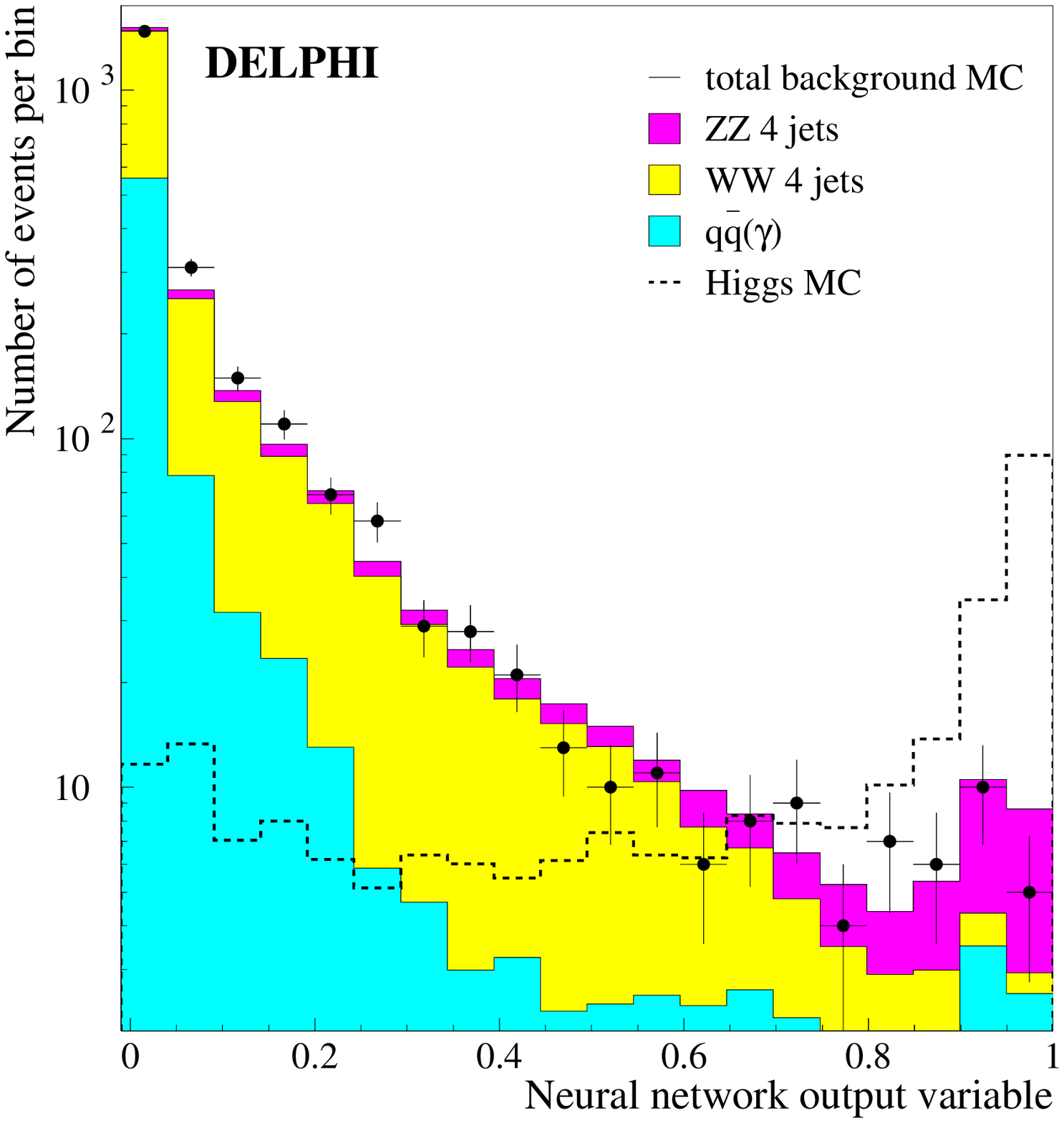,width=11cm} \\
\vspace{-0.5cm}
\epsfig{figure=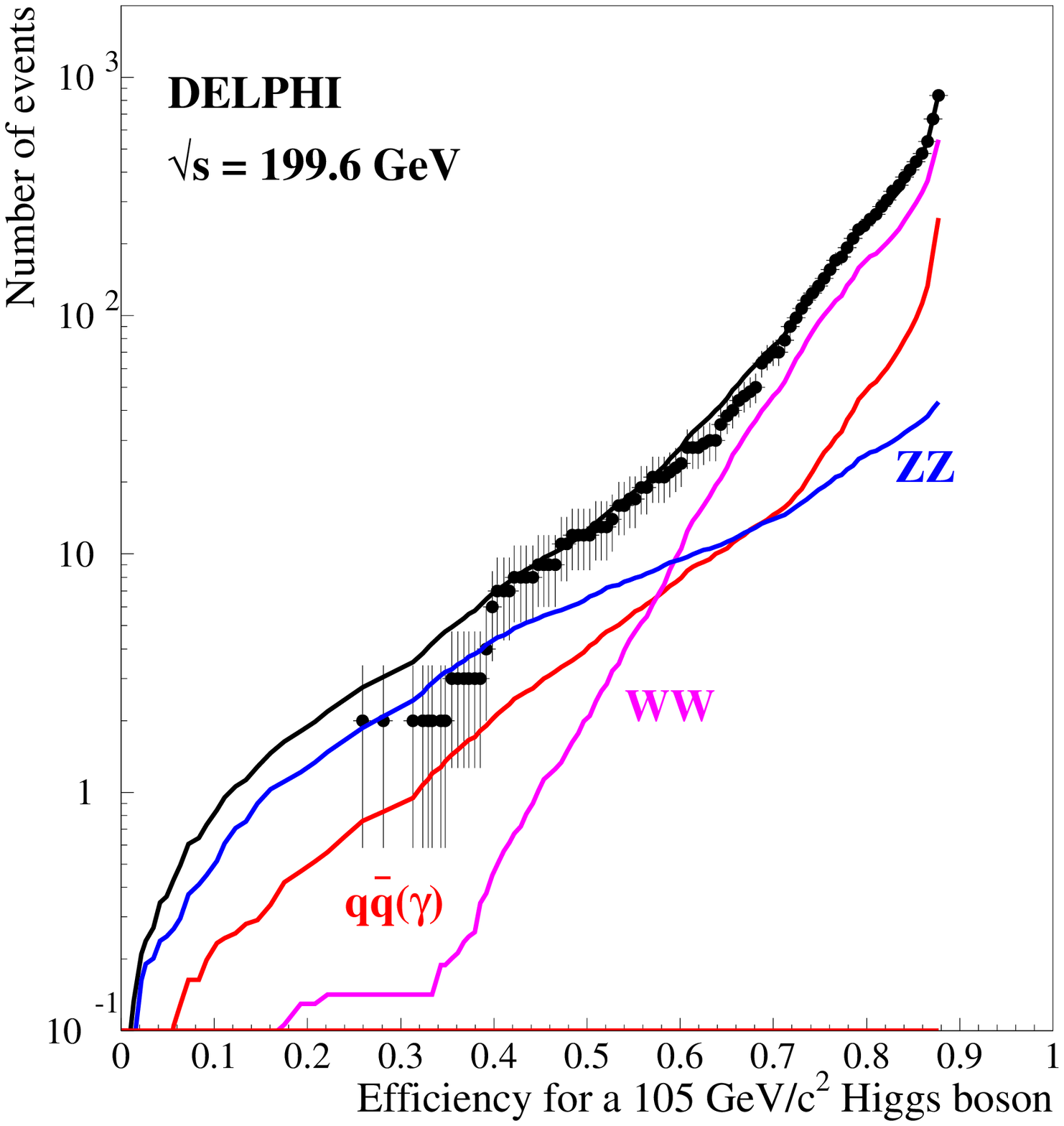,height=11cm}
\caption{\hqq\ channel: 
  Top: distributions of the neural network variable 
  for the expected  {\sc SM} backgrounds (full histograms), 
  191.6-201.7~\GeV\ data (dots) and the 
  expected Higgs signal at 105~\GeVcc\ (dashed histogram, 
  normalised to 20 times the expected rate).
  Bottom: curve of the expected  {\sc SM} background rate at \rs~=~199.6~\GeV\ as a 
  function of the efficiency for a 105~\GeVcc\ Higgs signal when
  varying the cut on the neural network variable. 
  The different background contributions are shown separately. 
  The dots represent the data. }
\label{fig:hqq_disc}
\end{center}
\end{figure}

The agreement between data and background simulation after the four-jet
preselection is illustrated 
in  Fig.~\ref{fig:hqq} which shows the distributions 
of three analysis variables and of the recontructed Higgs boson mass 
obtained as explained below. Fig.~\ref{fig:hqq_disc} shows the 
distribution of the final neural network output variable and, as an example,
the expected background rate and the data at \rs~=~199.6~GeV,
as a function of the efficiency 
for a 105~\GeVcc\ signal when varying the cut on the neural network
output variable.
As a final selection, a minimal value of 0.3 is required. This
suppresses the most background-like events, leaving 161 events in data and
a total expected background of \mbox{$175.4 \pm 1.3 (stat.)$}. 
%coming from \qqg\ and four-fermion processes.
  The effect of the selections on data and simulated samples  
are detailed in Tables~\ref{ta:hzsum192} to~\ref{ta:hzsum202},
while the efficiencies at the end of the analysis are reported in
Tables~\ref{ta:hzeff} and~\ref{ta:hzeff_lowm} as a function of \MH\ 
for the \hqq\ channel
and in Table~\ref{ta:aaqqeff} as a function of \MH\ and \MA\ for
the (\hAA) \qqbar\ channels. 
Since these two channels, specific to the {\sc MSSM}, are covered 
by the same analysis as that of the \hqq\ channel,
the three channels cannot be considered as independent in the
confidence level computation when testing {\sc MSSM} models. 
For this computation,
they are combined into one global \hqq\ channel:
at each test point, the signal expectations (rate and two-dimensional 
distribution as defined below) in this channel are obtained by summing 
the contributions from the three original signals weighted by their 
expected rates.

 The two-dimensional calculation of the confidence levels uses the
final neural network variable and the recontructed H boson mass 
estimated as follows. For each of the six possible pairings of jets
into an \ZH\ pair, a kinematic fit is applied, requiring energy and 
momentum 
conservation and one di-jet to be at the nominal \Zz\ mass.
The pairing of jets defining the Higgs boson and \Zz\ candidates is then 
that which maximises the probability~\cite{pap97} that both
the b-content of the different jets and the $\chi^2$ probability of the 
five-constraint fit are compatible with the production of an \ZH\ pair.

      The systematic uncertainties from the imperfect modelling of the
detector response were estimated by repeating the selection procedure on
the distribution of the neural network variable obtained by smearing, 
in turn, each of
the distributions of the three input variables according to the resolution 
in the variable. 
%The smearing was assumed to be gaussian with a width given by the
%the resolution in the variable and was applied on an event-by-event basis.
This leads to
relative uncertainties of $\pm 6.0\% $ related to b-tagging, 
$\pm 3.0\%$ related to the 
anti-QCD variable and $\pm 2.5\%$ related to the WW ideogram probability. This
results in an overall relative uncertainty of $\pm 7.2\%$ in the background
and efficiency estimates at each centre-of-mass energy.

\subsection{The hA four-b channel}

  The analysis is very similar to that published in~\cite{pap98}.
  After the common four-jet preselection, events are preselected further,
requiring a visible energy greater than 120~\GeV, 
%an effective centre-of-mass energy 
 \mbox{$ \sqrt{s'}$} greater than 150~\GeV, 
a missing momentum component along the beam direction lower than 30~\GeVc\
and at least two charged particles per jet. 
A four-constraint kinematic fit requiring energy and momentum
conservation is then applied, and the two di-jet masses are calculated
for each of the three different jet pairings. As the possible
production
of {\sc MSSM} Higgs bosons through the \hA\ mode 
dominates at large \tbeta\
where the two bosons are almost degenerate in mass,
the pairing defining the Higgs boson candidates is chosen as that
which minimizes the mass difference between the two di-jets.  
The final discrimination between background and signal is then based on
a multidimensional variable which combines the following eight variables
with a likelihood ratio method:
the event thrust, 
the second and fourth Fox-Wolfram moments, 
%the minimal difference between the two di-jet masses when considering the 
%three possible pairings of jets,
the difference between the Higgs boson candidate masses as given by the 
kinematic fit,
the production angle of the Higgs boson candidates, 
%identified as the di-jets
%in the pairing corresponding to this minimal mass difference, 
the sum of the four jet b-tagging variables,
the minimum di-jet b-tagging variable and 
the number of secondary vertices.
%For each event, the measured value of each of these discriminant variables is 
%compared with probability density functions obtained from simulated events.
For each variable, probability density functions (p.d.f.s) were obtained from
simulated events, using fractions of the statistics available 
in the \qqg\ background (40\%) and four-fermion background (80\%)
at \rs~=~195.6~GeV and 199.6~GeV and in
signal events with \MA~=~85, 90~\GeVcc\ and \tbeta~=20 (50\%) 
at \rs~=~195.6~GeV. The whole samples were used to derive the 
final results.

\begin{figure}[htbp]
\begin{center}
\begin{tabular}{cc}
\epsfig{figure=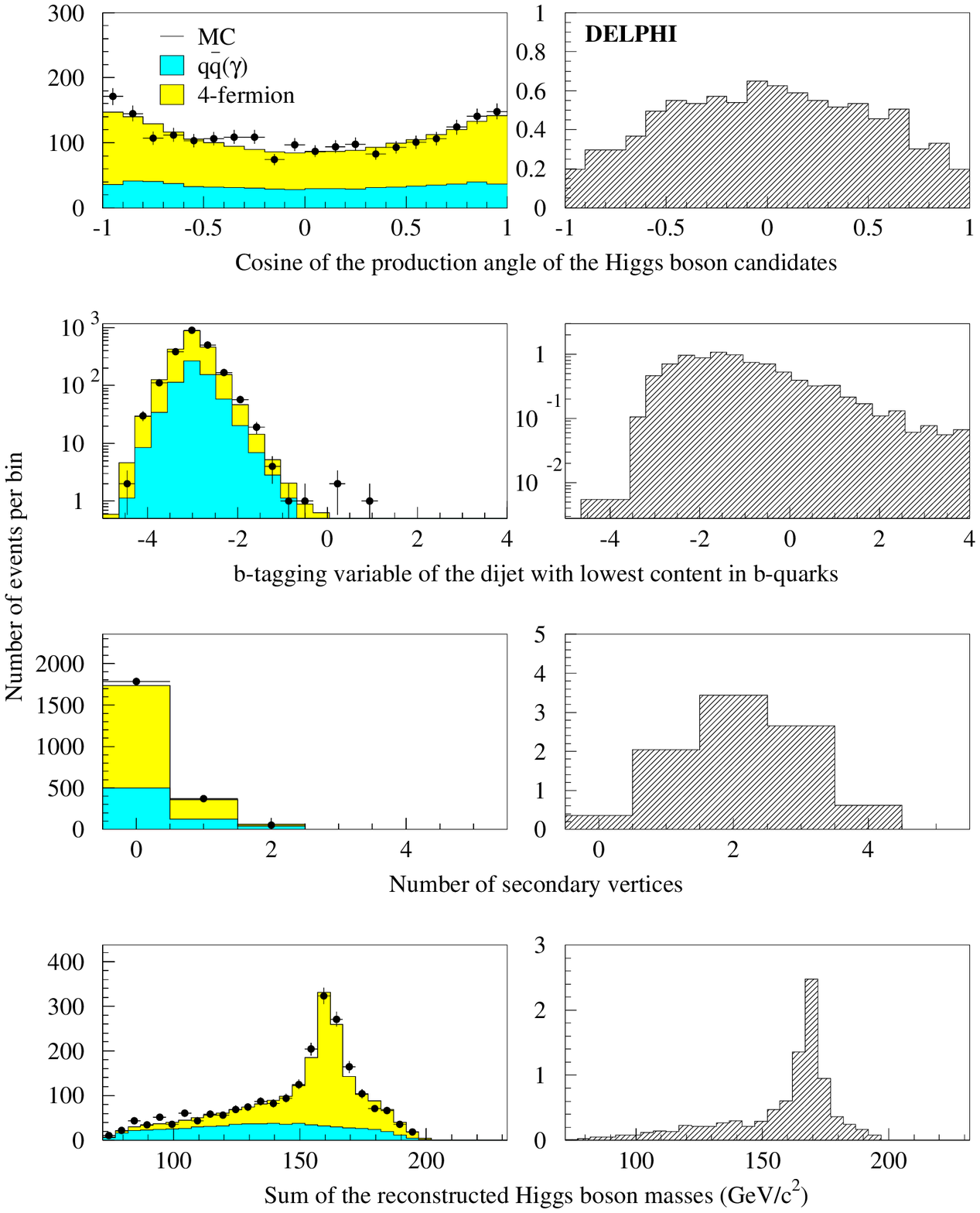,width=17cm} 
\end{tabular}
\caption[]{hA hadronic channel: 
  distributions of four analysis variables,
  as described in the text, at preselection level. Data at 
  \rs~=~191.6-201.7~\GeV\ (dots) are compared with  {\sc SM} background 
  expectations (left-hand side histograms) and with the expected distribution
  for a 85~\GeVcc\ signal at \tbeta~=~20 (right-hand side histogram,
  normalised to the expected rate).}
\label{fig:4b}
\end{center}
\end{figure}

\begin{figure}[htbp]
\begin{center}
\epsfig{figure=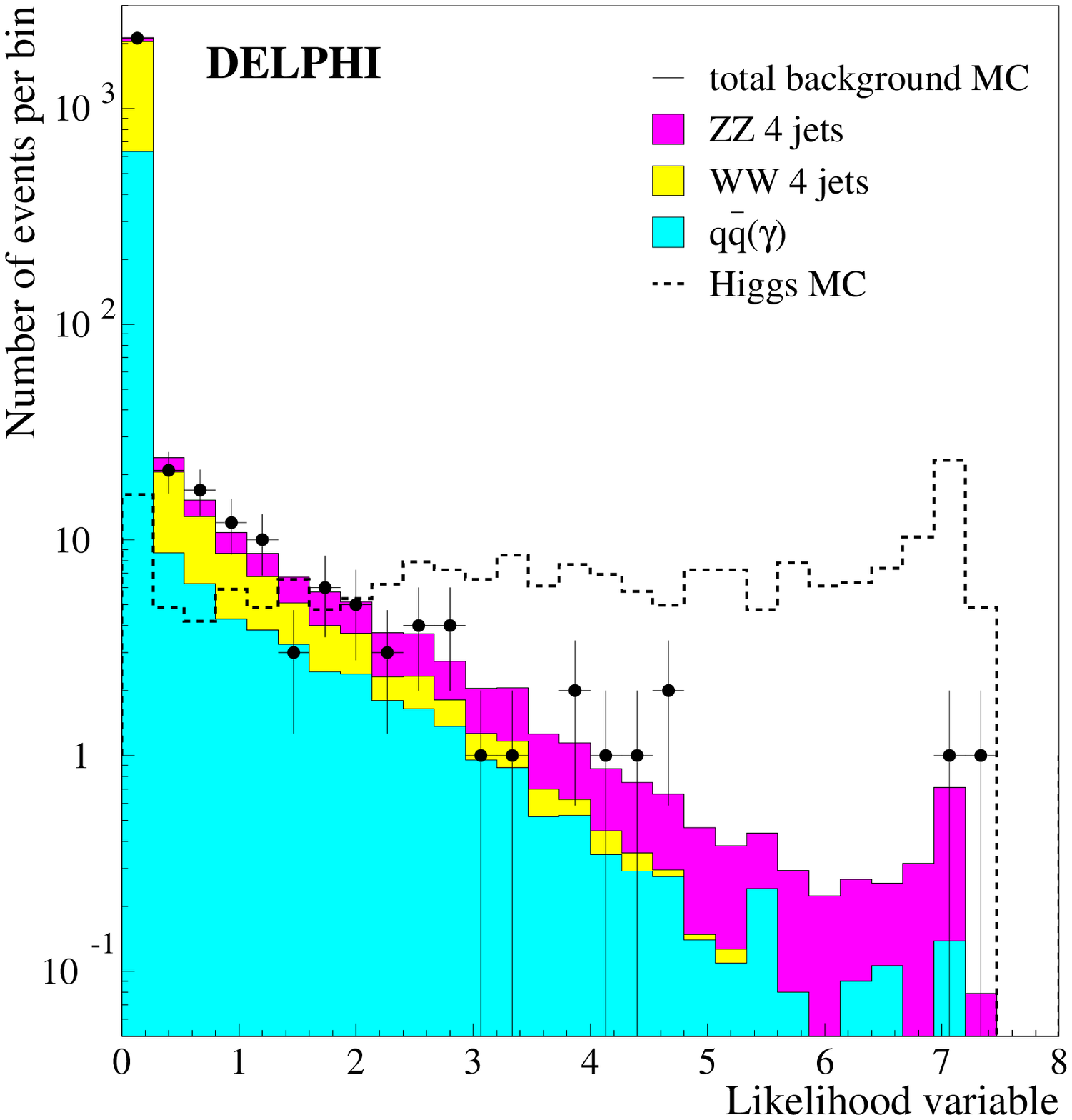,width=11.cm} \\
\vspace{-0.5cm}
\epsfig{figure=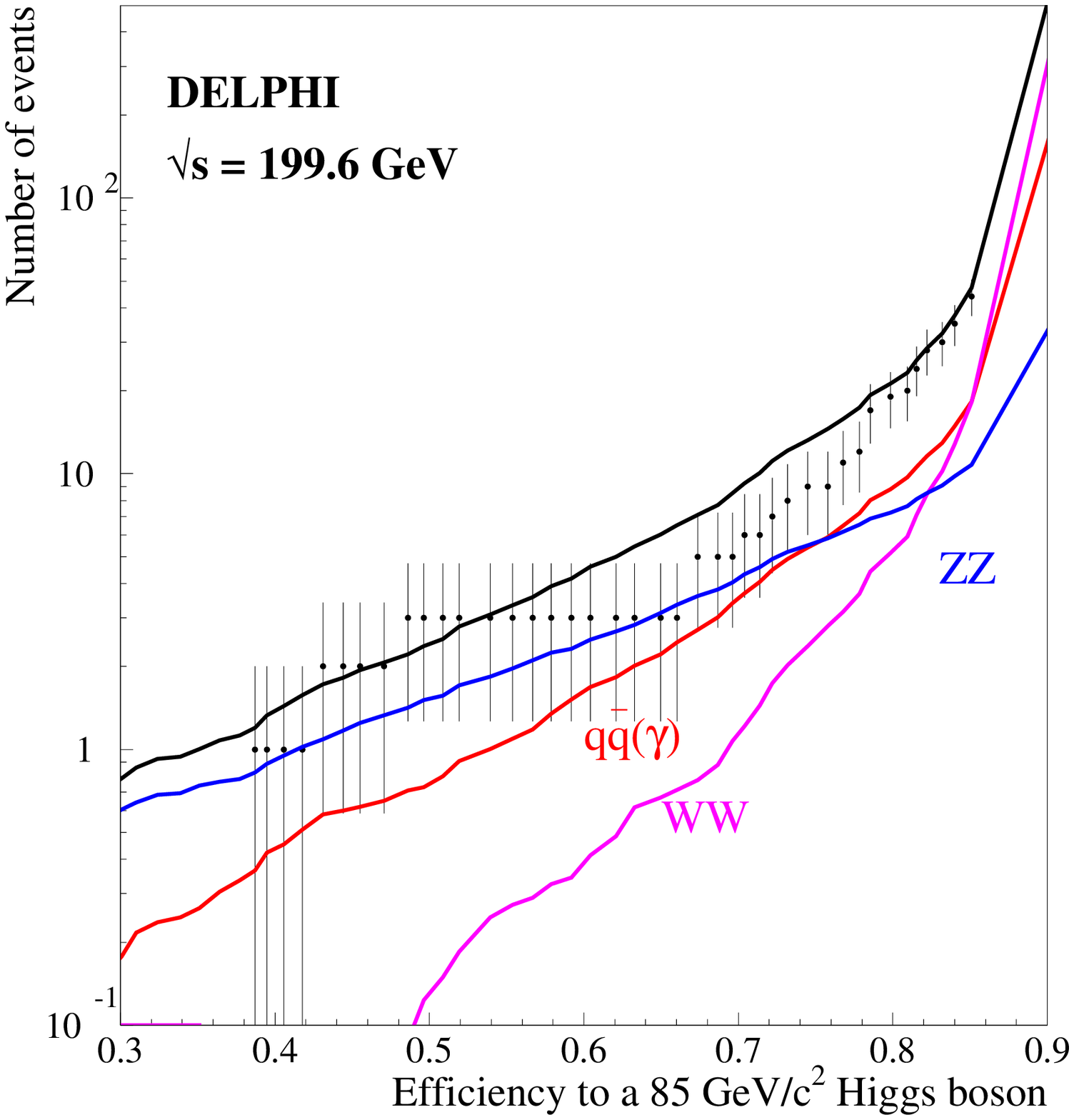,width=11.cm}
\caption{hA hadronic channel: 
Top: distributions of the likelihood variable 
  for the expected  {\sc SM} backgrounds (full histograms), 
  191.6-201.7~\GeV\ data (dots) and the 
  expected 85~\GeVcc\ Higgs signal at \tbeta~=~20
  (dashed histogram, normalised to 20 times the expected rate).
  Bottom: curve of the expected  {\sc SM} background rate at \rs~=~199.6~\GeV\ as a 
  function of the efficiency for a 85~\GeVcc\ Higgs signal 
  at \tbeta~=~20 when varying the cut on the likelihood variable. 
  The different background contributions are shown separately. 
  The dots show the data. }
\label{fig:4b_disc}
\end{center}
\end{figure}

The agreement between data and background simulation after the 
preselection 
is illustrated in  Fig.~\ref{fig:4b} which shows the distributions 
of three input variables and of the sum of the recontructed Higgs boson masses
%obtained as explained below. 
as given by the kinematic fit.
Fig.~\ref{fig:4b_disc} shows the 
distribution of the final discriminant variable and, as an example,
the expected background rate and the data at \rs~=~199.6~GeV,
as a function of the efficiency for a signal with \MA = 85 \GeVcc\ 
and \tbeta~=~20, when varying the cut on the discriminant variable.
As a final selection, a minimal value of 0.1 is required, leading to
136 events in data, for a total expected background of 
\mbox{$137.8 \pm 1.2 (stat.)$}.
%coming equally from \qqg\ and four-fermion processes.
  The effect of the selections on data and simulated samples  
 are detailed in Tables~\ref{ta:hzsum192} to~\ref{ta:hzsum202},
while the efficiencies at the end of the analysis are reported in
Tables~\ref{ta:haeff} and~\ref{ta:scaneff} as functions of 
\MA\ and \tbeta\ and of \MA\ and \mh.

 The two-dimensional calculation of the confidence levels uses the
likelihood variable and the sum of the reconstructed Higgs boson masses
as given by the kinematic fit.
%as provided by a four-constraint kinematic fit requiring energy and momentum
%conservation. The pairing is that which minimizes the mass difference
%between the two di-jets.  

   Systematic uncertainties due to the use of non-independent samples
in the definition of the p.d.f.s and in the final result derivation were
estimated at the level of  $\pm 4.0\%$ relative, by repeating the whole
procedure with two independent samples of lower size. Systematic
uncertainties due to the imperfect modelling of the detector response
were derived as in the previous section. The uncertainty 
related to b-tagging amounts to  $\pm 5.0\%$ and that related to shape
variables to  $\pm 3.0\%$, resulting in an overall 
relative uncertainty of  $\pm 7.0\%$ on background and efficiency 
estimates 
at each centre-of-mass energy.

\section{Results}
\label{sec:results}

The results of the searches presented in the previous
sections can be translated into exclusion limits on the masses of the 
neutral Higgs bosons in the {\sc SM} and {\sc MSSM}.

\subsection{Reconstructed mass spectra}
%  For each analysis of the \ZH\ and \hA\ channels at 192-202~\GeV,
% the integrated luminosity, the expected backgrounds and their errors,
% and the number of observed events at various levels of the analyses are
% summarised in Tables~\ref{ta:hzsum192} to~\ref{ta:hzsum202}. 
% Within each channel, the penultimate line represents the inputs for the 
% confidence level calculations (``final selection''), while 
% the last line gives the result of a tighter selection. 
% The efficiencies versus 
% Higgs mass at the final selection level can be seen in Table~\ref{hzeff} 
% ( {\sc SM} channels) and Table~\ref{haeff} ( {\sc MSSM} channels). 

\begin{figure}[htbp]
\begin{center}
\epsfig{figure=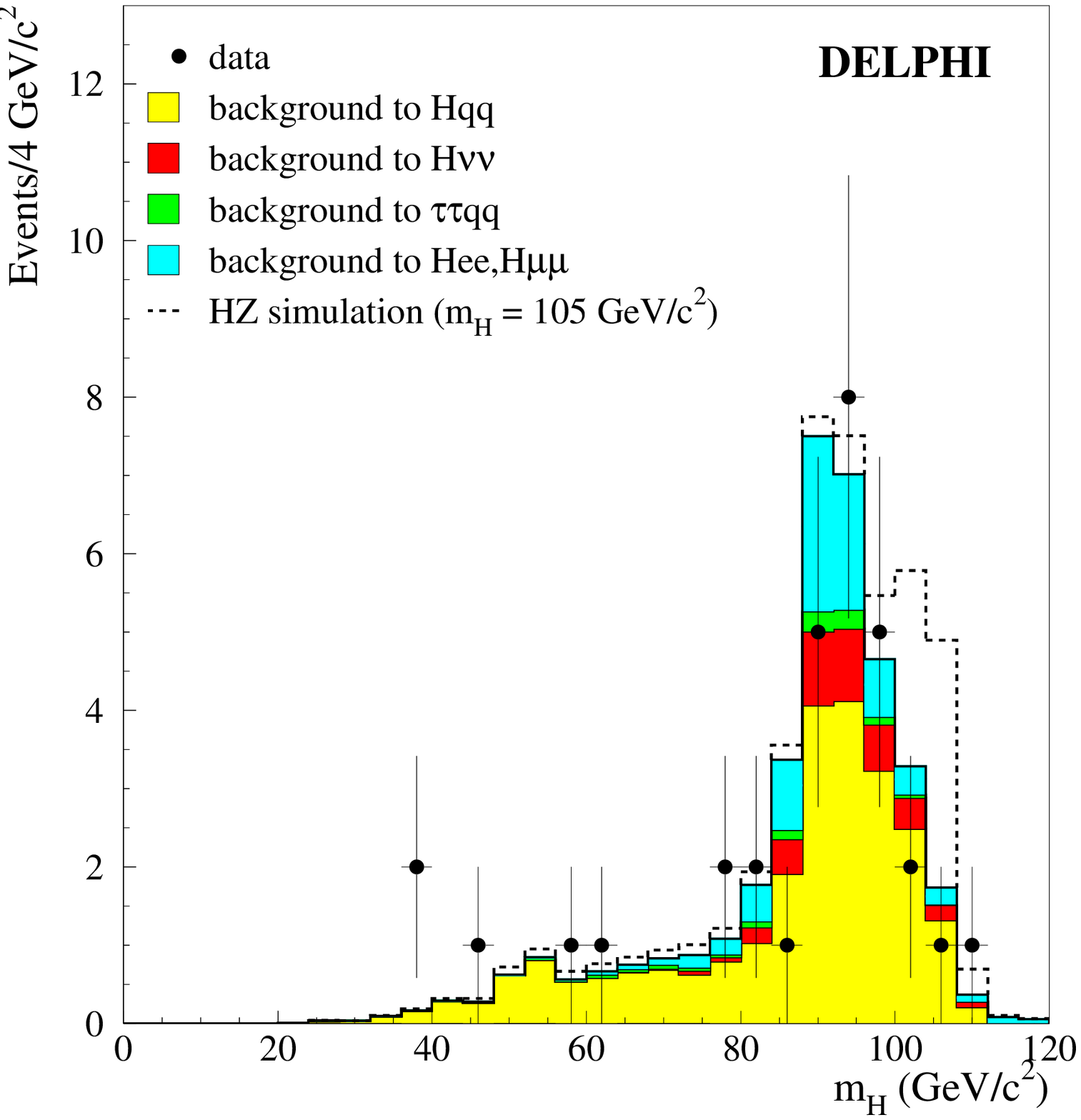,height=11.cm} \\
\vspace{-0.5cm}
\epsfig{figure=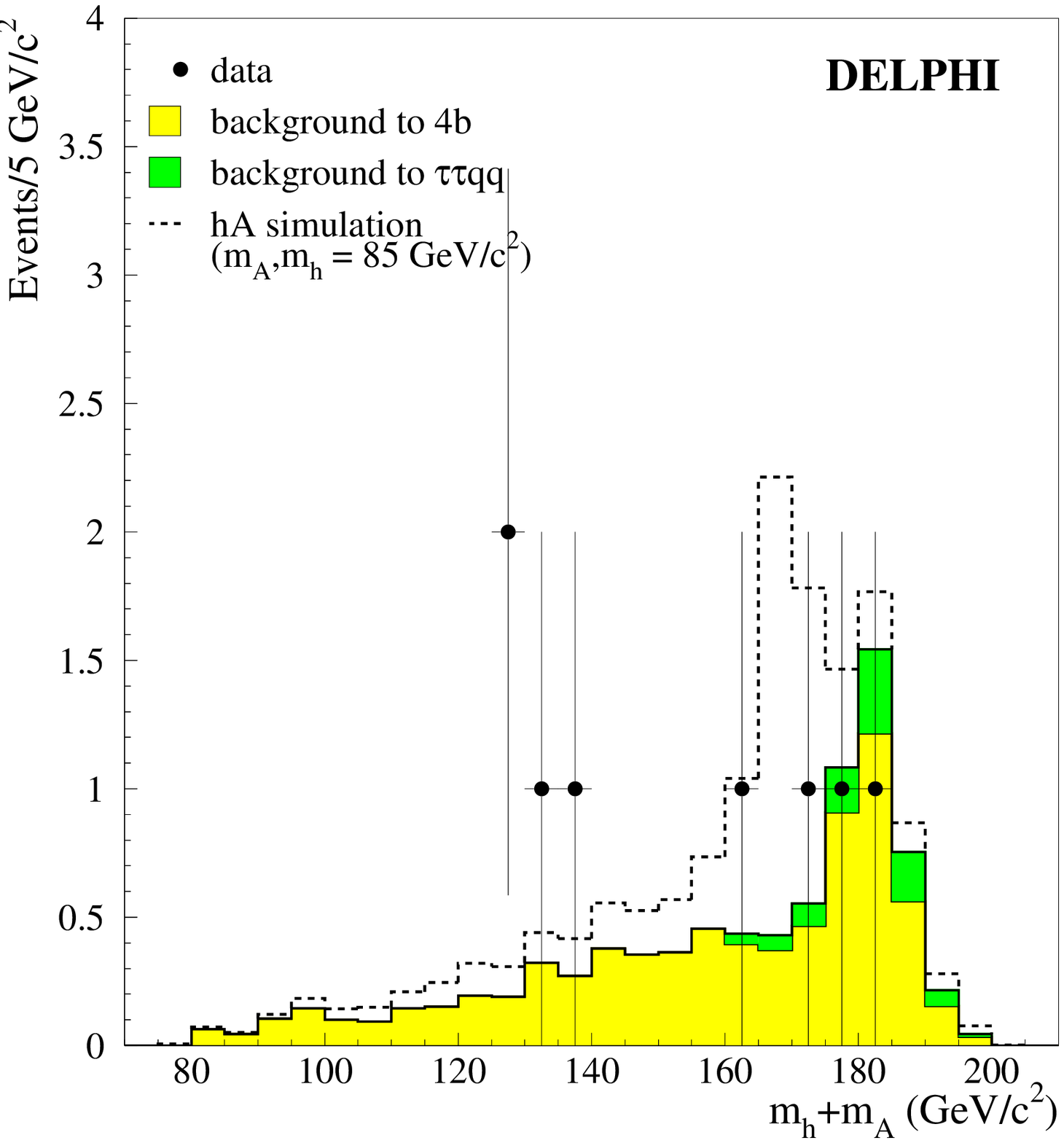,height=11.cm}
\caption[]{
Distribution of the reconstructed Higgs boson mass(es)
when combining all \ZH\ or \hA\ analyses at 191.6-201.7~\GeV. 
Data (dots) are compared with  {\sc SM} background expectations (full 
histograms)
and with the normalised signal spectrum added to the background 
contributions (dashed histogram). Mass hypotheses for the simulated signal
spectra are 
\MH~=~105~\GeVcc\ in the \ZH\ channel and 
\MA~=~85~\GeVcc, \tbeta~=~20 in the \hA\ channel.}
\label{fig:mass_pl}
\end{center}
\end{figure}

   As an illustration of the discrimination achieved against the 
residual  {\sc SM} background,
%in the signal-like region
distributions of the reconstructed Higgs boson mass(es) after tight selections 
are presented in Fig.~\ref{fig:mass_pl} in the \ZH\ and 
\hA\ channels. The selections correspond to requiring a minimal 
b-tagging value of -1.8 in the \hee\ and \hmm\ channels, minimal
likelihood values of 0.8, 7.0 and 3.5 in the \ttqq, \hnn\ and 4b
channels, respectively, and a minimal neural network output of 0.85 in the
\hqq\ channel. The corresponding 
observed and expected rates at each of the four
centre-of-mass energies are summarized in Table~\ref{ta:summ}.
%The errors are obtained by
%summing the statistical and systematic uncertainties quadratically.

\subsection{The  {\sc SM} Higgs boson}
\label{sec:smresults}

We proceed to set a limit on the  {\sc SM} Higgs boson mass, combining
the data analysed in the previous sections with those taken at 
lower energies, namely 
161.0, 172.0~\GeV~\cite{pap96}, 182.7~\GeV~\cite{pap97} and 
188.7~\GeV~\cite{pap98}.
The expected cross-sections and branching ratios are taken 
from the database provided by the LEP Higgs working group,
using the {\tt HZHA}~\cite{hzha} package, Version 3,
%from~\cite{gross,spira}, 
with the top mass set to 174.3~\GeVcc. 
%with the top mass set to 175~\GeVcc. 

\begin{figure}[htbp]
\begin{center}
\epsfig{figure=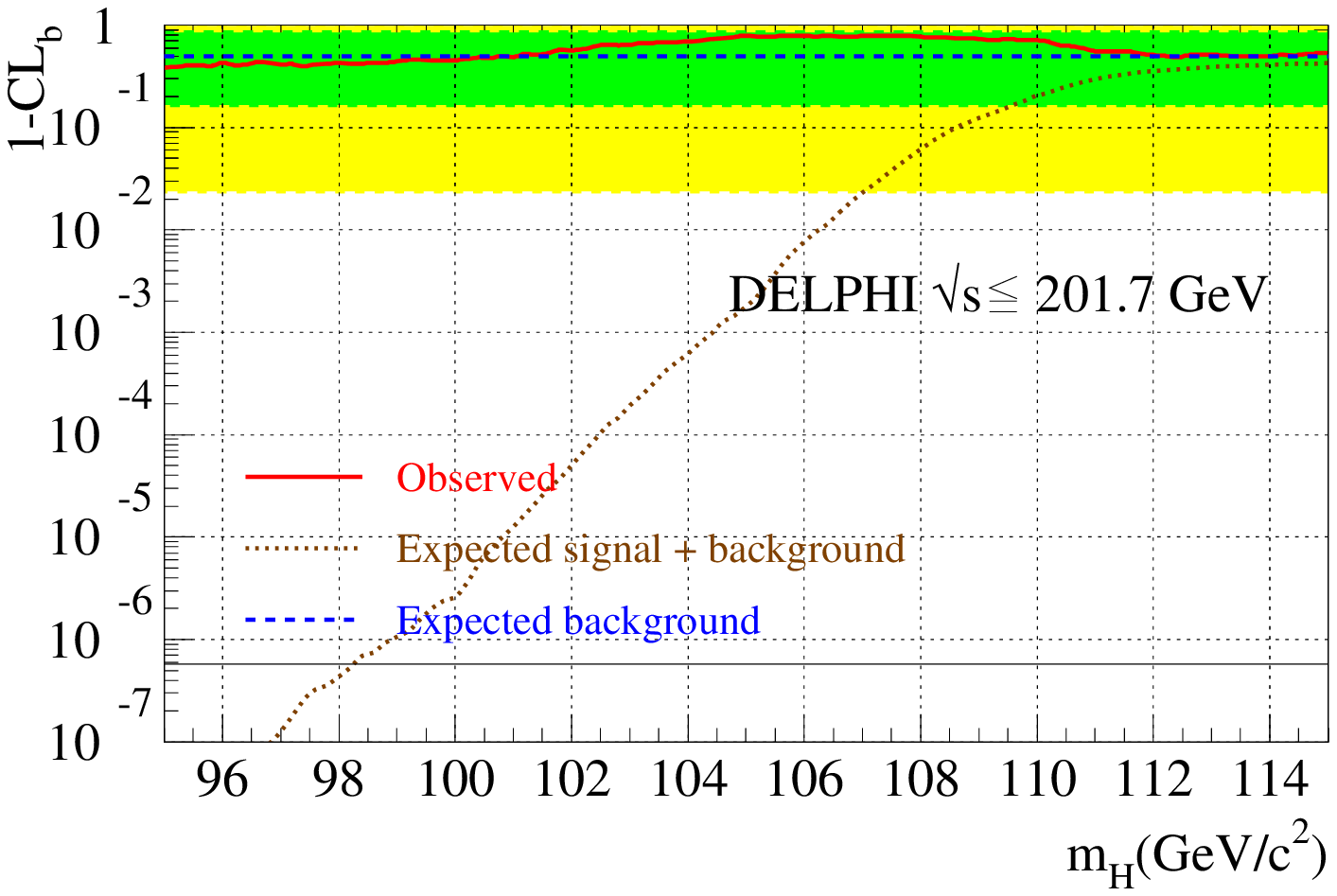,width=14.cm}\\
\epsfig{figure=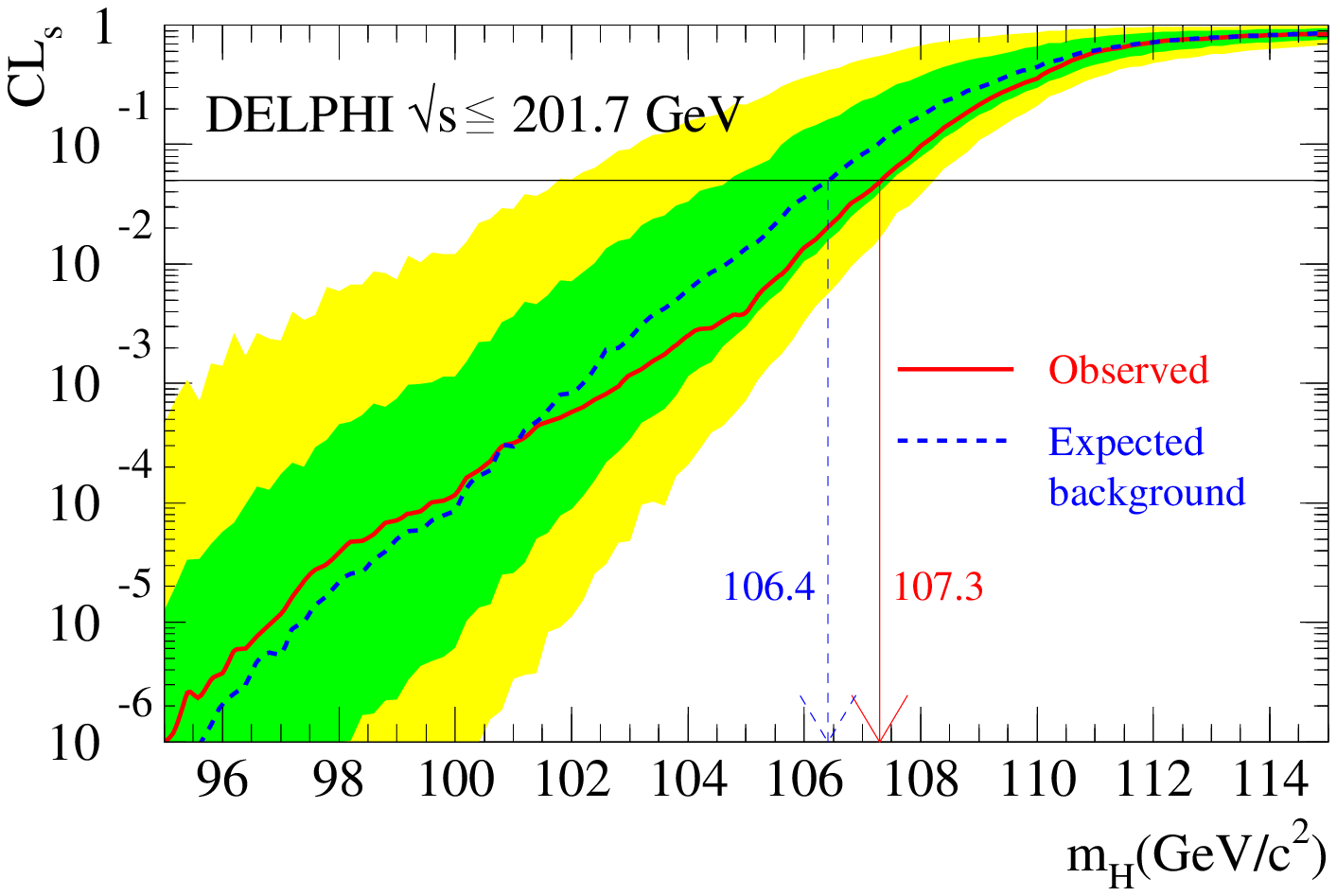,width=14.cm}
\caption[]{
 {\sc SM} Higgs boson: confidence levels as a function of \MH.
Curves are the observed (solid) and expected median 
(dashed) confidences from background-only experiments
while the bands correspond to the 68.3\% and 95.0\% confidence intervals 
from background-only experiments.
Top: 1- CL$_{\rm b}$ for the background hypothesis. Also shown here
is the curve of the median confidence as expected for a signal of mass given
in the abscissa (dotted line). The sensitivity for a 5$\sigma$
discovery, defined by the horizontal line at 5.7~10$^{-7}$, is 
for Higgs masses up to 98.2~\GeVcc.
Bottom: CL$_{\rm s}$, the pseudo-confidence level 
for the signal hypothesis. The intersections of the 
curves with the horizontal line at 5\% define the 
expected and observed 95\% CL lower limits on \MH.}
\label{fig:cl_sm}
\end{center}
\end{figure}

Curves of the confidence level $CL_b$ and $CL_s$ as a function of
the test mass \MH\ are shown in Fig.~\ref{fig:cl_sm}.
In the presence of a sizeable Higgs signal,
the value of the observed $CL_b$ (top of Fig.~\ref{fig:cl_sm}) 
would approach one, since it measures the fraction of background-only
experiments which are more background-like than the observation. 
Here the compatibility between the observation and the expectation 
from background-only is well within one standard deviation over 
the range of masses tested. 
Moreover, the mass giving an expected 5$\sigma$ discovery, defined by 
the intersection of the curve for signal plus background experiments with 
the horizontal line at $1-CL_b = 5.7\times 10^{-7}$, is 98.2~\GeVcc. 
The pseudo-confidence level in the signal is shown in Fig.~\ref{fig:cl_sm} 
(bottom).
The observed 95\% {\sc CL} lower limit on the mass is 107.3~\GeVcc\
while  the expected median limit is $106.4~\GeVcc$.

\begin{figure}[htbp]
\begin{center}
\epsfig{figure=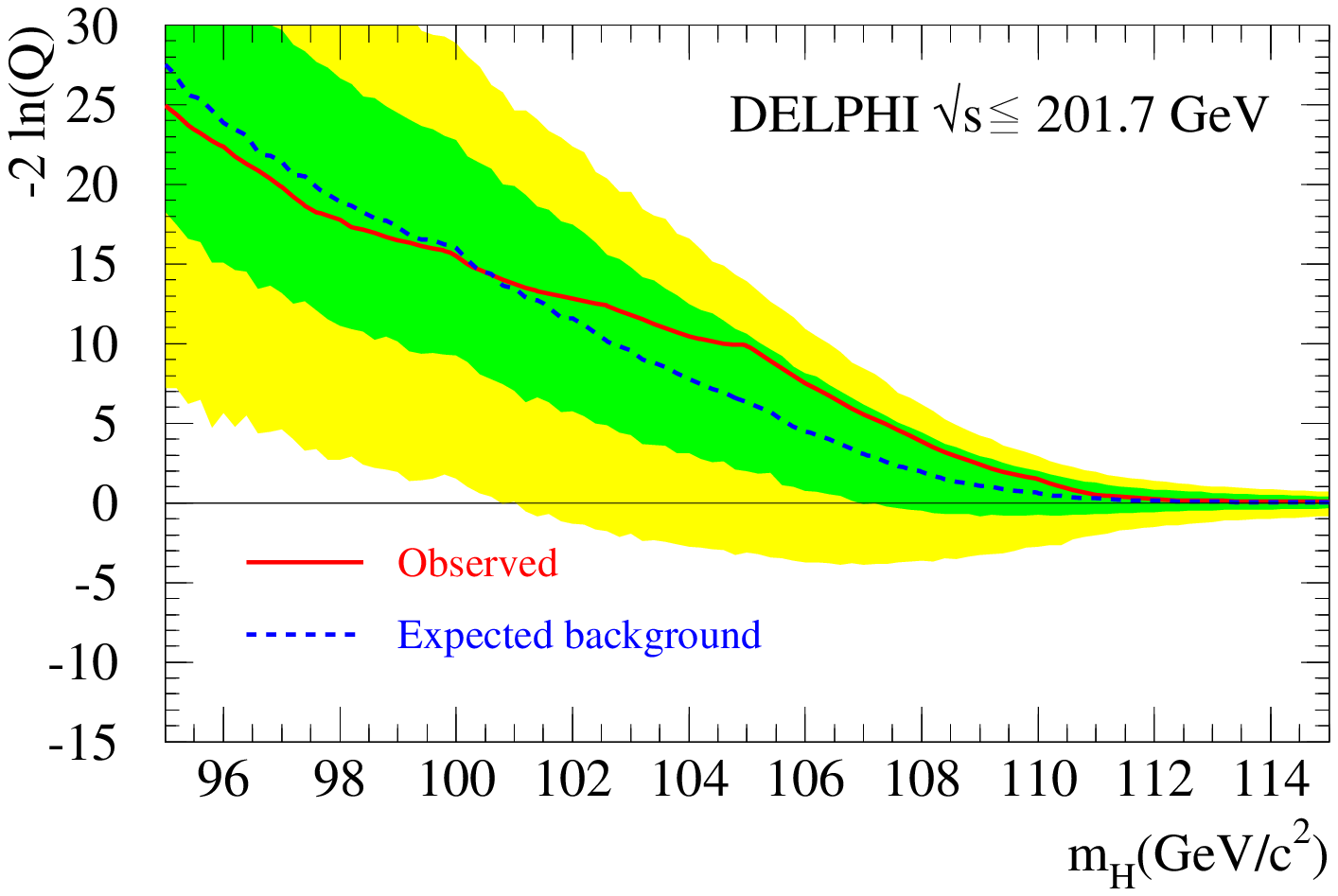,width=14.cm} \\
\epsfig{figure=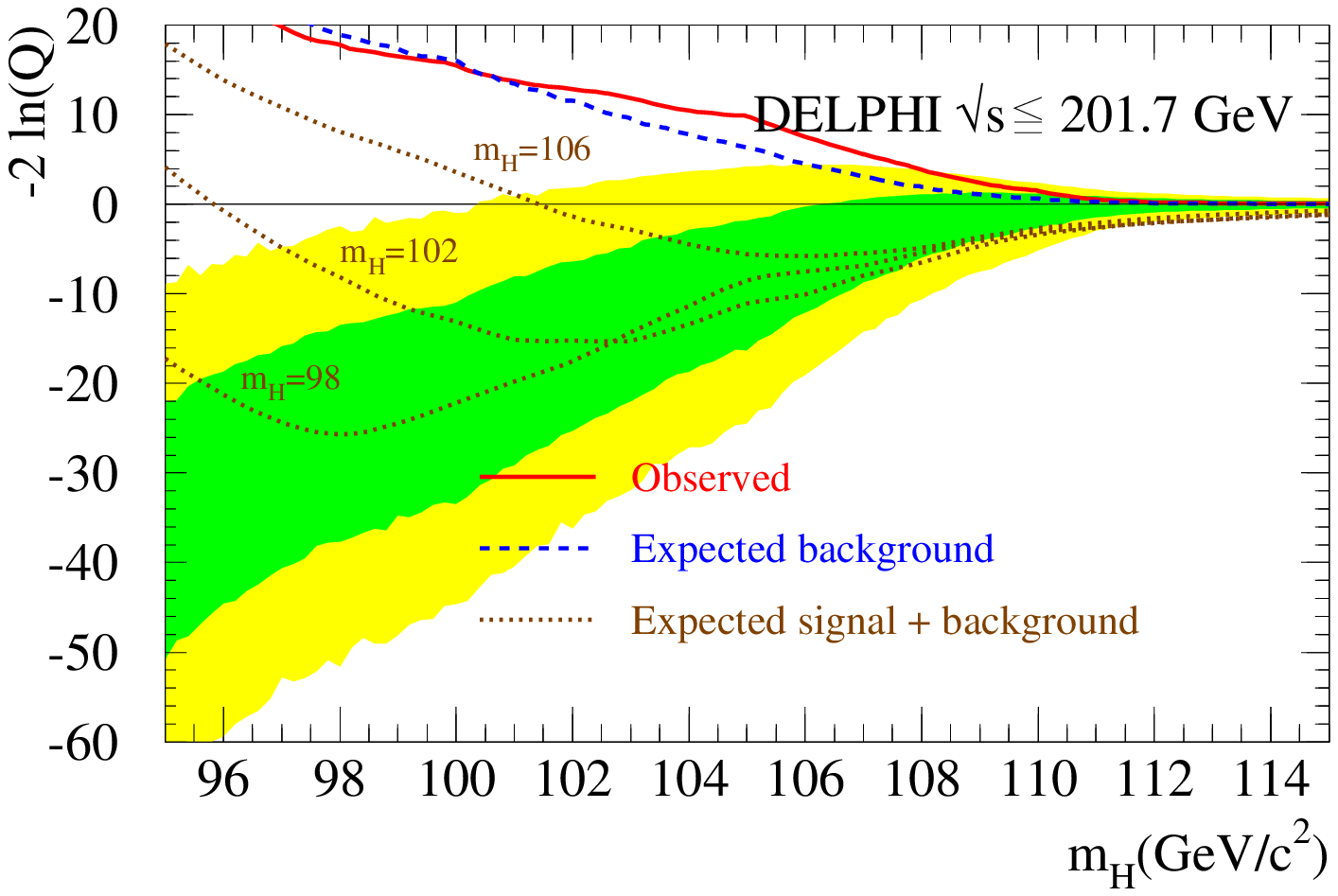,width=14.cm}
\caption[]{
 {\sc SM} Higgs boson: test-statistic \likear\ for each \MH\ hypothesis
in data (solid) and its
expected median value in background-only experiments (dashed). 
Top: the bands correspond to the 68.3\% and 95.0\% confidence intervals
from background-only experiments.
Bottom: the bands represent the confidence intervals around the minima 
of the $-2ln\cal Q$ curves expected for a signal of mass given in the 
abscissa
while the dotted lines show the median curves expected for three mass 
hypotheses.}
\label{fig:xi_sm}
\end{center}
\end{figure}

  The curve of the test-statistic \likear\ as a function of 
the mass hypothesis is 
shown in Fig.~\ref{fig:xi_sm}, where the observation is compared with the
expectations from background-only experiments (top) and from
signal plus background experiments (bottom). 
Over the whole range of masses, 
the test-statistic remains positive, while in the event of a discovery  
%$\Delta \chi^2$ 
it would be negative for mass hypotheses close to the actual mass of the 
signal.
%and could be used to extract the mass and its error,
%, as can be seen on the bottom plot of Fig.~\ref{fig:xi_sm}.

\subsection{Cross-section limit}

\begin{figure}[htbp]
\begin{center}
\epsfig{figure=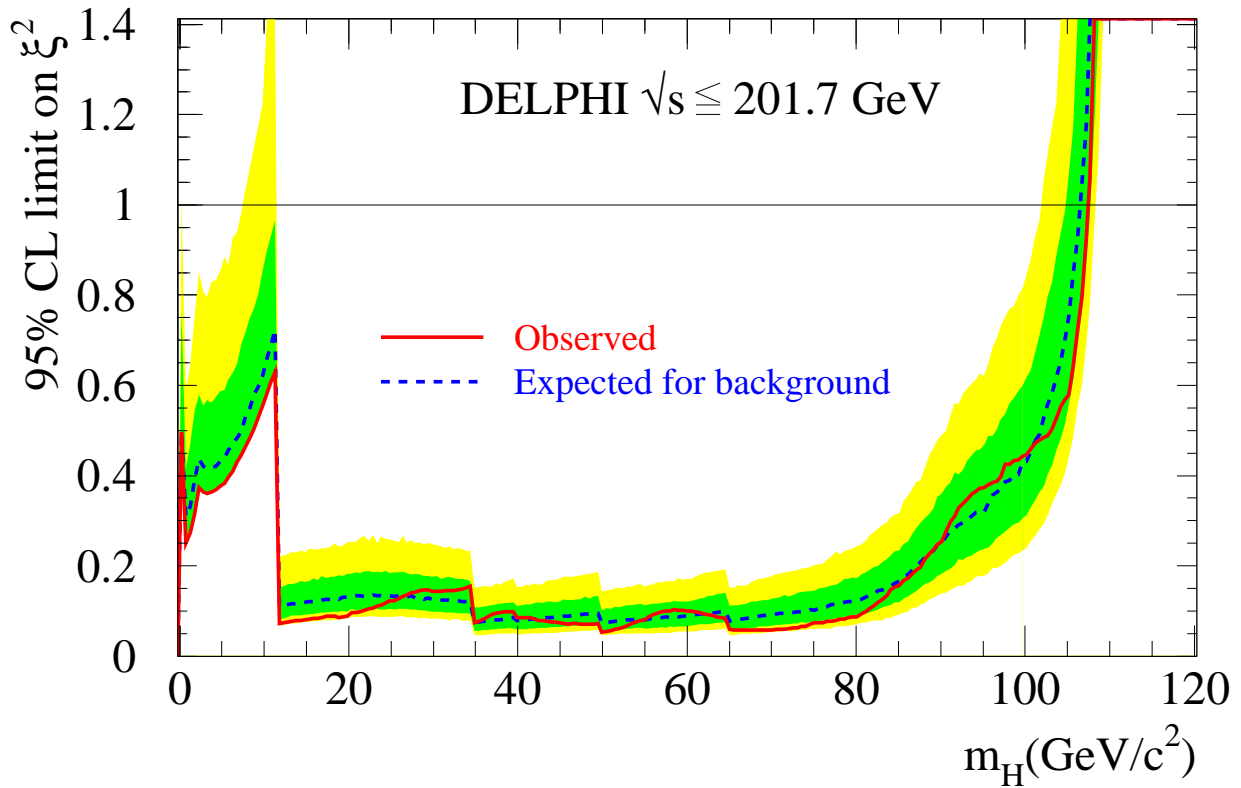,height=11cm}
\caption[]{
95\% CL upper bound on $\xi^2$, 
where $\xi$ is the HVV (V=\W\ or \Zz) coupling normalised to that
in the {\sc SM}, assuming {\sc SM} branching fractions for the Higgs 
boson. The limit
observed in data (solid line) is shown together with the expected median limit
in background-only experiments (dashed line).  The bands correspond to the 
68.3\% and 95.0\% confidence intervals from background-only experiments.
Most of the jumps below 40~\GeVcc\ correspond to the different analyses 
applied to
different subsets of LEP1 data to cover the various topologies expected
from a Higgs boson.
%whose decays vary rapidly below the \bbbar\ threshold.
The jumps at 12, 50 and 65~\GeVcc\ indicate where the LEP2 results at 
191.6-201.7, 182.7 and 188.7~\GeVcc\ respectively start to contribute.}
\label{fig:ghvv}
\end{center}
\end{figure}

  In a more general approach, the results of the searches for a 
{\sc SM} Higgs boson can be used to set a 95\% CL upper bound 
on the Higgs boson production cross-section,
assuming that the Higgs boson decay properties are identical to those in 
the {\sc SM} but that the Higgs boson couplings to pairs of 
\Zz\ and \W\ bosons (the latter arising in the \WW\ fusion production
mechanism) may be smaller. To achieve the best sensitivity over
the widest range of mass hypotheses, the results described in this paper 
are combined with those obtained at lower energies at 
LEP2~\cite{pap98,pap97,pap96}, as well as with those obtained at 
LEP1~\cite{paplep1} which covered masses up to 60~\GeVcc.
Both sets of results are treated with the same statistical procedure  
as for the {\sc SM}. For each mass hypothesis, the production 
cross-section is decreased with respect to its {\sc SM} value
until a pseudo-confidence level $CL_s$ of 5\% is obtained.  
The result is shown in Fig.~\ref{fig:ghvv} as an upper bound on the 
production cross-section, normalised to that in the {\sc SM}, for
masses of the Higgs boson from 0 to 110~\GeVcc. 
The {\sc SM} result 
described in the previous section corresponds to a ratio of 1.

\subsection{Neutral Higgs bosons in the  {\sc MSSM}}

  The results in the \hZ\ and \hA\ channels reported in
the previous sections are combined with the same statistical 
method as for the {\sc SM}, also using earlier results at LEP2 
energies~\cite{pap98,pap97,pap96,pap95}. The exclusion limits
obtained at LEP1~\cite{mssmlep1} (\mh$>$44~(46)~\GeVcc\ when \mh\ 
is above (below) the ${\mathrm{AA}}$ threshold) are
used as external constraints to limit the number of points in
the scans.

\subsubsection{The benchmark scenarios}

   At tree level, the production cross-sections and the 
Higgs branching fractions in the  {\sc MSSM}
depend on two free parameters, \tbeta\ and one Higgs 
boson mass, or, alternatively, two Higgs boson masses, eg \MA\ and \mh. 
%The properties of the {\sc MSSM} Higgs bosons are modified by 
%Radiative corrections introduce a large number of additional 
%parameters, whose number is reduced by the introduction of
%universality assumptions. 
Radiative corrections introduce additional parameters, related
to supersymmetry breaking.
%To limit the number of degrees of freedom, 
Hereafter, we make the usual assumption that some of them
are identical at a given energy scale: hence, 
the SU(2) and U(1) gaugino mass terms are assumed to be
unified at the so-called GUT scale, 
while the sfermion mass terms or the squark trilinear
couplings are assumed to be unified at the EW scale.
Within these assumptions, the parameters beyond tree level are:
the top quark mass, the Higgs mixing parameter, $\mu$, 
the common sfermion mass term at the EW scale, $M_{susy}$,
the SU(2) gaugino mass term at the EW scale, $M_2$,
the gluino mass, $m_{\tilde{g}}$,
and the common squark trilinear coupling at the EW scale, $A$.
The U(1) gaugino mass term at the EW scale, $M_1$, is 
related to $M_2$ through the GUT relation
$M_1 = (5/3) {\rm \tan}^2\theta_W M_2$.
The radiative corrections affect the relationships between the masses of 
the Higgs bosons, with the largest contributions arising from the 
top/stop loops. 
As an example, the h boson mass, which is below that of the Z boson
at tree level, increases by a few tens of \GeVcc\ in some regions
of the {\sc MSSM} parameter space due to radiative corrections.

 In the following, we consider three benchmark scenarios, as suggested 
in~\cite{new_pres}. The first two schemes, called the
\mbox{$ m_{\mathrm h}^{max}$} scenario and the no mixing scenario,
rely on radiative corrections computed at two-loop order 
%in the Feynman-diagrammatic approach
as in~\cite{FDradco}. The values of the underlying parameters are
quoted in Table~\ref{ta:benchmarks}.
The two scenarios differ only by the value of $X_t = A - \mu \cot \beta$, 
the parameter which controls the mixing in the stop sector, and hence
has the largest impact on the mass of the h boson.
The \mbox{$ m_{\mathrm h}^{max}$} scenario leads to the maximum
possible h mass as a function of \tbeta. The no mixing
scenario is its counterpart with vanishing mixing, leading to upper
bounds on \mh\ which are at least 15~\GeVcc\ lower than in the
\mbox{$ m_{\mathrm h}^{max}$} scheme.

 The third scenario, called the large $\mu$ scenario, predicts at least 
one scalar Higgs boson with a mass within kinematic reach at LEP2 
in each point of the MSSM parameter space. However, there are regions 
for which the Higgs bosons fall below detectability because of  
vanishing branching fractions into b quarks due to large radiative
corrections. In this scenario, the radiative corrections are
computed  
%in the renormalization group equation approach
as in~\cite{radco}.
The values of the underlying parameters are given in 
Table~\ref{ta:benchmarks}. The main difference with the
two previous schemes is the large and positive value of $\mu$
and the relatively small value of $m_{\tilde{g}}$.

\subsubsection{The procedure} \label{procedure}

%  In the three benchmark scenarios,  a scan is made over the {\sc MSSM} 
%parameters \tbeta\ and \MA, with \MA\ from 12~\GeVcc\ to 1~TeV/$c^2$ 
%(400~\GeVcc\ in the large $\mu$ scenario) and \tbeta\ between 0.4 
%(0.7 for the large $\mu$ scenario) and 50.  
  In the three benchmark scenarios, a scan is made over the {\sc MSSM} 
parameters \tbeta\ and \MA. The range in \MA\ spans from 12~\GeVcc, 
the minimal value which has been searched for at LEP2 in the DELPHI
analyses, up to the maximal value allowed by each scenario~\cite{new_pres},
%. This maximal value~\cite{new_pres} is identical to the supersymmetric 
%scale,
that is up to $M_{susy}$, which is 1~TeV/$c^2$ in the 
\mbox{$ m_{\mathrm h}^{max}$} 
and no mixing schemes, and 400~\GeVcc\ in the large $\mu$ scenario
(see Table~\ref{ta:benchmarks}).
The range in \tbeta\ goes from the minimal value allowed in each 
scenario (0.7 in the large $\mu$ scenario and 0.4 in the other two 
schemes) up to 50, a value chosen in the vicinity of the ratio of 
the top- and b-quark masses, which is an example of the large \tbeta\
hypothesis favored in some constrained {\sc MSSM} models~\cite{susygut}.
The scan steps are 1~\GeVcc\ in \MA\ and 0.1 in \tbeta\ in the regions 
where \mh\ varies rapidly with these parameters. 

  At each point of the parameter space, the \hZ\ and \hA\ cross-sections 
and the Higgs branching fractions 
%in the three scenarios defined in the previous section 
are taken from theoretical databases provided by
the LEP Higgs working group~\cite{hwg} on the basis of the
theoretical calculations in~\cite{FDradco,radco}.
The signal expectations in each channel are then derived from the 
theoretical cross-sections and branching fractions,
the experimental luminosity and the efficiencies. A correction
is applied to account for differing branching fractions of the
Higgs bosons into \bbbar and \toto\ between the test point 
and the simulation (e.g.~for the \hZ\ process, the simulation is done in the 
{\sc SM} framework). 
   For the hA channels, to account for the difference between 
the masses of the h and A bosons at low \tbeta\ as well as for
the non-negligible width of the h and A bosons at large \tbeta, 
the set of efficiencies as a function of \MA\
obtained from the simulations at \tbeta~=~50 are applied above 30 in \tbeta, 
while the efficiencies derived from the \tbeta~=~20 
(or \tbeta~=~2) simulations are applied between 2.5 and 30 (or below 2.5)
provided the difference between \mh\ and \MA\ at the test point is
below 25~\GeVcc; otherwise the set of efficiencies as a function of
\mh\ and \MA\ derived from the additional simulations corresponding
to large mass differences between the two bosons is preferred.
The same holds for the discriminant information. Finally, as there is 
a large overlap in the backgrounds selected by the analyses in the 
\hqq\ and 4b channels, only one channel is selected at each input point 
and at each centre-of-mass energy, 
on the basis of the best expected $CL_s$ from background-only experiments.
%confidence level in the signal hypothesis at each energy.
This ensures that the channels which are then
combined in the global confidence level computations are independent.

\subsubsection{Results}
 
\begin{figure}[htbp]
\begin{center}
\vspace{-0.9cm}
\epsfig{file=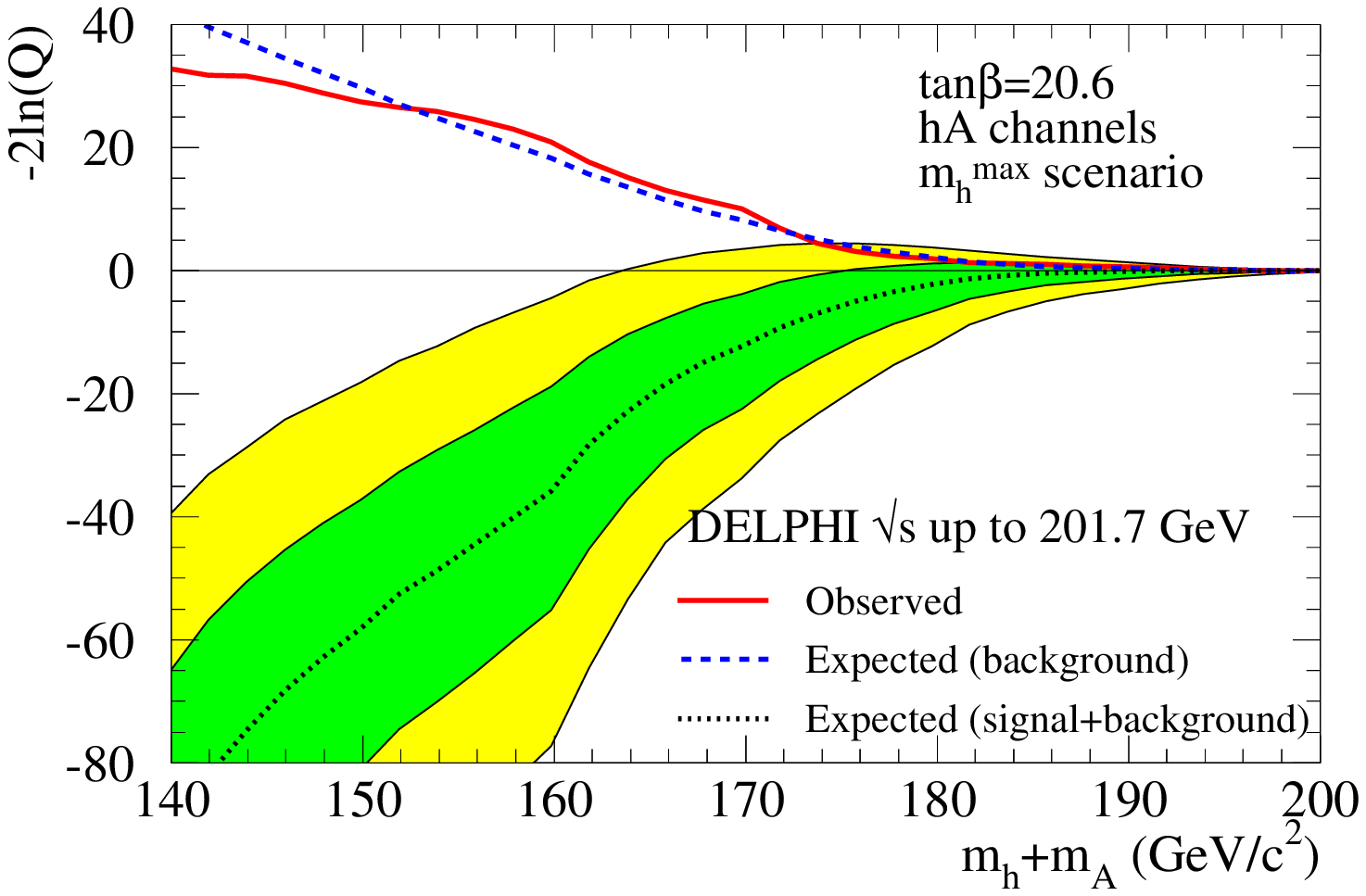,height=8cm} \\
\vspace{-0.3cm}
\epsfig{file=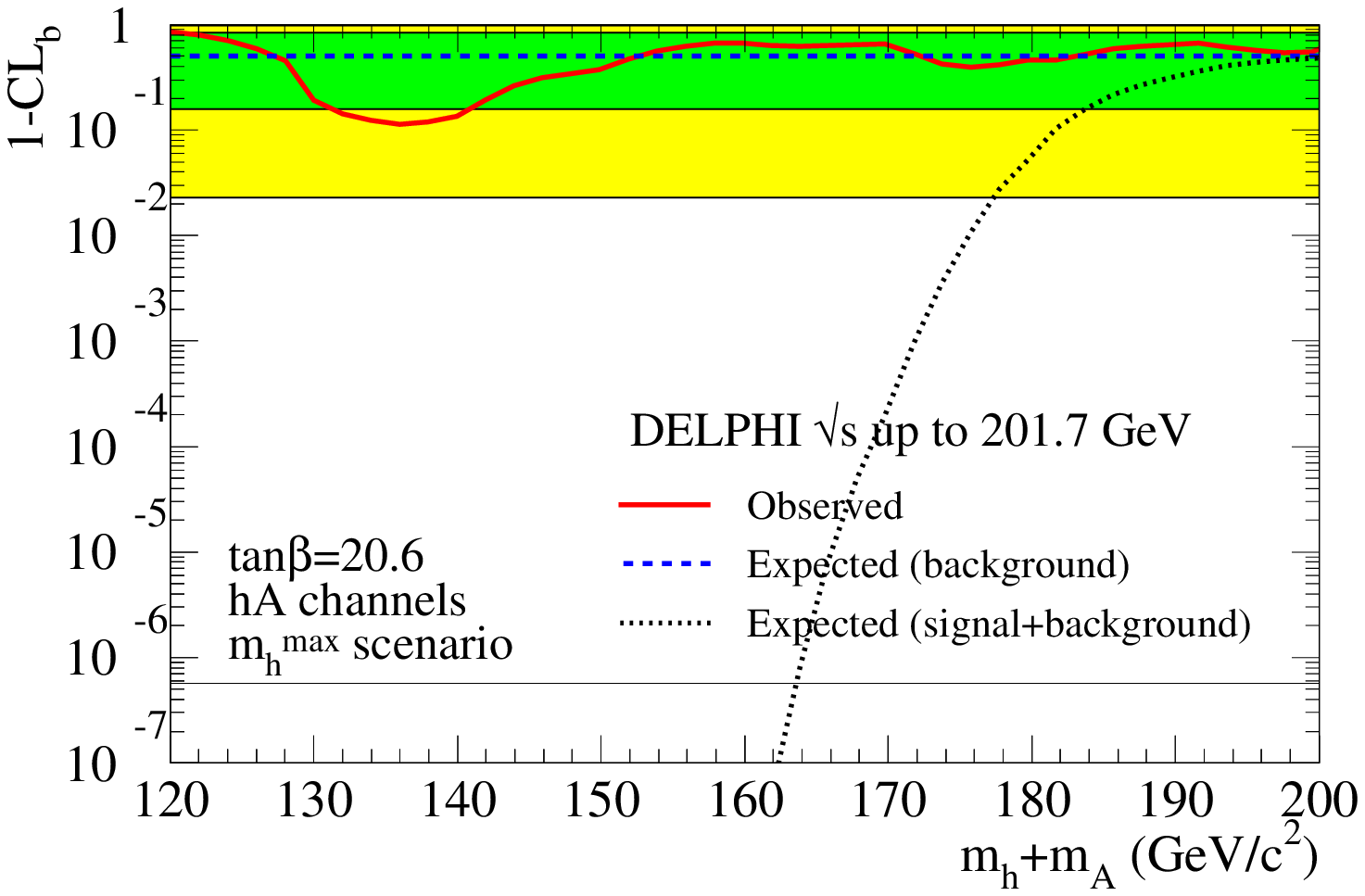,height=8cm} \\
\vspace{-0.3cm}
\epsfig{file=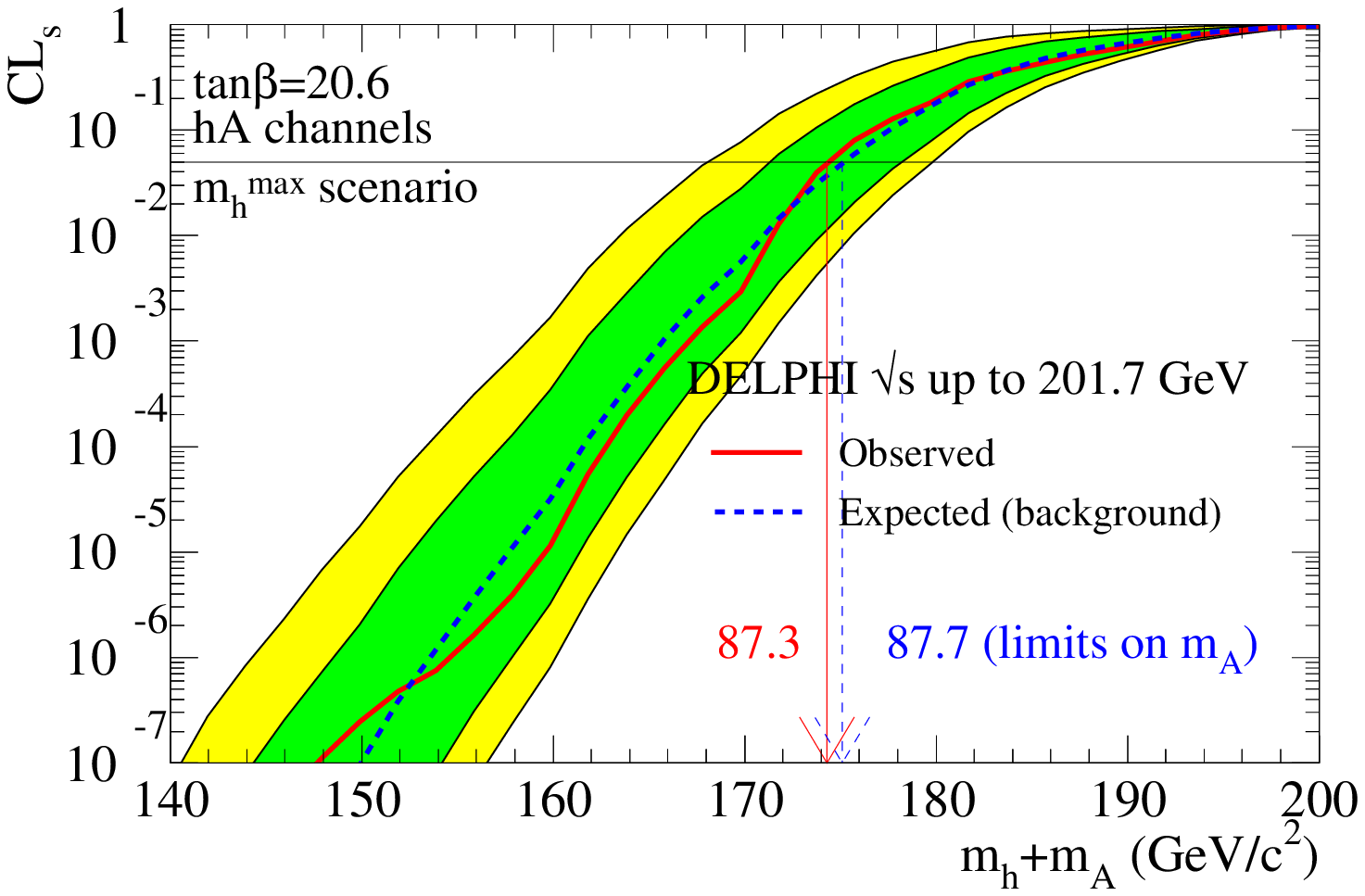,height=8cm}
\caption[]{
hA analyses: test-statistic (top) and 
confidence levels in background hypothesis (middle) and
in signal hypothesis (bottom) as  functions of \mh+\MA.
Curves are the observed (solid) and  median expected 
(dashed) confidences from background-only experiments
while the bands correspond to the 68.3\% and 95.0\% confidence intervals 
from background+signal experiments for a signal of mass given in the abscissa
(top) and from background-only experiments (middle and bottom).}
\label{fig:cls}
\end{center} 
\end{figure}

  To illustrate the compatibility tests of data with background only
and with signal plus background hypotheses in the \hA\ channels, 
Fig.~\ref{fig:cls} shows the curves of the test-statistic 
\likear\ and of the confidence levels $CL_b$ and $CL_s$ as a function of 
the test mass \mh+\MA, when using only the results in the 
two \hA\ channels.
The signal cross-sections are from the \mbox{$ m_{\mathrm h}^{max}$} 
scenario at \tbeta\ around 20.
Over the whole range of test masses,
data are in reasonable agreement with the background expectations. The 
largest deviation, slightly over one standard deviation, is observed
for test masses \mh+\MA\ around 135~\GeVcc\ and is due to
the small excess of events in the 4b channel with reconstructed 
masses in that region, as seen in Fig.~\ref{fig:mass_pl}.

\begin{figure}[htbp]
\begin{center}
\begin{tabular}{cc}
\hspace{-1.2cm}
\epsfig{figure=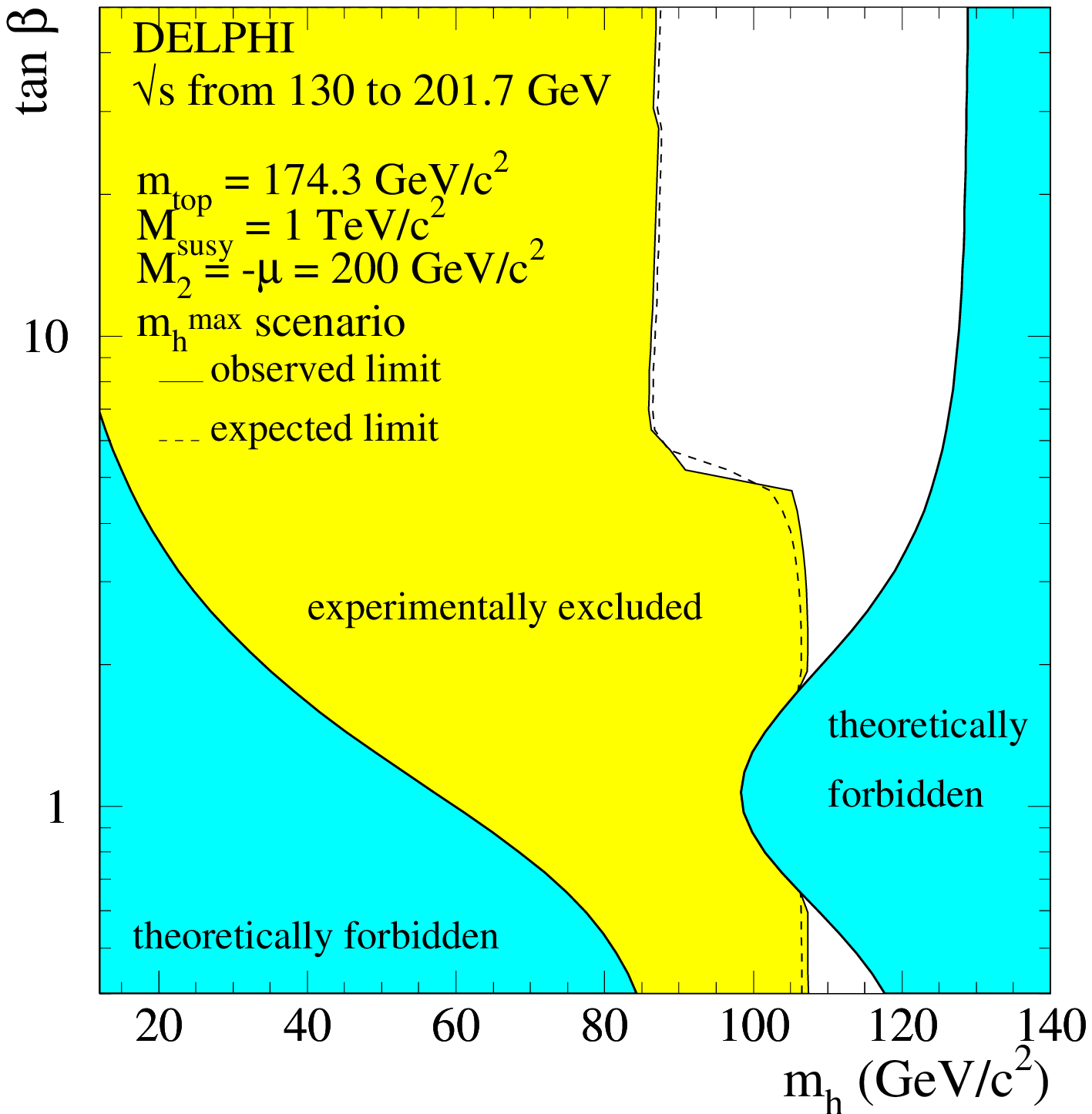,height=9.5cm} &
\hspace{-1cm}
\epsfig{figure=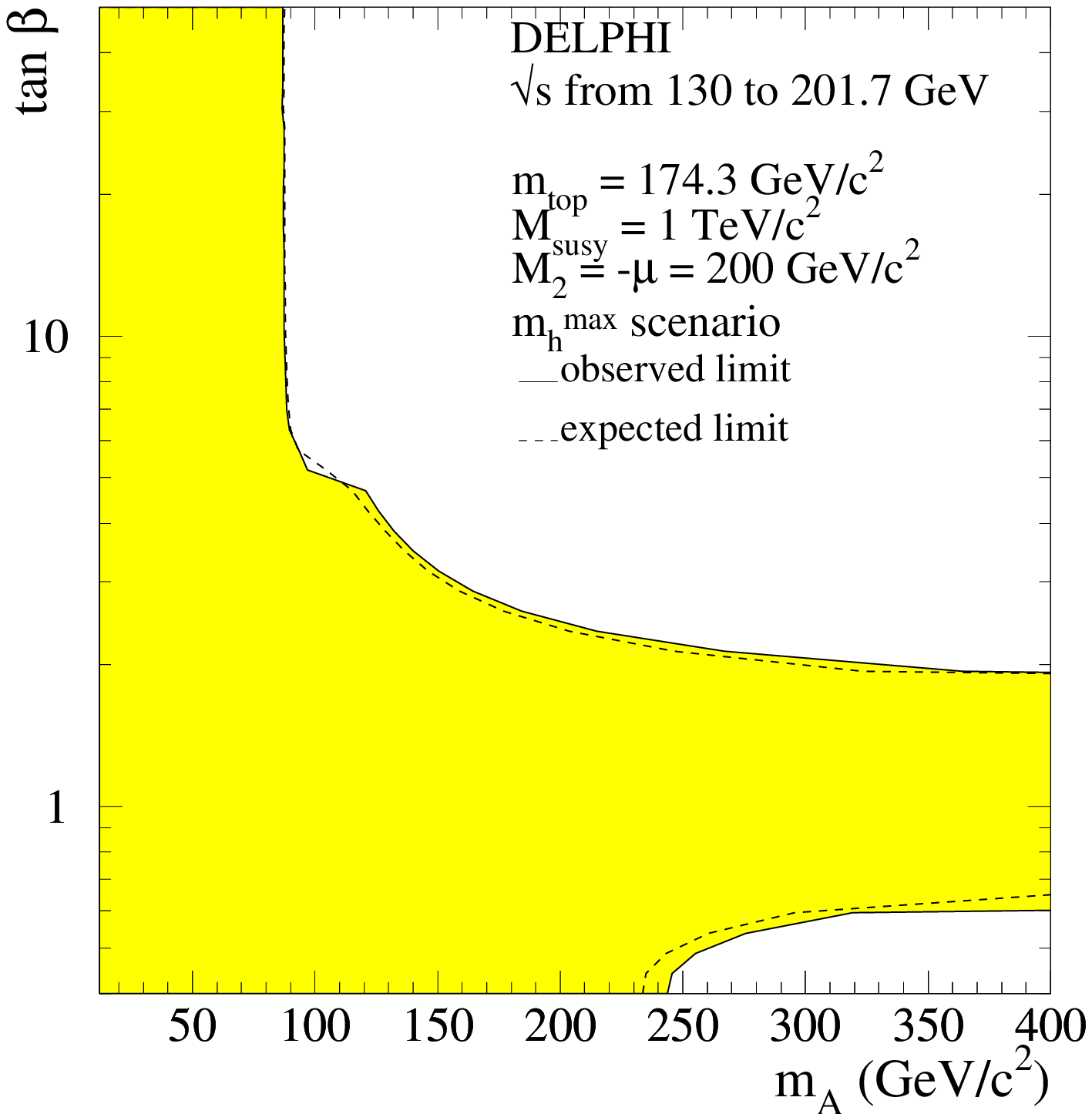,height=9.5cm} \\
\hspace{-1.2cm}
\epsfig{figure=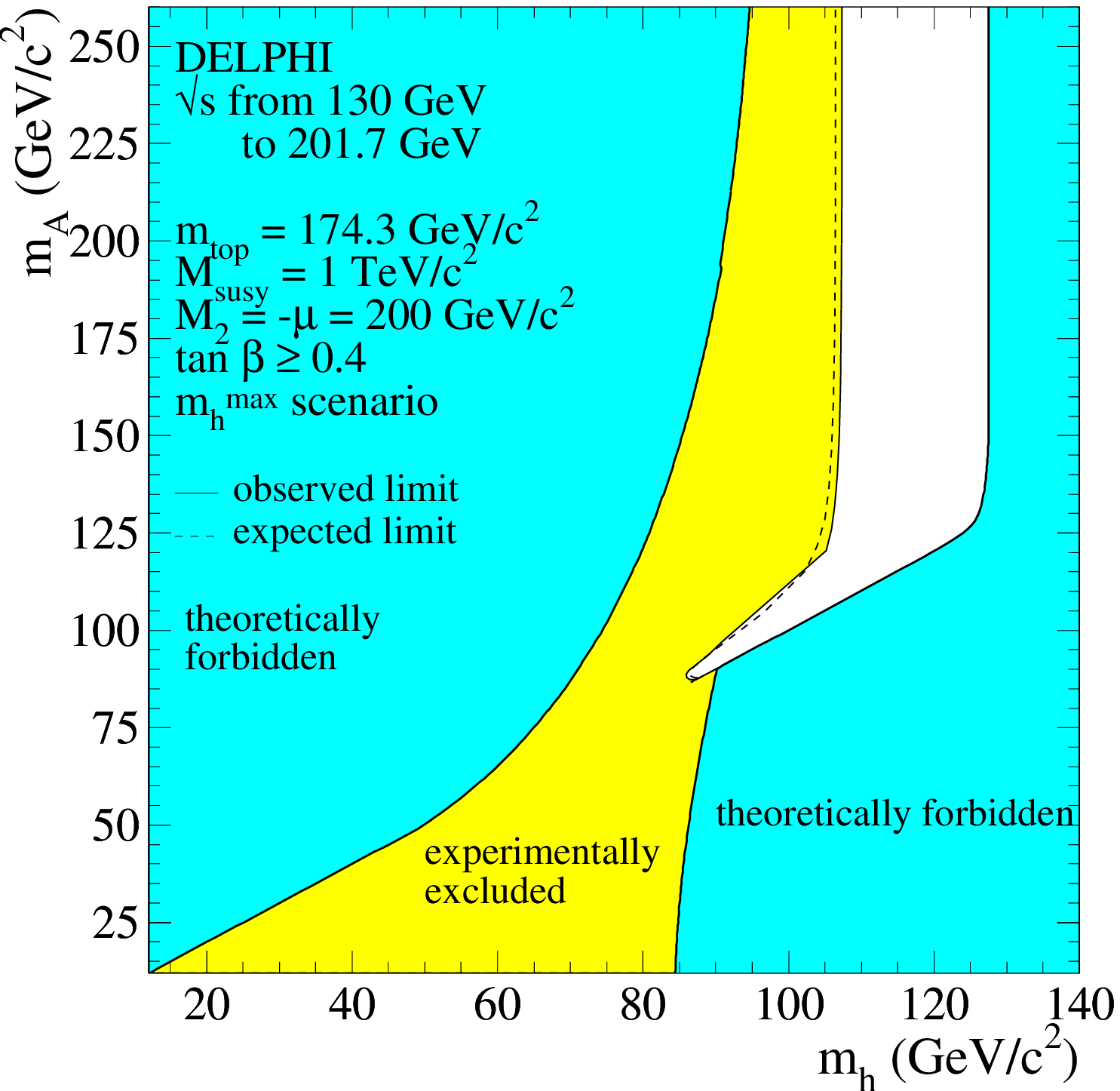,height=9.5cm}
\end{tabular}
\caption[]{
    {\sc MSSM} Higgs bosons: regions excluded at 95\% CL
   by the searches in the \hZ\ and \hA\ channels up to \rs~=~201.7~\GeV,
   in the \mbox{$ m_{\mathrm h}^{max}$} scenario.
   The dark shaded areas are the regions not allowed by the 
    {\sc MSSM} model in this scenario.
   The dashed lines show the median expected limits.}
\label{fig:limit_max}
\end{center}
\end{figure}

\begin{figure}[htbp]
\begin{center}
\begin{tabular}{cc}
\hspace{-1.2cm}
\epsfig{figure=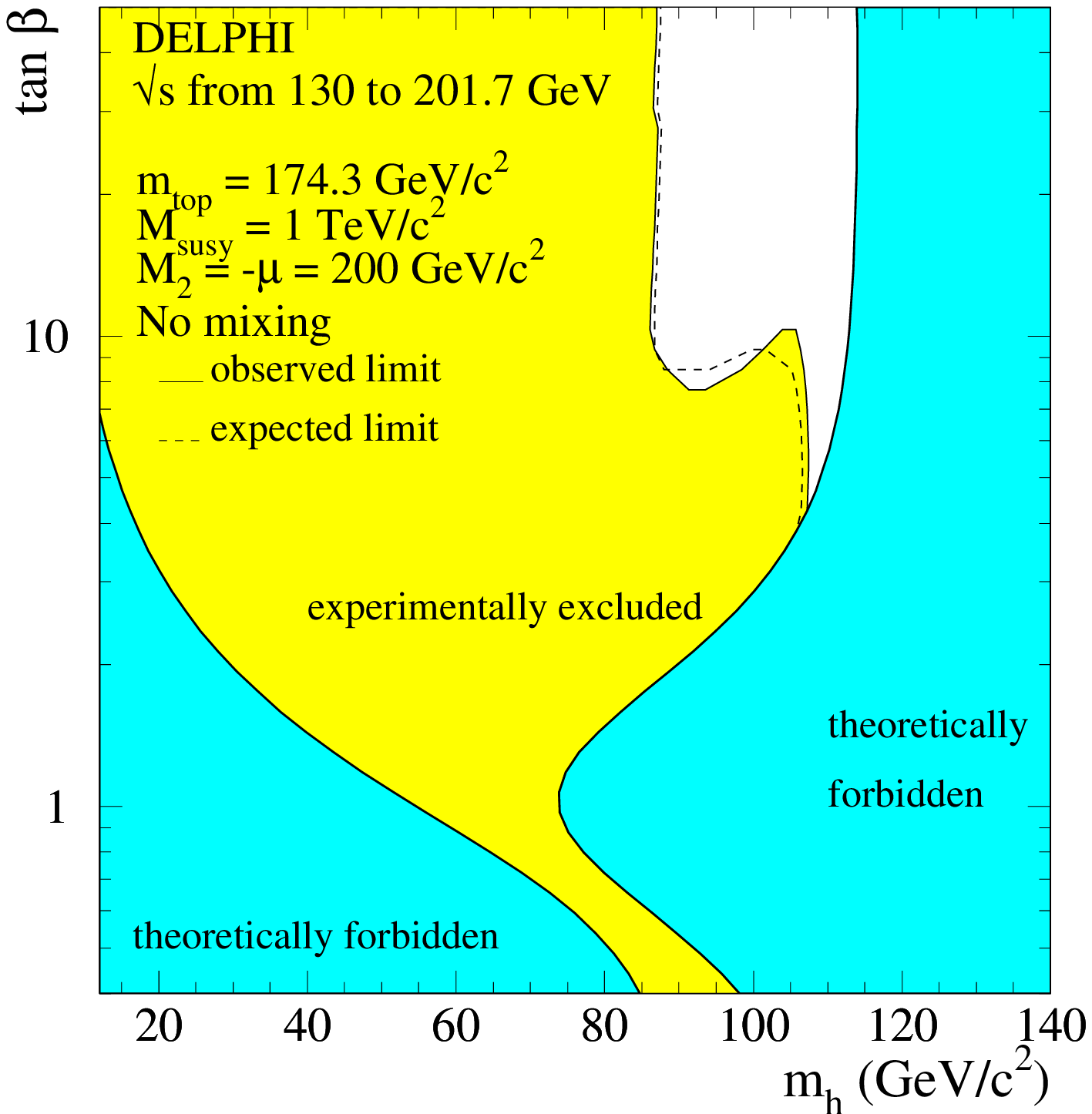,height=9.5cm} &
\hspace{-1cm}
\epsfig{figure=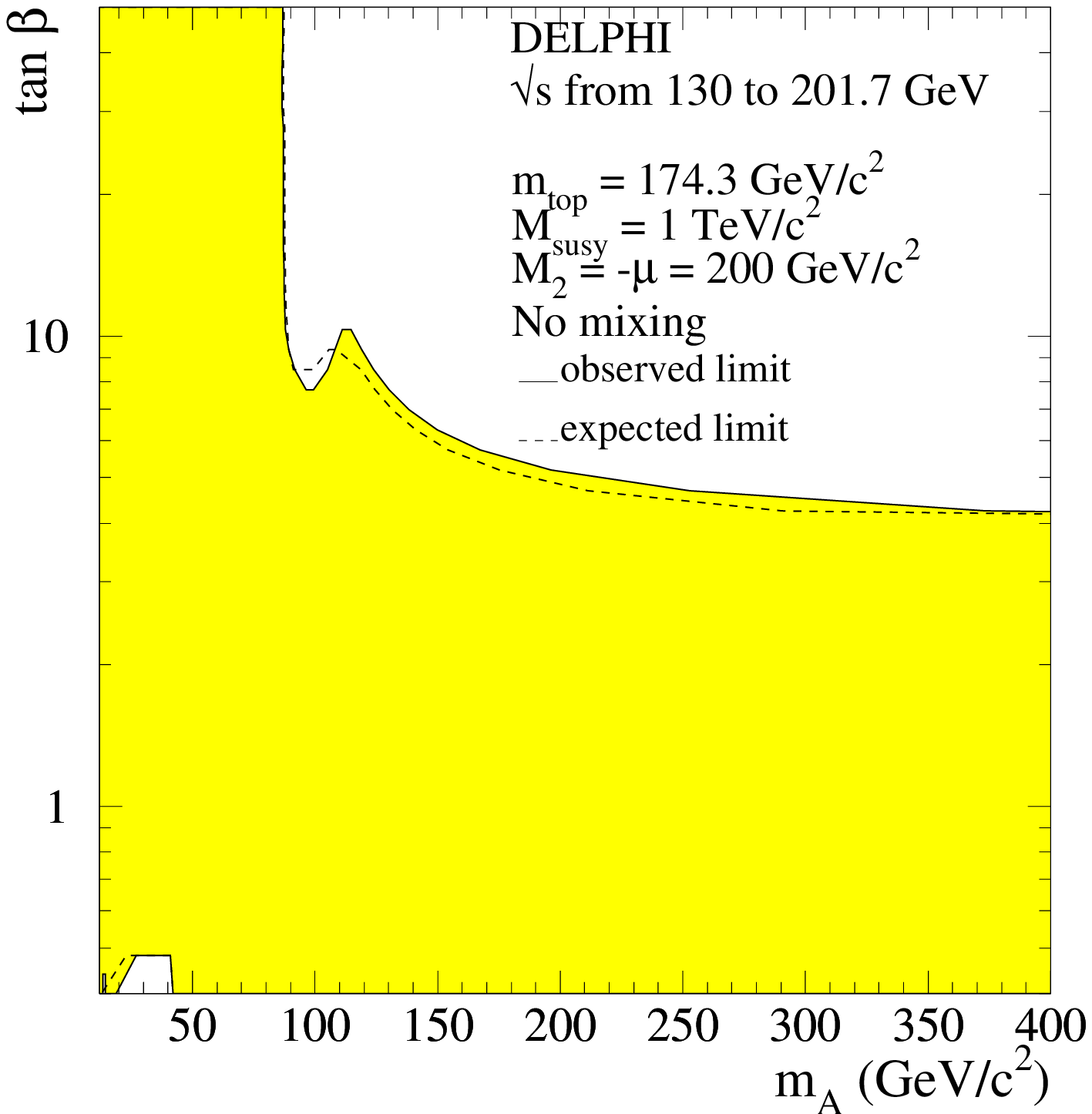,height=9.5cm} \\
\hspace{-1.2cm}
\epsfig{figure=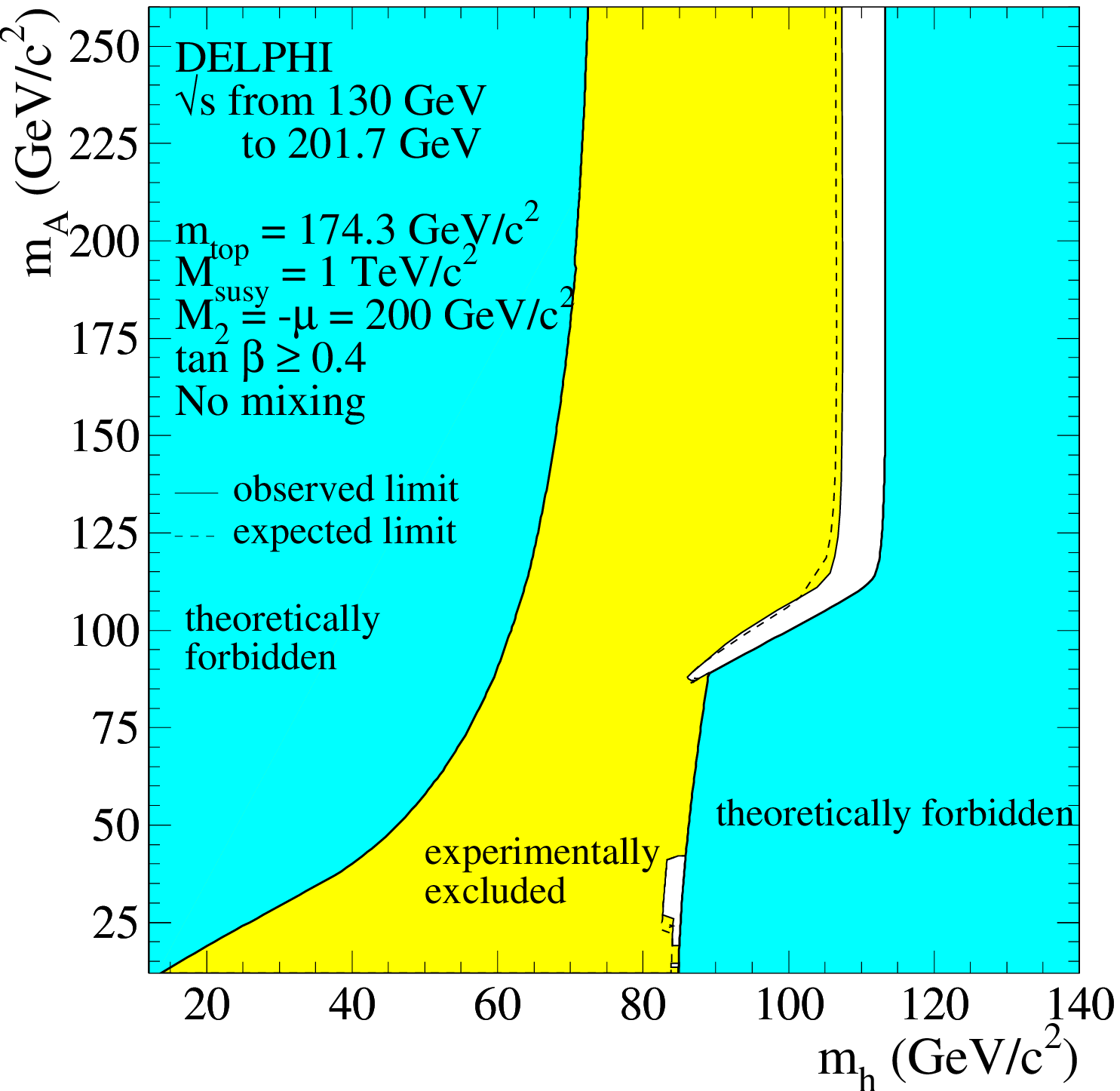,height=9.5cm}
\end{tabular}
\caption[]{
    {\sc MSSM} Higgs bosons: regions excluded at 95\% CL
   by the searches in the \hZ\ and \hA\ channels up to \rs~=~201.7~\GeV,
   in the no mixing scenario.
  There is a  region at \mh\ around of 85~\GeVcc\ and small \tbeta\ 
  that is not excluded, but is too small to be visible in the top
  left-hand plot.
   The dark shaded areas are the regions not allowed by the 
    {\sc MSSM} model in this scenario.
%for m$_{\rm{top}}=175$~\GeVcc, 
%   M$_{\rm{SUSY}}=1$~\TeVcc, M$_2 = - \mu = 200$~\GeVcc\  
%   and \MA $< 1$~\MeVcc\ or \MA $> 1$~\TeVcc\ 
   The dashed lines show the median expected limits.}
\label{fig:limit_no}
\end{center}
\end{figure}

  Combining the results in the \hZ\ and \hA\ channels gives
regions of the {\sc MSSM} parameter space which are excluded at 
95\% CL or more. The excluded regions in the (\mh, \tbeta), 
(\MA, \tbeta) and (\mh, \MA) planes are presented 
in Fig.~\ref{fig:limit_max} for the  
\mbox{$ m_{\mathrm h}^{max}$} scenario 
and in Fig.~\ref{fig:limit_no} for the no mixing scenario.
Basically,
the exclusion is made by the results in the \hZ\ (hA) channels in the
low (large) \tbeta\ region while they both contribute at intermediate
values.
For \MA\ below the kinematic threshold \mh~=~2\MA, which occurs at
low \tbeta\ only, the decay h$\rightarrow$AA opens, in which case it
supplants the h$\rightarrow$\bbbar\ decay.
%, especially in
%the no mixing scenario, where the h$\rightarrow$AA 
%branching ratio can be as large as 97\%. 
However, in most of the
region, the A$\rightarrow$\bbbar\ branching fraction remains large
which explains why the results in the two (\hAA) \qqbar\ channels 
reported in section~\ref{sec:1}, combined with studies of the \hAA\ decay 
at lower energies~\cite{pap97,pap96}, exclude most of this region.
An unexcluded hole remains in the no mixing scenario 
at \tbeta~$\sim$~0.4, \MA\ between 20 and 40~\GeVcc\  and
\mh\ around 85~\GeVcc\
(visible only in the (\MA, \tbeta) and (\mh, \MA) projections).
In that area, the A$\rightarrow$\ccbar\ decay dominates over the
A$\rightarrow$\bbbar\ decay but the branching fractions in both
modes are no longer large enough to give the necessary sensitivity
for an exclusion.

The above results establish 95\% {\sc CL} lower limits on \mh\ and \MA,
for either assumption on the mixing in the stop sector and
for all values of \tbeta\ above 0.49:

\[ \mh > 85.9~\GeVcc \hspace{1cm}
   \MA > 86.5~\GeVcc  .\]

\noindent
The expected median limits are 86.4~\GeVcc\ for \mh\ and 87.0~\GeVcc\ 
for \MA. The limit in \MA\ is reached in the no mixing scenario at \tbeta\
around 30 and thus is due to the non-negligible widths of the Higgs bosons, 
while the limit in \mh\ is obtained in the \mbox{$ m_{\mathrm h}^{max}$} 
scenario at \tbeta\ around 7, in a region where both the \hZ\ and 
\hA\ processes contribute.
Furthermore, there are excluded ranges in \tbeta\ between 
0.49 and 3.86 (expected [0.49-3.86]) in the no mixing case and between 
0.65 and 1.75 (expected [0.72-1.75]) in the
\mbox{$ m_{\mathrm h}^{max}$} scenario. 

\begin{figure}[htbp]
\begin{center}
\begin{tabular}{c}
\epsfig{figure=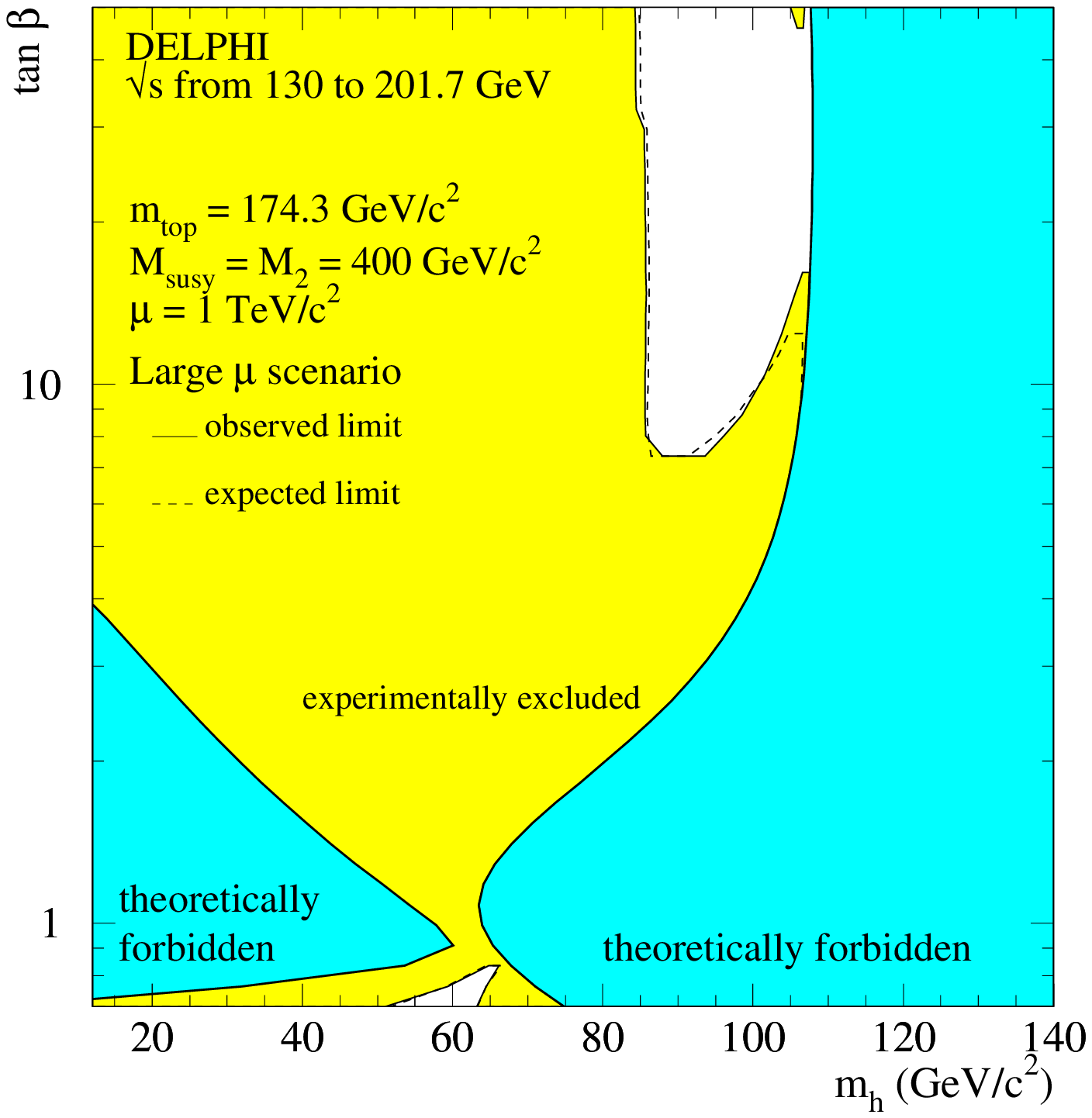,height=11.5cm}\\
\vspace{-0.2cm}
\epsfig{figure=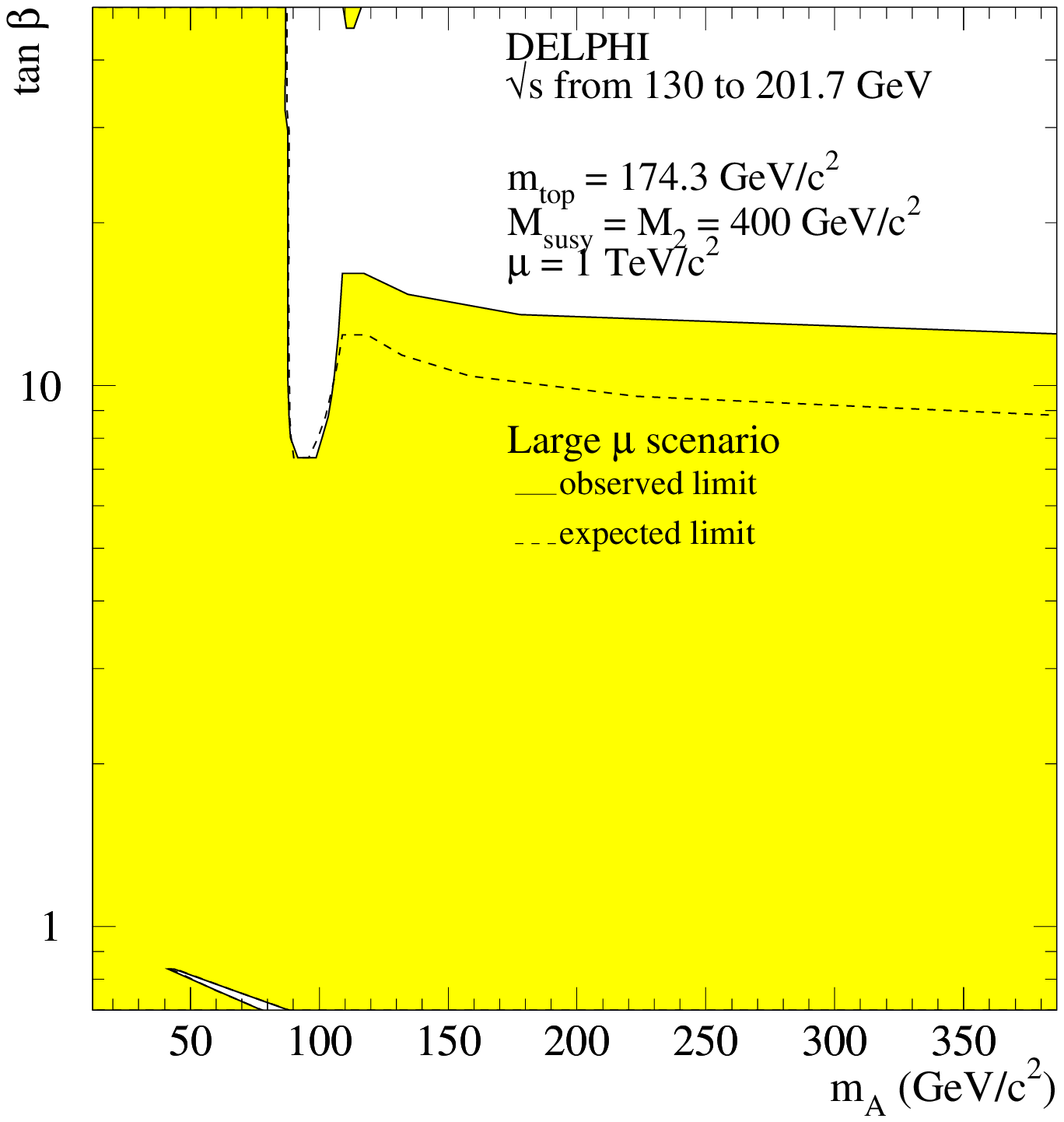,height=11.5cm}
\end{tabular}
\caption[]{
    {\sc MSSM} Higgs bosons: 
   regions excluded at 95\% CL, by the searches in the \hZ\ and \hA\ 
   channels up to \rs~=~201.7~\GeV, in the large $\mu$ scenario.
   The dark shaded areas are the regions not allowed 
   by the  {\sc MSSM} model in this scenario.
%   The region where the \hAA\ decay may occur is indicated.
   The dashed lines show the  median expected limits.}
\label{fig:limit_mu}
\end{center}
\end{figure}

  The excluded regions in the large $\mu$ scenario are presented in the 
(\mh, \tbeta) and (\MA, \tbeta) planes in Fig.~\ref{fig:limit_mu}.
A large fraction of the allowed domain is excluded by the present results 
in the \hZ\ and \hA\ channels. In particular, given that the 
theoretical upper bound
on the h boson mass in that scenario is slightly above 107~\GeVcc, 
the sensitivity of the \hZ\ channels is high even at large \tbeta, which
explains why the excluded region reaches the theoretically forbidden
area for values of \tbeta\ up to 13.5. On the other hand,
there is an unexcluded hole in the low \tbeta\ region at \mh\ around
60~\GeVcc\ which is due to a loss of sensitivity because of
vanishing h$\rightarrow$\bbbar\ branching fractions in that region.

%The unexcluded area at large \tbeta\ is mostly due to low expected rates
%in these channels (the \hA\ kinematical limit is close and the ZZh coupling
%is low) rather than to vanishing branching fractions into b's. 
%However, at these unexcluded points, 
%the production cross-section for the second scalar boson, H, is high 
%(its mass is around 108~\GeVcc\ and the ZZH coupling is enhanced at 
%large \tbeta) as well as its branching fraction into b quarks. 
%The reinterpretation of the present results in the \hZ\ channels
%in terms of \HZ\ production is thus expected to improve the
%coverage of that region. The unexcluded area contains also
%points with vanishing branching fractions 
%of the h boson into b quarks. At these points, the second scalar boson, H, is 
%kinematically inaccessible but the branching fractions of the h boson
%into gluons, c quarks, pairs of W are sizeable. Dedicated analyses
%for these decays are expected to give some sensitivity there.

\subsubsection{Extended scan of the parameter space}

\begin{figure}[htbp]
\begin{center}
\begin{tabular}{cc}
\hspace{-1cm}
\epsfig{figure=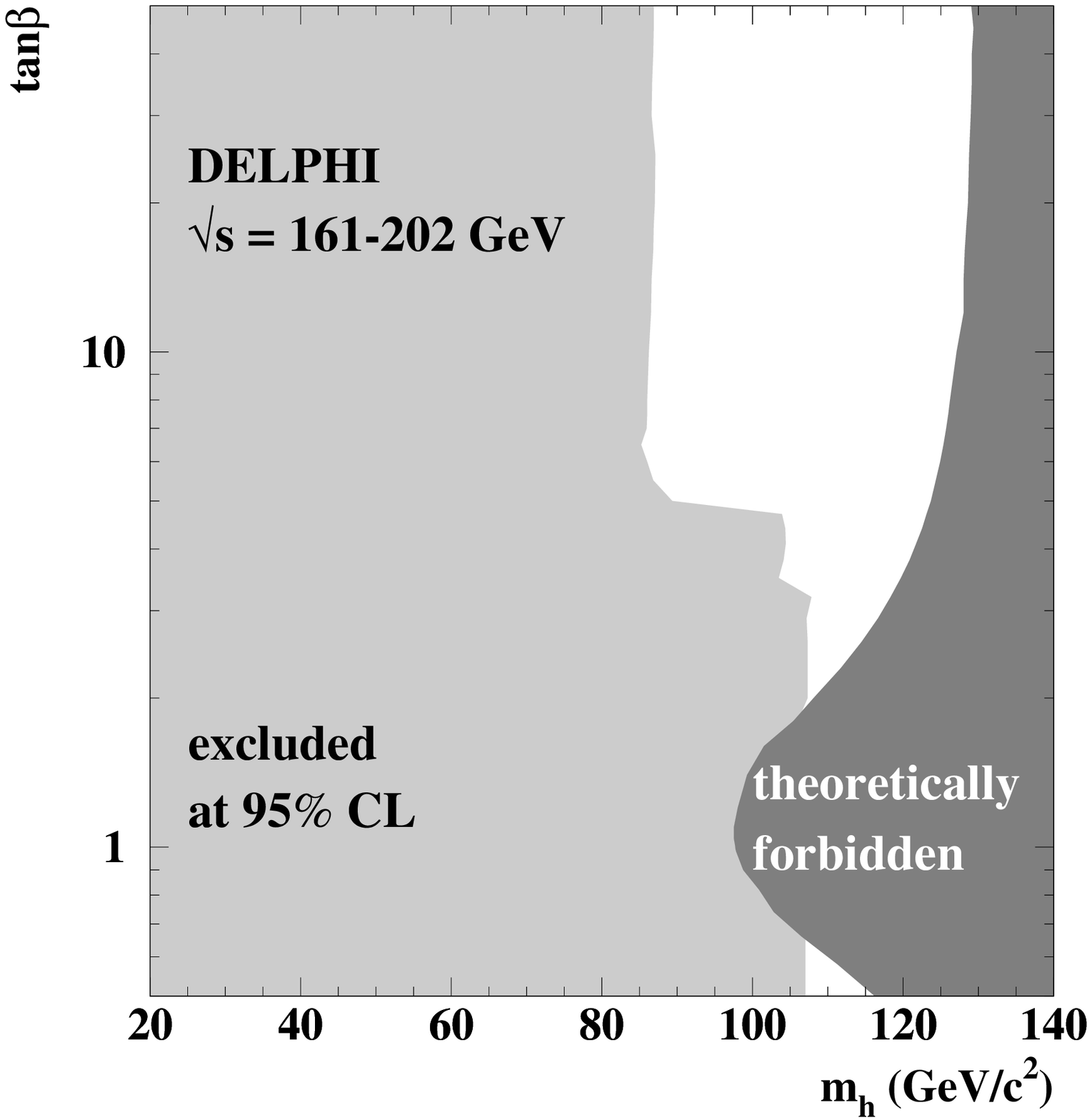,height=10.5cm} &
\hspace{-2cm}
\epsfig{figure=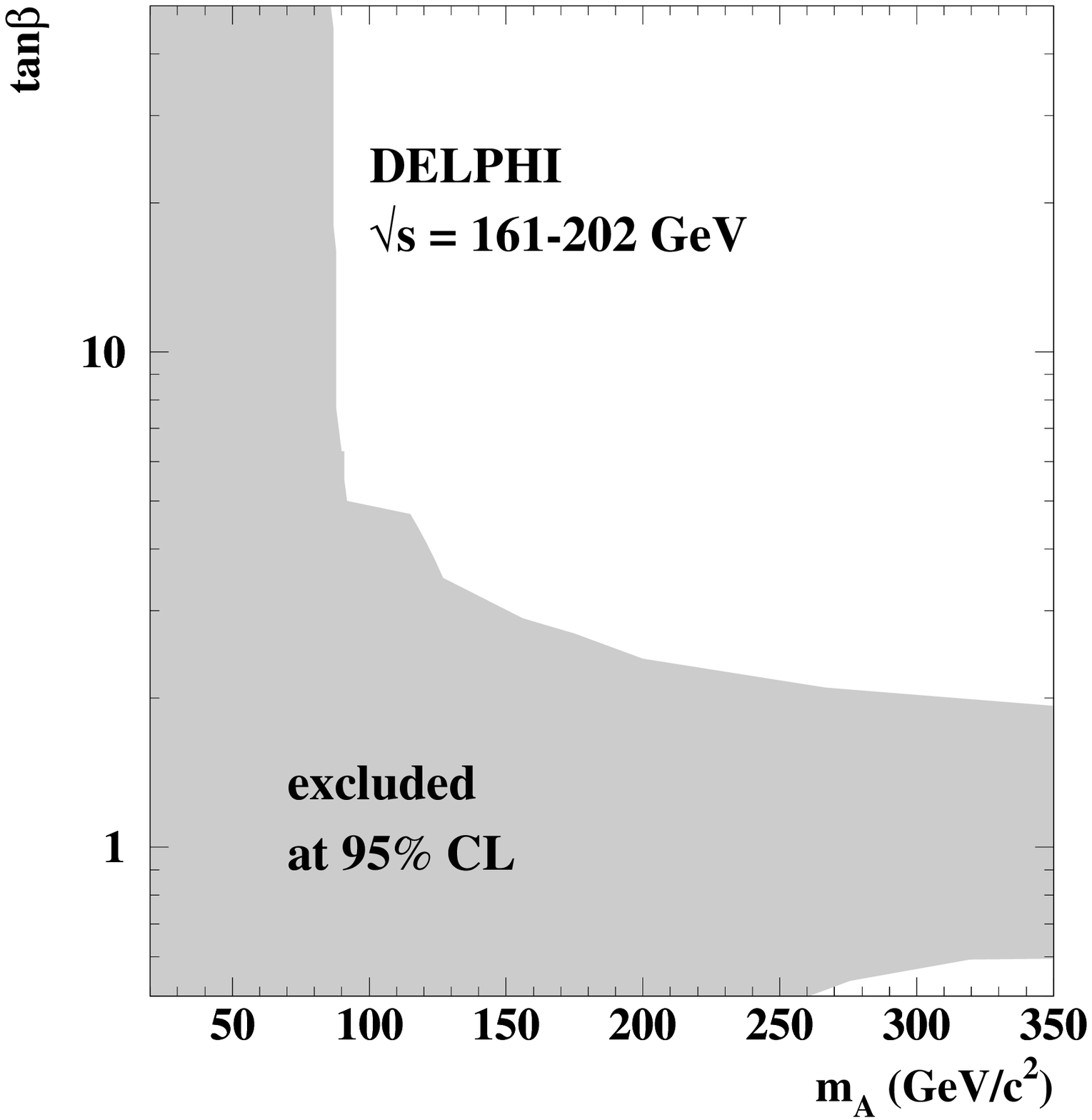,height=10.5cm} \\
\hspace{-1cm}
\epsfig{figure=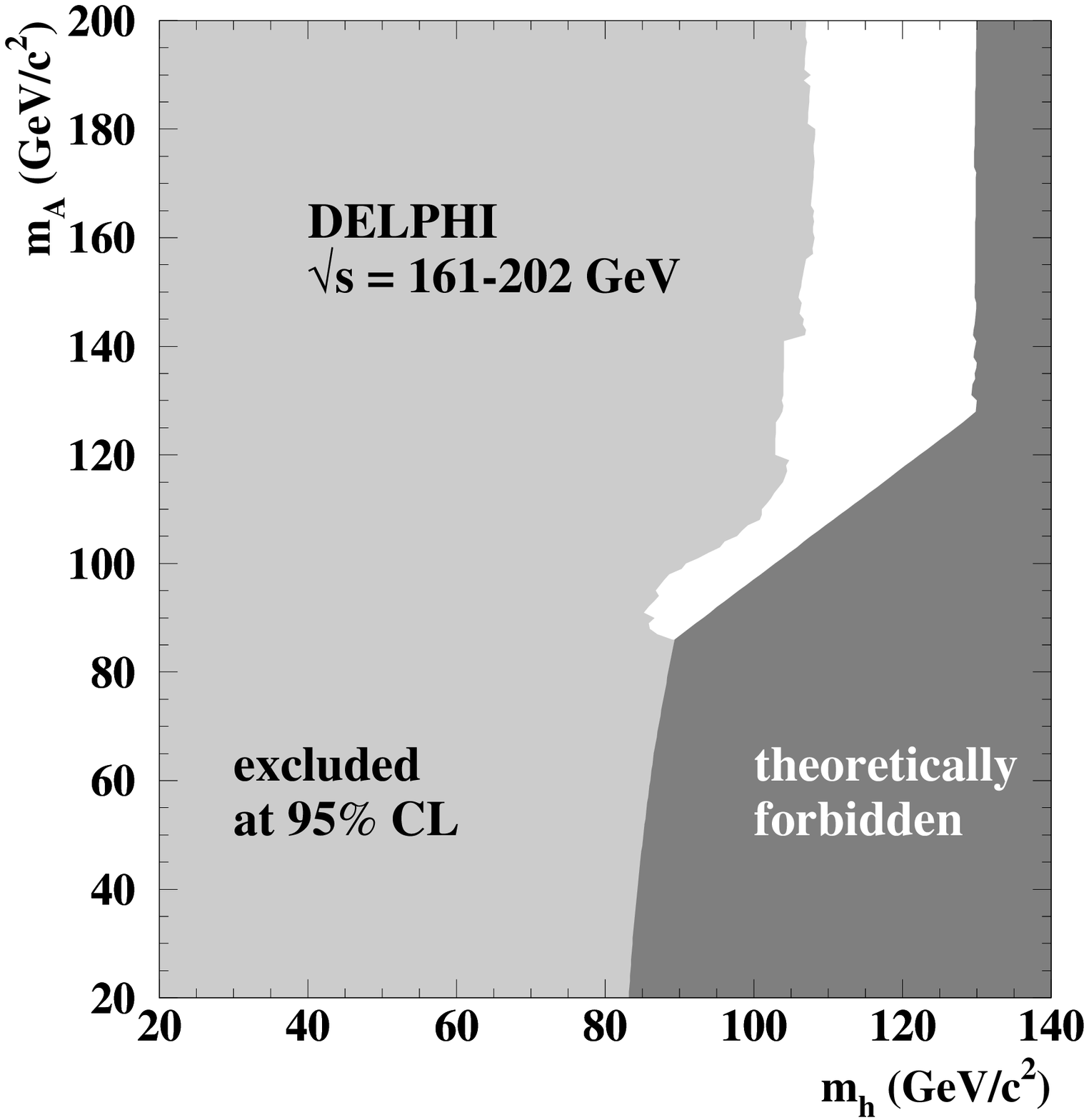,height=10.5cm} & \\
\end{tabular}
\caption[]{
    {\sc MSSM} Higgs bosons: 
   regions excluded at 95\% CL 
   by the searches in the \hZ\ and \hA\ channels up to \rs~=~201.7~\GeV,
   in an extended scan of the  {\sc MSSM} parameter space.
   The dark shaded area is the region at high \mh\ not allowed 
   by the {\sc MSSM} model in this scan.
%%   The region where the \hAA\ decay may occur is indicated.
%%   The dashed lines show the expected median limits
}
\label{fig:limit_scan}
\end{center}
\end{figure}

  The robustness of the limits obtained in the benchmark scenarios has
been tested in an extended scan of the {\sc MSSM} parameter space. The
Higgs bosons masses, cross-sections and branching fractions are computed
with radiative corrections at two-loop order 
%in the Feynman-diagrammatic approach
as in~\cite{FDradco}. The top mass is fixed at 175~\GeVcc\ while
the {\sc MSSM} parameters, \MA, \tbeta\ and the parameters governing the 
radiative corrections, $M_{susy}$, $M_2$,  $\mu$ and $A$ are varied within 
the ranges given in Table~\ref{tab:scan}. The values $\mu=\pm 1000$~\GeVcc\
have been studied in addition. As far as the granularity of the scan is 
concerned, steps of 1~\GeVcc\ are used for \MA\ up to 200~\GeVcc\ and 
larger steps between 200 and 1000~\GeVcc; for each value of \MA, 
up to 2700 parameter combinations are investigated. 
To limit the number of points in the scan, 
only points above 70~\GeVcc\ in \mh\ and \MA\ are considered, 
since all points below this limit have already been excluded by our 
previous extended search~\cite{scan98}.
The scan relies on the same
channels and data sets as the representative scans previously reported and
uses the same procedure to compute the confidence levels at each input
point.
% (see section~\ref{procedure}). 
%There is however one difference: 
However, for some parameter sets, 
the branching ratio of the neutral Higgs bosons into neutralinos is
dominant, which is never the case in the benchmark scenarios. In such a case,
the cross-section limits obtained in the search for invisible decays of 
a neutral Higgs boson~\cite{inv} are applied to check whether these 
points are excluded or not. Any other point in a given plane 
(e.g. the (\mh, \MA) plane) is excluded if the observed $CL_s$ at that 
point is below 5\% for all sets 
of values of the parameters governing the radiative corrections that 
correspond to that point. 
The results of this extended scan are presented in 
Fig.~\ref{fig:limit_scan} in the three projections (\mh, \tbeta), 
(\MA, \tbeta) and (\mh, \MA).
The extension of the {\sc MSSM} parameter 
ranges in the scan
leads to 95\% {\sc CL} lower limits of 85~\GeVcc\ on \mh\ and 
86~\GeVcc\ on \MA, thus only about 1~\GeVcc\ below the limits obtained in the
\mbox{$ m_{\mathrm h}^{max}$} and no mixing scenarios.
%These limits are valid for all values of \tbeta\ above 0.5 and assume
%\MA\ to be greater than 20~\GeVcc.

\section{Conclusions}

The 228~\pbinv\ of data taken by DELPHI at 191.6-201.7~\GeV,
combined with our lower energy data, sets the lower limit  
at 95\% CL on the mass of the Standard Model Higgs boson at:

\[ \MH  > 107.3~\GeVcc  .\]

These data sets  also allow studies of the representative 
\mbox{$ m_{\mathrm h}^{max}$} and no mixing scenarios. The 
95\% CL limits on the masses of the lightest neutral scalar and
neutral pseudoscalar are:

\[ \mh > 85.9~\GeVcc \hspace{1cm} \MA > 86.5~\GeVcc  .\]

\noindent
for all values of \tbeta\ above 0.49 and assuming \MA $>$ 12~\GeVcc.
These limits have been proved to be robust in an extended scan
of the {\sc MSSM} parameter space.

%         Modified on 04-06-1999 by dimartino
%-------------------------------------------------------------------
\subsection*{Acknowledgements}
\vskip 3 mm
 We are greatly indebted to our technical 
collaborators, to the members of the CERN-SL Division for the excellent 
performance of the LEP collider, and to the funding agencies for their
support in building and operating the DELPHI detector.\\
We acknowledge in particular the support of \\
Austrian Federal Ministry of Science and Traffics, GZ 616.364/2-III/2a/98, \\
FNRS--FWO, Flanders Institute to encourage scientific and technological 
research in the industry (IWT), Belgium,  \\
FINEP, CNPq, CAPES, FUJB and FAPERJ, Brazil, \\
Czech Ministry of Industry and Trade, GA CR 202/96/0450 and GA AVCR A1010521,\\
Commission of the European Communities (DG XII), \\
Direction des Sciences de la Mati$\grave{\mbox{\rm e}}$re, CEA, France, \\
Bundesministerium f$\ddot{\mbox{\rm u}}$r Bildung, Wissenschaft, Forschung 
und Technologie, Germany,\\
General Secretariat for Research and Technology, Greece, \\
National Science Foundation (NWO) and Foundation for Research on Matter (FOM),
The Netherlands, \\
Norwegian Research Council,  \\
State Committee for Scientific Research, Poland, 2P03B06015, 2P03B11116 and
SPUB/P03/DZ3/99, \\
JNICT--Junta Nacional de Investiga\c{c}\~{a}o Cient\'{\i}fica 
e Tecnol$\acute{\mbox{\rm o}}$gica, Portugal, \\
Vedecka grantova agentura MS SR, Slovakia, Nr. 95/5195/134, \\
Ministry of Science and Technology of the Republic of Slovenia, \\
CICYT, Spain, AEN96--1661 and AEN96-1681,  \\
The Swedish Natural Science Research Council,      \\
Particle Physics and Astronomy Research Council, UK, \\
Department of Energy, USA, DE--FG02--94ER40817, \\
%Flanders Institute to encourage sientific and technological research in the
%industry (IWT), Belgium. \\   
%=========================================================================%

%=========================================================================%

%\newpage

%%%%%%%%%%%%%%%%%%%%%%%%%%%%%%%%%%%%%%%%%%%%%%%%%%%%%%%%%%%%%%%%

\clearpage

\begin{table}[htbp]
\begin{center}
\begin{tabular}{cccccc}     \hline
Selection & Data & Total  & \qqg  & 4 fermion & Efficiency (\%)\\
          &      & background        &       &           &    \\ 
\hline \hline
\multicolumn{6}{c} {Electron channel 25.2 \pbinv } \\ \hline
Preselection     &  152 &  158.6$\pm$ 2.5 &   111.7 &  43.6 &    81.0 \\
%lepton id        &   48 &   45.3$\pm$ 1.4 &    24.1 &  18.2 &    71.0 \\
cuts on leptons   &   15 &   9.7$\pm$ 0.6 &     2.5 &   6.2 &    65.7 \\
5C fit prob.     &    3 &    3.7$\pm$ 0.3 &    1.1  &   2.3 &    61.7 \\
final selection  &    1 &  1.19$\pm$ 0.08 &    0.05  &   1.1 &   55.0 \\ 
\hline
\multicolumn{6}{c} {Muon channel 25.9 \pbinv } \\ \hline
Preselection     &  336   & 364.8  $\pm$ 3.8  & 274.5 & 86.7 & 78.8\\
cuts on leptons  &    2   &  2.30  $\pm$ 0.12 &  0.15 & 2.15 & 72.4\\ 
final selection  &    1   &  0.93  $\pm$ 0.02 &  0.0    & 0.93 & 57.8\\
\hline
\multicolumn{6}{c} {Tau channel 25.9 \pbinv } \\ \hline
Preselection     &  1209&  1127$\pm$ 2.6&   747&   380&    98.3 \\
\llqq\           &    0 &   2.1$\pm$ 0.1&  0.09 &  2.0&    18.5 \\
final selection  &    0 &   0.76$\pm$0.08& 0.04 &  0.72&   16.7 \\
\hline
\multicolumn{6}{c} {Missing energy channel 24.9 \pbinv } \\ \hline
Anti \gaga\      & 2378 & 2368.0$\pm$ 4.5 & 1904.1 & 427.3 & 86.0 \\
Preselection     & 139  & 130.7 $\pm$ 2.0 & 81.3  & 45.7  & 73.9 \\
\like$ > 1.0$    &  11  & 12.6$\pm$ 0.6 &  7.7 & 4.9 & 59.8 \\ 
\hline
\multicolumn{6}{c} {Four-jet channel 25.9 \pbinv } \\ \hline
Preselection     & 302 & 280.6$\pm$ 2.8&   91.8&  188.8&    89.4 \\
ANN $ > 0.3$     &  16 &  19.1$\pm$ 0.5&  4.3&  14.8&    60.4 \\
\hline
\hline
\multicolumn{6}{c} {hA four-jet channel 25.9 \pbinv } \\ \hline
Preselection     & 273 & 255.1$\pm$ 2.3   &   79.7&  175.4 &    90.7 \\
\like$ > 0.1 $   &  18 & 16.8 $\pm$ 0.6&  7.7&  9.1&    85.0 \\
\hline
\end{tabular}
\caption[]{
%\it {
 Effect of the selection cuts on data, 
 simulated  background and simulated signal events at \rs~=~191.6~\GeV.
 Efficiencies are given for a signal with  \MH~=~105~\GeVcc\ 
 for the  {\sc SM} and \MA = 85 \GeVcc, \tbeta\ = 20 for the  {\sc MSSM}. 
 The quoted errors are statistical only. For each channel, the
 first line shows the integrated luminosity used; the
 last line gives the inputs for the limit derivation.}
%}
\label{ta:hzsum192}
\end{center}
\end{table}

\begin{table}[htbp]
\begin{center}
\begin{tabular}{cccccc}     \hline
Selection & Data & Total  & \qqg  & 4 fermion & Efficiency (\%)\\
          &      & background        &       &           &    \\ 
\hline \hline
\multicolumn{6}{c} {Electron channel 76.2 \pbinv } \\ \hline
Preselection     &  452   & 428.8$\pm$ 3.5 & 284.7  & 133.7 & 79.1  \\
cuts on leptons  &   33   &  29.0$\pm$ 1.3 &   7.7  &  18.5 & 64.2  \\
5C fit prob.     &   18   &  11.5$\pm$ 0.7 &   3.15 &   7.5 & 60.6  \\
final selection  &    5   &  3.88$\pm$0.18 &   0.13 &   3.5 & 57.4  \\ 
\hline
\multicolumn{6}{c} {Muon channel 76.9 \pbinv } \\ \hline
Preselection     & 1081  & 1092.5 $\pm$ 4.2  & 801.5 & 280.2 & 79.5 \\
cuts on leptons  &   3   &  7.19  $\pm$ 0.17 &  0.46 & 6.73  & 71.4 \\ 
final selection  &   2   &  3.02  $\pm$ 0.07 &  0.02 & 3.0  & 67.4 \\
\hline
\multicolumn{6}{c} {Tau channel 76.9 \pbinv } \\ \hline
Preselection     &  3479& 3215 $\pm$ 4.7&  2056&  1159&    98.3 \\
\llqq\           &    7 &  6.6$\pm$ 0.2&   0.5&  6.1&    18.5 \\
final selection  &    3 &  2.38$\pm$0.12&  0.07&  2.31&    18.3 \\
\hline
\multicolumn{6}{c} {Missing energy channel 75.0 \pbinv } \\ \hline
Anti \gaga\      & 7005 & 6757.8 $\pm$ 5.2 & 5343.0 & 1309.1 & 86.1 \\
Preselection     & 403  &  384.7 $\pm$ 2.2 & 238.8  & 134.1  & 74.4 \\
\like$ > 1.0$    &  38  & 34.0 $\pm$ 0.6 & 20.2 & 13.8 & 62.1 \\ 
\hline
\multicolumn{6}{c} {Four-jet channel 76.9 \pbinv } \\ \hline
Preselection     & 839 & 827.1$\pm$ 5.1&   260.0&  567.1&    87.5 \\
ANN $ > 0.3$   &  51 &  58.7$\pm$ 0.7&  11.3& 47.1 &    67.7 \\
\hline
\hline
\multicolumn{6}{c} {hA four-jet channel 76.9 \pbinv } \\ \hline
Preselection     & 747 & 757.3$\pm$ 4.7 &   232.3&  525.0 &    90.9 \\
\like$ > 0.1 $   &   47 & 48.0$\pm$ 0.7 &  19.6&  28.5&    86.8 \\
\hline
\end{tabular}
\caption[]{
%\it {
 As in Table~\ref{ta:hzsum192}, but for
% Effect of the selection cuts on data, simulated  background and 
% simulated signal events at 
 \rs~=~195.6~\GeV.}
% Efficiencies (in \%) are given for the signal, ie. \MH~=~105~\GeVcc\ 
% for the SM and \MA = 85 \GeVcc, \tbeta\ = 20 for the  {\sc MSSM}. 
% The quoted errors are statistical only.}
%}
\label{ta:hzsum196}
\end{center}
\end{table}

\begin{table}[htbp]
\begin{center}
\begin{tabular}{cccccc}     \hline
Selection & Data & Total  & \qqg  & 4 fermion & Efficiency (\%)\\
          &      & background        &       &           &    \\ 
\hline \hline
\multicolumn{6}{c} {Electron channel 82.8 \pbinv } \\ \hline
Preselection     &  489   & 453.3$\pm$ 3.6 & 294.5  & 148.4 & 79.3 \\
cuts on leptons  &   30   &  31.1$\pm$ 1.4 &   7.0  & 21.3 & 63.3 \\
5C fit prob.     &   11   &  12.8$\pm$1.1  &   2.6   &  8.4 & 59.5  \\
final selection  &    4   &  4.22$\pm$0.19 &   0.05 & 3.9 & 56.0 \\ 
\hline
\multicolumn{6}{c} {Muon channel 84.3 \pbinv } \\ \hline
Preselection     &  1141  & 1148.4 $\pm$ 4.3  & 807.3 & 329.8 & 78.9 \\
cuts on leptons  &    11  &  8.20  $\pm$ 0.21 & 0.54  & 7.66  & 72.1 \\ 
final selection  &    5   &  3.59  $\pm$ 0.08 & 0.02  & 3.57  & 69.7 \\
\hline
\multicolumn{6}{c} {Tau channel 84.3 \pbinv } \\ \hline
Preselection     &  3629& 3434 $\pm$ 9.7&  2152&  1282&    98.1 \\
\llqq\           &   8 &  7.7$\pm$ 0.2&   0.5&  7.2&    18.9 \\
final selection  &   3 &  2.60$\pm$0.13&  0.14&  2.46&    17.9 \\
\hline
\multicolumn{6}{c} {Missing energy channel 82.2 \pbinv } \\ \hline
Anti \gaga\      & 7211 & 7112.4$\pm$ 5.8& 5566.8 & 1450.8 & 85.5 \\
Preselection     & 421  & 425.7 $\pm$ 2.5& 260.6  & 151.8  & 75.8 \\
\like$ > 1.0$    &  38  & 40.5$\pm$ 0.7 & 24.4 & 16.1 & 64.0 \\ 
\hline
\multicolumn{6}{c} {Four-jet channel 84.3 \pbinv } \\ \hline
Preselection     & 882 & 896.8$\pm$ 2.9&   273.6&  623.3&    87.7 \\
ANN $ > 0.3$   &  61 &  65.3$\pm$ 0.8&  13.1&  52.2&    70.4 \\
\hline
\hline
\multicolumn{6}{c} {hA four-jet channel 84.3 \pbinv } \\ \hline
Preselection     & 783 & 817.9$\pm$ 2.7   &   243.1&  574.8 &    90.9 \\
\like$ > 0.1 $   &   44 &49.1$\pm$ 0.7&  18.9&  30.2&    85.5 \\
\hline
\end{tabular}
\caption[]{
%\it {
 As in Table~\ref{ta:hzsum192}, but for
% Effect of the selection cuts on data, 
% simulated  background and simulated signal events at 
 \rs~=~199.6~\GeV.}
% Efficiencies (in \%) are given for the signal, ie. \MH~=~105~\GeVcc\ 
% for the SM and \MA = 85 \GeVcc, \tbeta\ = 20 for the  {\sc MSSM}. 
% The quoted errors are statistical only.}
%}
\label{ta:hzsum200}
\end{center}
\end{table}

\begin{table}[htbp]
\begin{center}
\begin{tabular}{cccccc}     \hline
Selection & Data & Total  & \qqg  & 4 fermion & Efficiency (\%)\\
          &      & background        &       &           &    \\ 
\hline \hline
\multicolumn{6}{c} {Electron channel 40.4 \pbinv } \\ \hline
Preselection     &   232   & 214.3$\pm$ 2.0 & 136.5  & 72.7 & 79.6 \\
cuts on leptons   &   18   &  15.7$\pm$ 0.7 &  3.83  & 10.6 & 64.3 \\
5C fit prob.     &     3   & 6.47$\pm$ 0.55 &   1.51 &  4.1 & 60.2  \\
final selection  &     1   &  2.18$\pm$0.10 &   0.09 & 1.97 & 56.8 \\
\hline
\multicolumn{6}{c} {Muon channel 41.1 \pbinv } \\ \hline
Preselection     &  574 & 561.9  $\pm$ 2.9  & 391.1 & 165.2 & 81.0 \\
cuts on leptons  &    0 &  4.31  $\pm$ 0.14 & 0.28  & 4.03  & 73.5 \\ 
final selection  &    0 &  1.83  $\pm$ 0.04 &  0.0    & 1.83  & 71.0 \\
\hline
\multicolumn{6}{c} {Tau channel 41.1 \pbinv } \\ \hline
Preselection     &  1716&  1648$\pm$ 5.8&  1019&   629&    98.5 \\
\llqq\           &    0 &   3.6$\pm$ 0.1&   0.2&   3.4&    23.1 \\
final selection  &    0 &  1.17$\pm$ 0.05&  0.03&  1.14&   22.1 \\
\hline
\multicolumn{6}{c} {Missing energy channel 40.4 \pbinv } \\ \hline
Anti \gaga\      & 3305 & 3401.1 $\pm$ 3.7 & 2632.3 & 715.3 & 85.8 \\
Preselection     & 209  & 204.9  $\pm$ 1.7 & 125.2  & 73.3 & 77.8 \\
\like$ > 1.0$    &  21  & 18.6 $\pm$ 0.5 & 11.0 & 7.5   & 65.7 \\ 
\hline
\multicolumn{6}{c} {Four-jet channel 41.1 \pbinv } \\ \hline
Preselection     & 442 & 432.4$\pm$ 3.2&   129.8&  302.6&    87.0 \\
ANN $ > 0.3$     &  33 &  32.3$\pm$ 0.5&  6.5&  25.8&    69.7 \\
\hline
\hline
\multicolumn{6}{c} {hA four-jet channel 41.1 \pbinv } \\ \hline
Preselection     & 405 & 393.9$\pm$ 2.1   &   115.6&  278.3 &    90.4 \\
\like$ > 0.1 $   &  27 &  23.9$\pm$ 0.4&  8.9&  14.9&    84.6 \\
\hline
\end{tabular}
\caption[]{
%\it {
 As in Table~\ref{ta:hzsum192}, but for 
% Effect of the selection cuts on data, 
% simulated  background and simulated signal events at 
 \rs~=~201.7~\GeV.}
% Efficiencies (in \%) are given for the signal, ie. \MH~=~105~\GeVcc\ 
% for the SM and \MA = 85 \GeVcc, \tbeta\ = 20 for the  {\sc MSSM}. 
% The quoted errors are statistical only.}
%}
\label{ta:hzsum202}
\end{center}
\end{table}

%\newpage

\begin{table} [htbp]
\begin{center}
\begin{tabular}{ccccccc}  \hline
 \MH\     & Electron & Muon    & H\toto  & \toto Z & Mis. Energy & Four-jet \\
(\GeVcc)& channel & channel & channel & channel & channel & channel \\ \hline
\multicolumn{7}{c}{\rs~=~191.6~\GeV} \\ \hline 
  70.0      & 56.1 $\pm$ 1.1  & 65.0 $\pm$ 1.1
& 21.6 $\pm$ 0.9 & 22.5 $\pm$ 0.9  & 49.0 $\pm$ 1.1 & 45.1 $\pm$ 1.1  \\
  80.0      & 57.4 $\pm$ 1.1  & 70.6 $\pm$ 1.0
& 19.6 $\pm$ 0.9 & 24.0 $\pm$ 1.0  & 57.0 $\pm$ 1.1 & 57.5 $\pm$ 1.1  \\
  85.0      & 59.4 $\pm$ 0.8  & 68.8 $\pm$ 1.0
& 19.6 $\pm$ 0.9 & 23.3 $\pm$ 0.9  & 61.9 $\pm$ 1.1 & 62.3 $\pm$ 1.1  \\
  90.0      & 58.4 $\pm$ 0.8  & 69.0 $\pm$ 1.0
& 20.6 $\pm$ 0.9 & 22.8 $\pm$ 0.9  & 65.0 $\pm$ 1.1 & 69.1 $\pm$ 1.0 \\
  95.0      & 58.1 $\pm$ 0.8  & 70.0 $\pm$ 0.8 
& 19.8 $\pm$ 0.9 & 22.0 $\pm$ 0.9  & 66.5 $\pm$ 1.1 & 69.1 $\pm$ 1.0  \\
  100.0     & 55.2 $\pm$ 0.8  & 67.4 $\pm$ 1.0 
& 19.5 $\pm$ 0.9 & 19.9 $\pm$ 0.9  & 61.9 $\pm$ 1.1 & 69.4 $\pm$ 1.0 \\
  105.0     & 55.0 $\pm$ 0.8  & 57.8 $\pm$ 1.1 
& 16.7 $\pm$ 0.8 & 19.3 $\pm$ 0.9  & 59.8 $\pm$ 1.1 & 60.5 $\pm$ 1.1 \\
  110.0     & 52.5 $\pm$ 1.1  & 47.4 $\pm$ 1.1 
& 14.0 $\pm$ 0.8 & 18.4 $\pm$ 0.9  & 59.1 $\pm$ 1.1 & 53.7 $\pm$ 1.1 \\ \hline
\multicolumn{7}{c}{\rs~=~195.6~\GeV} \\ \hline 
  70.0      & 53.9 $\pm$ 1.1  & 63.3 $\pm$ 1.1
& 20.0 $\pm$ 0.9 & 23.0 $\pm$ 0.9  & 47.9 $\pm$ 1.1 & 42.7 $\pm$ 1.1  \\
  80.0      & 58.3 $\pm$ 1.1  & 68.3 $\pm$ 1.0
& 22.3 $\pm$ 0.9 & 21.8 $\pm$ 0.9  & 55.3 $\pm$ 1.1 & 56.4 $\pm$ 1.1  \\
  85.0      & 59.8 $\pm$ 1.1  & 71.1 $\pm$ 1.0
& 21.3 $\pm$ 1.3 & 24.4 $\pm$ 1.4 & 60.8 $\pm$ 1.1& 60.4 $\pm$ 1.1  \\
  90.0      & 57.9 $\pm$ 1.1 & 67.7 $\pm$ 1.1
& 20.8 $\pm$ 1.3 & 23.6 $\pm$ 1.3 & 64.2 $\pm$ 1.1 & 66.1 $\pm$ 1.0 \\
  95.0      & 59.8 $\pm$ 1.1  & 69.7 $\pm$ 1.1 
& 21.3 $\pm$ 1.3 & 23.4 $\pm$ 1.3 & 67.8 $\pm$ 1.0 & 68.5 $\pm$ 1.0  \\
  100.0     & 59.0 $\pm$ 0.9  & 71.3 $\pm$ 0.8 
& 18.2 $\pm$ 1.2 & 21.8 $\pm$ 1.3 &  65.8 $\pm$ 0.7 & 70.0  $\pm$ 0.6 \\
  105.0     & 57.4 $\pm$ 1.1  & 67.4 $\pm$ 1.1 
& 18.3 $\pm$ 1.2 & 20.2 $\pm$ 1.3 &  62.1 $\pm$ 1.1 & 67.7  $\pm$ 1.0 \\ 
  110.0     & 56.0 $\pm$ 1.1  & 55.5 $\pm$ 1.1 
& 14.4 $\pm$ 0.8 & 20.4 $\pm$ 0.9 &  55.6 $\pm$ 1.1 & 57.3  $\pm$ 1.1 \\
  115.0     & 53.5 $\pm$ 1.1  & 45.9 $\pm$ 1.1 
& 13.8 $\pm$ 0.8 & 16.9 $\pm$ 0.8 &  53.2 $\pm$ 1.1 & 52.4  $\pm$ 1.1 \\ \hline
\multicolumn{7}{c}{\rs~=~199.6~\GeV} \\ \hline 
  70.0      & 54.6 $\pm$ 1.1  & 62.5 $\pm$ 1.1
& 20.0 $\pm$ 0.9 & 23.3 $\pm$ 0.9  & 44.9 $\pm$ 1.1 & 44.2 $\pm$ 1.1  \\
  80.0      & 56.9 $\pm$ 1.1  & 68.1 $\pm$ 1.0
& 20.7 $\pm$ 0.9 & 23.6 $\pm$ 0.9  & 52.5 $\pm$ 1.1 & 55.1 $\pm$ 1.1  \\
  85.0      & 55.8 $\pm$ 1.6  & 70.2 $\pm$ 1.0
& 22.0 $\pm$ 1.5 & 23.6 $\pm$ 1.5 & 59.5 $\pm$ 1.1 & 59.5 $\pm$ 1.1  \\
  90.0      & 59.6 $\pm$ 1.1  & 69.5 $\pm$ 1.0
& 19.2 $\pm$ 1.4 & 23.9 $\pm$ 1.5 & 62.2 $\pm$ 1.1& 63.6 $\pm$ 1.0 \\
  95.0      & 59.6 $\pm$ 1.1  & 70.9 $\pm$ 1.1 
& 19.9 $\pm$ 1.4 & 21.7 $\pm$ 1.5 & 64.7 $\pm$ 1.1& 67.3 $\pm$ 1.0  \\
  100.0     & 57.9 $\pm$ 1.1  & 72.1 $\pm$ 1.0 
& 20.0 $\pm$ 1.4 & 22.1 $\pm$ 1.5 & 67.7 $\pm$ 1.0 & 69.0  $\pm$ 1.0 \\
  105.0     & 56.0 $\pm$ 0.9  & 69.7 $\pm$ 0.6 
& 17.9 $\pm$ 1.4 & 19.5 $\pm$ 1.4 & 64.0 $\pm$ 1.0 & 70.4  $\pm$ 0.6 \\ 
  110.0     & 56.1 $\pm$ 1.1  & 66.7 $\pm$ 1.0 
& 18.3 $\pm$ 1.4 & 20.7 $\pm$ 1.4 & 61.7 $\pm$ 0.8 & 68.2  $\pm$ 1.0 \\
  115.0     & 53.5 $\pm$ 1.1  & 58.3 $\pm$ 1.2 
& 14.4 $\pm$ 1.1 & 18.8 $\pm$ 1.2 & 60.4 $\pm$ 0.9 & 60.2  $\pm$ 1.5 \\ \hline
\multicolumn{7}{c}{\rs~=~201.7~\GeV} \\ \hline 
  70.0      & 53.9 $\pm$ 1.1  & 62.5 $\pm$ 1.1
& 19.6 $\pm$ 0.9 & 24.2 $\pm$ 1.0  & 41.2 $\pm$ 1.1 & 42.5 $\pm$ 1.1  \\
  80.0      & 57.1 $\pm$ 1.1  & 67.2 $\pm$ 1.0
& 19.1 $\pm$ 0.9 & 24.2 $\pm$ 1.0  & 50.5 $\pm$ 1.1 & 53.0 $\pm$ 1.1  \\
  85.0      & 61.4 $\pm$ 1.1  & 69.5 $\pm$ 1.1
& 22.0 $\pm$ 1.3 & 23.6 $\pm$ 1.3 & 59.1 $\pm$ 1.1 & 58.5 $\pm$ 1.1  \\
  90.0      & 58.6 $\pm$ 1.1  & 68.3 $\pm$ 1.1
& 23.0 $\pm$ 1.3 & 23.2 $\pm$ 1.3 & 61.9 $\pm$ 1.1 & 64.5 $\pm$ 1.1 \\
  95.0      & 60.2 $\pm$ 1.1  & 70.0 $\pm$ 1.1 
& 21.2 $\pm$ 1.3 & 21.4 $\pm$ 1.3 & 64.3 $\pm$ 1.1 & 63.1 $\pm$ 1.1  \\
  100.0     & 60.2 $\pm$ 1.1  & 67.5 $\pm$ 1.1 
& 21.7 $\pm$ 1.2 & 22.0 $\pm$ 1.3 & 66.9 $\pm$ 1.0 & 69.9 $\pm$ 1.0 \\
  105.0     & 56.8 $\pm$ 0.9  & 71.0 $\pm$ 1.0 
& 22.1 $\pm$ 1.2 & 22.8 $\pm$ 1.3 & 65.7 $\pm$ 1.1 & 69.7 $\pm$ 1.1 \\ 
  110.0     & 56.9 $\pm$ 1.1  & 69.1 $\pm$ 1.0 
& 15.0 $\pm$ 1.1 & 19.3 $\pm$ 1.2 & 62.1 $\pm$ 1.1 & 69.4 $\pm$ 1.0 \\
  115.0     & 57.3 $\pm$ 1.1  & 58.3 $\pm$ 1.2
& 14.7 $\pm$ 1.1 & 18.8 $\pm$ 1.2 & 63.4 $\pm$ 1.1 & 61.6 $\pm$ 1.1 \\ \hline
\end{tabular}
\caption[]{\ZH\ channels: 
%\it {
 efficiencies (in \%) of the selection at \rs~=~191.6-201.7~\GeV\
 as a function of the mass of the Higgs boson, for masses above
 70~\GeVcc. The quoted errors are statistical only.}
%}
\label{ta:hzeff}
\end{center}
\end{table}

\begin{table} [htbp]
\begin{center}
\begin{tabular}{ccccc}  \hline
 \MH\     & \multicolumn{4}{c}{\rs~(\GeV)} \\
(\GeVcc) & 191.6 & 195.6 & 199.6 & 201.7 \\ \hline
\multicolumn{5}{c}{Electron channel} \\ \hline 
  50.0      & 5.8 $\pm$ 0.5  & 8.5 $\pm$ 0.6
& 8.8 $\pm$ 0.6 & 8.8 $\pm$ 0.6   \\
  60.0      & 37.0 $\pm$ 1.1  & 37.0 $\pm$ 1.1
& 36.2 $\pm$ 1.1 & 38.6 $\pm$ 1.1   \\\hline
\multicolumn{5}{c}{Muon channel} \\ \hline  
  40.0      & 27.8 $\pm$ 1.0  & 25.0 $\pm$ 1.0
& 23.2 $\pm$ 0.9 & 23.4 $\pm$ 0.9  \\ 
  50.0      & 44.4 $\pm$ 1.1  & 41.6 $\pm$ 1.1
& 38.4 $\pm$ 1.1 & 37.4 $\pm$ 1.1   \\
  60.0      & 55.0 $\pm$ 1.1  & 55.2 $\pm$ 1.1
& 52.8 $\pm$ 1.1 & 49.4 $\pm$ 1.1   \\\hline
\multicolumn{5}{c}{H\toto channel} \\ \hline
  50.0      & 12.5 $\pm$ 0.7  & 11.9 $\pm$ 0.7
& 12.7 $\pm$ 0.7 & 12.6  $\pm$ 0.7   \\
  60.0      & 18.9 $\pm$ 0.9  & 18.4 $\pm$ 0.9
& 18.8 $\pm$ 0.9 & 19.1 $\pm$ 0.9   \\\hline
\multicolumn{5}{c}{\toto Z channel} \\ \hline  
  50.0      &  7.9 $\pm$ 0.6  & 6.0 $\pm$ 0.5
& 6.3 $\pm$ 0.5 & 6.0 $\pm$ 0.5  \\
  60.0      & 21.2 $\pm$ 0.9  & 17.3 $\pm$ 0.8
& 14.1 $\pm$ 0.8 & 14.1 $\pm$ 0.8   \\\hline 
\multicolumn{5}{c}{Missing Energy channel} \\ \hline
  12.0      & 17.6 $\pm$ 0.8  & 21.7 $\pm$ 0.9  
& 25.2 $\pm$ 1.0 & 18.1 $\pm$ 0.9   \\
  18.0      & 23.8 $\pm$ 0.9  & 32.8 $\pm$ 1.0
& 34.4 $\pm$ 1.1 & 22.4 $\pm$ 0.9   \\  
  24.0      & 25.3 $\pm$ 0.9  & 32.9 $\pm$ 1.0
& 33.7 $\pm$ 1.1 & 22.2 $\pm$ 0.9  \\   
  30.0      & 25.6 $\pm$ 1.0  & 32.8 $\pm$ 1.0
& 31.5 $\pm$ 1.0 & 23.5 $\pm$ 0.9   \\ 
  40.0      & 29.0 $\pm$ 1.0  & 32.0 $\pm$ 1.0
& 32.0 $\pm$ 1.0 & 23.3 $\pm$ 0.9  \\ 
  50.0      & 33.5 $\pm$ 1.0  & 35.7 $\pm$ 1.1
& 34.0 $\pm$ 1.1 & 28.5 $\pm$ 1.0  \\
  60.0      & 36.0 $\pm$ 1.1  & 41.7 $\pm$ 1.1
& 38.1 $\pm$ 1.1 & 35.5 $\pm$ 1.1   \\\hline 
\multicolumn{5}{c}{Four-jet channel} \\ \hline 
  24.0      & 6.8 $\pm$ 0.6  & 8.0 $\pm$ 0.6
& 7.9 $\pm$ 0.6 & 8.6 $\pm$ 0.7   \\   
  30.0      & 13.5 $\pm$ 0.8  & 16.2 $\pm$ 0.8
& 16.6 $\pm$ 0.8 & 17.1 $\pm$ 0.8  \\ 
  40.0      & 29.8 $\pm$ 1.0  & 29.6 $\pm$ 1.0
& 28.1 $\pm$ 1.0 & 28.9 $\pm$ 1.0  \\ 
  50.0      & 38.6 $\pm$ 1.1  & 39.6 $\pm$ 1.1
& 37.8 $\pm$ 1.1 & 38.8 $\pm$ 1.0  \\
  60.0      & 42.9 $\pm$ 1.1  & 44.0 $\pm$ 1.1
& 42.4 $\pm$ 1.1 & 42.1 $\pm$ 1.1  \\\hline
\end{tabular}
\caption[]{\ZH\ channels: 
 efficiencies (in \%) of the selection at \rs~=~191.6-201.7~\GeV\
 as a function of the mass of the Higgs boson, for masses below
 70~\GeVcc. Only efficiencies higher than 5\% are shown.
 The quoted errors are statistical only.}
\label{ta:hzeff_lowm}
\end{center}
\end{table}

\begin{table} [htbp]
\begin{center}
\begin{tabular}{cccc}  \hline
&& \Abb & \Acc \\ \hline
% \MA & \mh & Efficiency & \MA & \mh & Efficiency \\
%(\GeVcc)   & (\GeVcc)  & (\%) & (\GeVcc) & (\GeVcc)  & (\%)\\ \hline
 \MA & \mh & Efficiency & Efficiency \\
(\GeVcc)   & (\GeVcc)  & (\%) & (\%) \\ \hline
  12.0 & 30.0   & 20.8 $\pm$ 1.2 & 6.5 $\pm$ 0.8\\
  12.0 & 50.0   & 48.5 $\pm$ 1.6 & 19.2 $\pm$ 1.2\\
  12.0 & 70.0   & 56.0 $\pm$ 1.6 & 24.4 $\pm$ 1.4\\
  12.0 & 90.0   & 80.2 $\pm$ 1.3 & 37.6 $\pm$ 1.5\\
  12.0 & 105.0  & 77.3 $\pm$ 1.3 & 54.3 $\pm$ 1.6\\
  20.0 & 50.0   & 46.0 $\pm$ 1.6 & 17.4 $\pm$ 1.2\\
  20.0 & 70.0   & 57.6 $\pm$ 1.6 & 24.3 $\pm$ 1.3\\
  20.0 & 90.0   & 75.6 $\pm$ 1.4 & 36.4 $\pm$ 1.5\\
  20.0 & 105.0  & 80.1 $\pm$ 1.3 & 59.5 $\pm$ 1.5\\
  30.0 & 70.0   & 63.4 $\pm$ 1.6 & 26.0 $\pm$ 1.4\\
  30.0 & 90.0   & 72.8 $\pm$ 1.5 & 31.8 $\pm$ 1.5 \\
  30.0 & 105.0  & 81.4 $\pm$ 1.3 & 55.6 $\pm$ 1.6\\ 
  40.0 & 90.0   & 74.3 $\pm$ 1.5 & 35.0 $\pm$ 1.6\\
  40.0 & 105.0  & 80.3 $\pm$ 1.3 & 42.4 $\pm$ 1.6\\
  50.0 & 105.0  & 82.4 $\pm$ 1.3 & 47.8 $\pm$ 1.7 \\ 
\hline\end{tabular}
\caption[]{(\hAA )(\Zqq ) channels with \Abb\ or \Acc:
%\it {
  efficiencies of the selection (in~\%)
  at \rs~=~199.6~\GeV\ as a function of the masses of the A and h bosons. 
  The quoted errors are statistical only.}
%}
\label{ta:aaqqeff}
\end{center}
\end{table}

%\newpage
\begin{table} [htbp]
\begin{center}
\begin{tabular}{c|cc|cc|cc}  \hline
 & \multicolumn{2}{c|}{\tbeta\ = 2} &
 \multicolumn{2}{c|}{\tbeta\ = 20}  &
 \multicolumn{2}{c}{\tbeta\ = 50} \\ \hline
 \MA & Four-jet & Tau   & Four-jet & Tau  & 
 Four-jet & Tau\\
(\GeVcc)   & channel & channel & channel & channel &
 channel & channel \\ \hline
\multicolumn{7}{c}{\rs~=~191.6~\GeV} \\ \hline 
% 40.0  &  39.9 $\pm$ 1.1 & $<$ 5\% 
 40.0  &  39.9 $\pm$ 1.1 & -
& 21.8 $\pm$ 0.9 & -
& 14.5 $\pm$ 0.8 & - \\  
 50.0  &  58.3 $\pm$ 1.1 & -  
& 54.8 $\pm$ 1.1 & -
& 49.5 $\pm$ 1.1& - \\
 60.0  &  66.1 $\pm$ 1.0 & 7.1 $\pm$ 0.6  
& 67.3 $\pm$ 1.0 & 6.2  $\pm$ 0.5
& 63.6 $\pm$ 1.1 & 6.5  $\pm$ 0.5 \\
 70.0 &  72.4 $\pm$ 1.0 & 14.6 $\pm$ 0.8 
& 75.2 $\pm$ 1.0 & 22.7 $\pm$ 0.9 
& 73.4 $\pm$ 1.0 & 16.9 $\pm$ 0.8 \\
 80.0 &  77.1 $\pm$ 0.9 & 22.3 $\pm$ 0.9  
& 81.1 $\pm$ 0.9 & 24.3 $\pm$ 1.0 
& 77.0 $\pm$ 0.9 & 19.9 $\pm$ 0.9 \\
 85.0 &  78.8 $\pm$ 0.9 & 23.4 $\pm$ 0.9 
& 85.0 $\pm$ 0.8 & 24.1 $\pm$ 0.9
& 80.8 $\pm$ 0.9 & 20.7 $\pm$ 0.9 \\
 90.0 &  83.1 $\pm$ 0.8 & 22.6 $\pm$ 0.9 
& 82.8 $\pm$ 0.8 & 22.1 $\pm$ 0.9
& 79.0 $\pm$ 0.9 & 19.7 $\pm$ 0.9\\
 95.0 &  82.2 $\pm$ 0.8 & 20.6 $\pm$ 0.9   
& 78.5 $\pm$ 0.9 & 19.0 $\pm$ 0.9
& 77.4 $\pm$ 1.0 & 15.4 $\pm$ 0.8\\\hline
\multicolumn{7}{c}{\rs~=~195.6~\GeV} \\ \hline
  40.0 & 38.4 $\pm$ 1.1 & -  
& 17.7 $\pm$ 0.8 & -
& 13.6 $\pm$ 0.8 & -\\
  50.0 & 58.0 $\pm$ 1.1 & -  
& 55.4 $\pm$ 1.1 & -
& 47.1 $\pm$ 1.1 & -\\
  60.0 & 67.0 $\pm$ 1.0 & 7.0 $\pm$ 0.6    
& 66.9 $\pm$ 1.0 & 7.2 $\pm$ 0.6
& 62.0 $\pm$ 1.1 & 6.0 $\pm$ 0.5\\
  70.0 & 71.1 $\pm$ 1.1 & 13.4 $\pm$ 0.8
& 72.6 $\pm$ 1.0 & 21.4 $\pm$ 0.9
& 71.3 $\pm$ 1.0 & 17.5 $\pm$ 0.8\\ 
  80.0 & 81.0 $\pm$ 0.8 & 15.9 $\pm$ 1.1  
& 83.9 $\pm$ 0.8 & 20.3 $\pm$ 1.3
& 77.4 $\pm$ 0.9 & 23.7 $\pm$ 0.9\\
  85.0 & 81.9 $\pm$ 0.8 & 22.2 $\pm$ 1.3 
& 86.8 $\pm$ 0.8 & 21.8 $\pm$ 1.3
& 80.9 $\pm$ 0.9 & 21.1 $\pm$ 0.9\\
  90.0 & 81.5 $\pm$ 0.8 & 25.3 $\pm$ 1.4
& 87.0 $\pm$ 0.6 & 22.1 $\pm$ 1.3
& 82.0 $\pm$ 0.9 & 20.1 $\pm$ 0.9\\
  95.0 & 82.6 $\pm$ 0.8 & 20.6 $\pm$ 0.9  
& 82.5 $\pm$ 0.8 & 19.0 $\pm$ 0.9
& 81.2 $\pm$ 0.9 & 17.6 $\pm$ 0.8 \\ \hline
\multicolumn{7}{c}{\rs~=~199.6~\GeV} \\ \hline
  40.0 & 38.6 $\pm$ 1.1 & -  
& 14.5 $\pm$ 0.8 & -
& 11.2 $\pm$ 0.7 & - \\
  50.0 & 58.3 $\pm$ 1.1 & -  
& 54.8 $\pm$ 1.1 & -
& 47.7 $\pm$ 1.1 & - \\
  60.0 & 67.0 $\pm$ 1.0 & 5.8 $\pm$ 0.5 
& 66.9 $\pm$ 1.0 & 6.8 $\pm$ 0.6
& 62.0 $\pm$ 1.1 & 5.5 $\pm$ 0.5 \\
  70.0 & 71.1 $\pm$ 1.1 & 11.4 $\pm$ 0.7
& 72.6 $\pm$ 1.0 & 19.1 $\pm$ 0.9
& 71.3 $\pm$ 1.0 & 15.1 $\pm$ 0.8\\  
  80.0 & 78.4 $\pm$ 0.9 & 15.9 $\pm$ 1.1  
& 82.2 $\pm$ 0.9 & 22.5 $\pm$ 1.3
& 77.4 $\pm$ 0.9 & 20.1 $\pm$ 0.9\\
  85.0 & 80.2 $\pm$ 0.8 & 20.8 $\pm$ 1.3 
& 85.5 $\pm$ 0.8 & 25.6 $\pm$ 1.4
& 80.9 $\pm$ 0.9 & 20.6 $\pm$ 0.9\\
  90.0 & 83.5 $\pm$ 0.8 & 24.4 $\pm$ 1.4 
& 85.6 $\pm$ 0.8 & 21.6 $\pm$ 1.3
& 82.0 $\pm$ 0.9 & 21.1 $\pm$ 0.9\\
  95.0 & 84.5 $\pm$ 0.7 & 20.6 $\pm$ 1.3 
& 83.5 $\pm$ 0.8 & 22.7 $\pm$ 1.3
& 81.2 $\pm$ 0.9 & 18.9 $\pm$ 0.9 \\ \hline
\multicolumn{7}{c}{\rs~=~201.7~\GeV} \\ \hline 
  40.0 & 33.3 $\pm$ 1.0 & -  
& 13.6 $\pm$ 0.8 & -
& 10.5 $\pm$ 0.7 & - \\ 
  50.0 & 60.1 $\pm$ 1.1 & -  
& 54.6 $\pm$ 1.1 & -
& 47.1 $\pm$ 1.1 & - \\ 
  60.0 & 67.0 $\pm$ 1.0 & 5.8 $\pm$ 0.5   
& 66.9 $\pm$ 1.0 & 6.7 $\pm$ 0.6
& 62.0 $\pm$ 1.1 & 5.8 $\pm$ 0.5 \\
  70.0 & 71.1 $\pm$ 1.1 & 11.2 $\pm$ 0.7
& 72.6 $\pm$ 1.0 & 17.0 $\pm$ 0.8
& 71.3 $\pm$ 1.0 & 15.4 $\pm$ 0.8\\  
  80.0 & 77.0 $\pm$ 0.9 & 15.9 $\pm$ 1.1  
& 80.4 $\pm$ 0.9 & 22.5 $\pm$ 1.3
& 77.4 $\pm$ 0.9 & 20.5 $\pm$ 0.9\\
  85.0 & 80.2 $\pm$ 0.8 & 20.8 $\pm$ 1.3 
& 84.6 $\pm$ 0.8 & 25.6 $\pm$ 1.4
& 80.9 $\pm$ 0.9 & 22.4 $\pm$ 0.9\\
  90.0 & 82.5 $\pm$ 0.8 & 24.4 $\pm$ 1.4 
& 86.8 $\pm$ 0.8 & 21.6 $\pm$ 1.3
& 82.0 $\pm$ 0.9 & 20.8 $\pm$ 0.9\\
  95.0 & 85.8 $\pm$ 0.8 & 20.6 $\pm$ 1.3 
& 84.1 $\pm$ 0.8 & 22.7 $\pm$ 1.3 
& 81.2 $\pm$ 0.9 & 18.4 $\pm$ 0.9\\ \hline
\end{tabular}
\caption[]{\hA\ channels: 
%\it {
  efficiencies of the selection (in \%) 
  at \rs~=~191.6-201.7~\GeV\ as a function of the mass of the A boson and 
  \tbeta. Only efficiencies higher than 5\% are shown. 
  The quoted errors are statistical only.}
%}
\label{ta:haeff}
\end{center}
\end{table}

\begin{table} [htbp]
\begin{center}
\begin{tabular}{ccc|ccc}  \hline
 \MA & \mh & Efficiency & \MA & \mh & Efficiency \\
(\GeVcc)   & (\GeVcc)  & (\%) & (\GeVcc) & (\GeVcc)  & (\%)\\ \hline
  12.0 & 70.0   & 27.2 $\pm$ 1.4 & 30.0 & 130.0 & 67.7 $\pm$ 1.5 \\
  12.0 & 90.0   & 35.8 $\pm$ 1.5 & 30.0 & 150.0 & 60.1 $\pm$ 1.5 \\
  12.0 & 110.0  & 59.8 $\pm$ 1.5 & 50.0 & 70.0  & 66.2 $\pm$ 1.5 \\
  12.0 & 130.0  & 63.4 $\pm$ 1.5 & 50.0 & 90.0  & 78.4 $\pm$ 1.3 \\
  12.0 & 150.0  & 52.0 $\pm$ 1.6 & 50.0 & 110.0 & 80.5 $\pm$ 1.3 \\
  12.0 & 170.0  & 32.4 $\pm$ 1.5 & 50.0 & 130.0 & 75.5 $\pm$ 1.4 \\
  30.0 & 50.0   & 19.9 $\pm$ 1.3 & 70.0 & 90.0  & 80.4 $\pm$ 1.3 \\
  30.0 & 70.0   & 54.7 $\pm$ 1.6 & 70.0 & 110.0 & 81.8 $\pm$ 1.2 \\
  30.0 & 90.0   & 65.2 $\pm$ 1.5 & 80.0 & 90.0  & 87.2 $\pm$ 1.1 \\
  30.0 & 110.0  & 68.9 $\pm$ 1.5 & 85.0 & 95.0  & 83.5 $\pm$ 1.2 \\
% orig. table with (almost) equal masses - removed here (kept in the inputs!)
%  12.0 & 70.0   & 27.2 $\pm$ 1.4 & 50.0 & 70.0  & 66.2 $\pm$ 1.5 \\
%  12.0 & 90.0   & 35.8 $\pm$ 1.5 & 50.0 & 90.0  & 78.4 $\pm$ 1.3\\
%  12.0 & 110.0  & 59.8 $\pm$ 1.5 & 50.0 & 110.0 & 80.5 $\pm$ 1.3\\
%  12.0 & 130.0  & 63.6 $\pm$ 1.5 & 50.0 & 130.0 & 75.5 $\pm$ 1.4\\
%  12.0 & 150.0  & 52.0 $\pm$ 1.6 & 68.0 & 75.0  & 78.3 $\pm$ 1.3 \\
%  12.0 & 170.0  & 32.4 $\pm$ 1.5 & 70.0 & 70.0  & 73.9 $\pm$ 1.4 \\
%  30.0 & 50.0   & 19.9 $\pm$ 1.3 & 70.0 & 90.0  & 80.4 $\pm$ 1.3\\
%  30.0 & 70.0   & 54.7 $\pm$ 1.6 & 70.0 & 110.0 & 81.8 $\pm$ 1.2\\
%  30.0 & 90.0   & 65.2 $\pm$ 1.5 & 72.0 & 80.0  & 78.9 $\pm$ 1.3\\
%  30.0 & 110.0  & 69.0 $\pm$ 1.5 & 77.0 & 85.0  & 80.9 $\pm$ 1.2 \\
%  30.0 & 130.0  & 67.7 $\pm$ 1.5 & 80.0 & 90.0  & 87.2 $\pm$ 1.1 \\
%  30.0 & 150.0  & 60.1 $\pm$ 1.5 & 85.0 & 95.0  & 83.5 $\pm$ 1.2\\
%  50.0 & 50.0   & 59.6 $\pm$ 1.5 & 90.0 & 90.0  & 85.6 $\pm$ 1.1\\ 
\hline\end{tabular}
\caption[]{\hA\ four-jet channel : 
%\it {
  efficiencies of the selection (in \%) 
  at \rs~=~199.6~\GeV\ as a function of the masses of the A and h bosons,
  from simulated samples corresponding
  to large mass differences between the two bosons. 
  The quoted errors are statistical only.}
%}
\label{ta:scaneff}
\end{center}
\end{table}

\begin{table}[htbp]
\begin{center}
\begin{tabular}{cccc|ccc}     \hline
\rs\    & \multicolumn{3}{c|} {\ZH\ channel} & 
          \multicolumn{3}{c} {\hA\ channel} \\
(\GeV)  & Data & Background  & Signal & Data & Background  & Signal \\
\hline
191.6 & 1  & 4.8$\pm$0.3 & 0.16$\pm$0.01 & 3 & 1.1$\pm$0.2 & 0.56$\pm$0.02 \\
195.6 & 13 & 12.4$\pm$0.3 & 1.18$\pm$0.02 & 1 & 2.9$\pm$0.2 & 2.00$\pm$0.04\\
199.6 & 13 & 12.9$\pm$0.3 & 4.70$\pm$0.06 & 3 & 2.9$\pm$0.2 & 2.31$\pm$0.05\\
201.7 & 5  &  6.9$\pm$0.2 & 2.87$\pm$0.06 & 1 & 1.5$\pm$0.1 & 1.24$\pm$0.02 \\
\hline
\end{tabular}
\caption[]{
%\it {
  Observed and expected rates after tight selections applied to data
  at \rs~=~191.6-201.7~\GeV. 
  Signal expectations are given for a signal with  \MH~=~105~\GeVcc\ 
  for the  {\sc SM} and \MA = 85 \GeVcc, \tbeta\ = 20 for the  {\sc MSSM}.
  The quoted errors 
%combine statistical and systematic uncertainties quadratically.}
  are statistical only.}
%}
\label{ta:summ}
\end{center}
\end{table}

\begin{table}[htbp]
\begin{center}
\begin{tabular}{c|cccccc}     \hline
scenario & \mtop\   & $M_{susy}$ & $M_2$ & $m_{\tilde{g}}$ & $\mu$ 
         & $X_t = A - \mu \cot \beta$ \\
& (\GeVcc) & (\GeVcc) & (\GeVcc) & (\GeVcc) & (\GeVcc) & (\GeVcc)\\
\hline
\mbox{$ m_{\mathrm h}^{max}$} scenario &
     174.3 & 1000 & 200 & 800 & -200 & 2 $M_{susy}$ \\
no mixing &
     174.3 & 1000 & 200 & 800 & -200 & 0 \\
large $\mu$ &
     174.3 & 400 & 400 & 200 & 1000 & -300 \\
\hline
\end{tabular}
\caption[]{
%\it {
Values of the underlying parameters for the three 
representative {\sc MSSM} scenarios scanned in this paper.}
%}
\label{ta:benchmarks}
\end{center}
\end{table}

\begin{table}[htb]
\begin{center}
\begin{tabular}{c|c|c|c|c|c|c}
\hline
Parameter & $m_{\mathrm{A}}$  & \tbeta\ & $M_{susy}$
&  $M_2$ & $\mu$ & $A$/$M_{susy}$ \\
& (\GeVcc) & & (\GeVcc) & (\GeVcc) & (\GeVcc) &\\
\hline
Range     & 20\,:\,1000 & $0.5$\,:\,50 & 200\,:\,1000 &
200\,:\,1000
& $-500$\,:\,+500 & $-2$\,:\,$+2$ \\
\hline
\end{tabular}
\end{center}
\caption[]{
Ranges of variation of the underlying parameters used in the extended
scan of the {\sc MSSM} parameter space described in this paper.}
\label{tab:scan}
\end{table}

\newpage

%%%%%%%%%%%%%%%%%%%%%%%%%%%%%%%%%%%%%%%%%%%%%%%%%%%%%%%%%%%%%%%%
%%%%%%%%%%%%% btagging figures

\clearpage
%%%%%%%%%%%%% ee qq figures

\clearpage
%%%%%%%%%%%%% mumu qq figures

\clearpage
%%%%%%%%%%%%%% tautau qq figures

%%%%%%%%%%%%%% hnunu figures

%%%%%%%%%%%%%% Four-jet figures

\clearpage

\clearpage

%%%%%%%%%%%%%%%%%%%%% Mass plots figures

%%%%%%%%%%%%%%%%%%%%%% Limits figures

%%%%%%%%%%%%%%% MSSM limits figures

\clearpage

\clearpage

\end{document}